\documentclass[manuscript]{aa}

\usepackage{graphicx}
\usepackage[intlimits,sumlimits]{amsmath}
\usepackage{deluxetable}
\usepackage{times,epsfig}
\usepackage{amssymb}
\usepackage{mathrsfs}
\usepackage{natbib}
\bibpunct{(}{)}{;}{a}{}{,} 
\usepackage{amsmath}
\usepackage[english]{babel}
\usepackage{longtable}
\usepackage{url}
\usepackage{epstopdf}
\usepackage{hyperref}

\begin{document}

\title{Metallicity at the explosion sites of interacting transients\thanks{Based on observations performed at the Nordic Optical Telescope (Proposal numbers: P45-004, P49-016; PI: F. Taddia), La Palma, Spain.}}

\author{F. Taddia\inst{1}
\and J. Sollerman\inst{1}
\and C. Fremling\inst{1}
\and A. Pastorello\inst{2}
\and G. Leloudas\inst{3,4}
\and C. Fransson\inst{1}
\and A. Nyholm\inst{1}
\and M.~D. Stritzinger\inst{5}
\and M. Ergon\inst{1}
\and R. Roy\inst{1}
\and K. Migotto\inst{1}}

\institute{The Oskar Klein Centre, Department of Astronomy, Stockholm University, AlbaNova, 10691 Stockholm, Sweden.\\ (\email{francesco.taddia@astro.su.se})
\and INAF-Osservatorio Astronomico di Padova, Vicolo dell'Osservatorio 5, 35122 Padova, Italy.
\and Department of Particle Physics \& Astrophysics, Weizmann Institute of Science, Rehovot 76100, Israel.
\and Dark Cosmology Centre, Niels Bohr Institute, University of Copenhagen, Juliane Maries Vej 30, 2100 Copenhagen, Denmark.
\and Department of Physics and Astronomy, Aarhus University, Ny Munkegade 120, 8000 Aarhus C, Denmark.}
\date{Received / Accepted}

\abstract
{ Some circumstellar-interacting (CSI) supernovae (SNe) are produced by
the explosions of massive stars that have lost mass shortly before the SN explosion. There is evidence
that the precursors of some SNe~IIn were luminous blue variable (LBV) stars. For a small number of CSI~SNe, outbursts have been observed before the SN explosion. Eruptive events of massive stars are named as SN impostors (SN~IMs) and whether they herald a forthcoming SN or not is still unclear. The large variety of observational properties of CSI~SNe suggests the existence of other progenitors, such as red supergiant (RSG) stars with superwinds. Furthermore, the role of metallicity in the mass loss of CSI~SN progenitors is still largely unexplored.}
{Our goal is to gain insight on the nature of the progenitor stars of CSI~SNe by studying their environments, in particular the metallicity at their locations.}
{We obtain metallicity measurements at the location of 60 transients (including SNe~IIn, SNe~Ibn, and SN~IMs), via emission-line diagnostic on optical spectra obtained at the Nordic Optical Telescope and through public archives. Metallicity values from the literature complement our sample. We compare the metallicity distributions among the different CSI~SN subtypes and to those of other core-collapse SN types. We also search for possible correlations between metallicity and CSI~SN observational properties.} 
{We find that SN~IMs tend to occur in environments with lower metallicity than those of SNe~IIn. Among SNe~IIn, SN~IIn-L(1998S-like) SNe show higher metallicities, similar to those of SNe~IIL/P, whereas long-lasting SNe~IIn (1988Z-like) show lower metallicities, similar to those of SN~IMs. The metallicity distribution of SNe~IIn can be reproduced by combining the metallicity distributions of SN~IMs (that may be produced by major outbursts of massive stars like LBVs) and SNe~IIP (produced by RSGs). The same applies to the distributions of the Normalized Cumulative Rank (NCR) values, which quantifies the SN association to \ion{H}{ii} regions. For SNe~IIn, we find larger mass-loss rates and higher CSM velocities at higher metallicities. The luminosity increment in the optical bands during SN~IM outbursts tend to be larger at higher metallicity, whereas the SN~IM quiescent optical luminosities tend to be lower.}
{The difference in metallicity between SNe~IIn and SN~IMs suggests that LBVs are only one of the progenitor channels for SNe~IIn, with 1988Z-like and 1998S-like SNe possibly arising from LBVs and RSGs, respectively. Finally, even though line-driven winds likely do not primarily drive the late mass-loss of CSI~SN progenitors, metallicity has some impact on the observational properties of these transients.}


\keywords{supernovae: general -- stars: evolution -- galaxies: abundances} 

\maketitle

\section{Introduction}
\label{sec:intro}


The study of the environment of supernovae (SNe) has become crucial to understand the link between different SN classes and their progenitor stars. Mass and metallicity are among the progenitor properties that can be investigated, and are fundamental to understand stellar evolution and explosions \citep[see][for a review]{anderson15}.
The study of SN host galaxy metallicity is now a popular line of investigation in the SN field. Metallicity can be obtained as a global measurement for a SN host galaxy \citep[e.g.][]{prieto08}, following the known luminosity-metallicity or color-luminosity-metallicity relations (e.g., \citealp{tremonti04}, \citealp{sanders13}).  It can also be estimated via strong line diagnostics when spectra of the host galaxies are obtained. In particular, it has been shown \citep{thoene09,anderson10,modjaz11,leloudas11,kelly12,sanders12,kunca13a,kunca13b,taddia13met,kelly14} 
that metallicity measurements at the exact SN location provide the most reliable estimates of the SN metal content. This is because galaxies are characterized by metallicity gradients \citep[e.g.,][hereafter P04]{pilyugin04} and small-scale ($<$kpc) variations \citep[e.g.,][]{niino15}.

In this work we study the environments and in particular the metallicity of supernovae that interact with their circumstellar medium (CSM). 
A large variety of these transients has been observed in the past years, and in the following we briefly describe their physics, their observational properties and the 
possible progenitor scenarios, and how the study of their environments can help us to constrain the nature of their precursors.

\subsection{CSI~SNe: subclassification and progenitor scenarios}

When the rapidly expanding SN ejecta reach the surrounding CSM a forward shock forms which propagates through the CSM, while a reverse shock travels backward into the SN ejecta \citep[see e.g.,][]{chevalier94}. The high energy (X-ray) photons produced in 
the shock region ionize the 
unshocked CSM, 
giving rise to the characteristic narrow 
(full-width-at-half-maximum FWHM$\sim$100~km~s$^{-1}$) 
emission lines of Type IIn SNe 
(SNe~IIn) 
\citep{schlegel90}. Their emission lines are furthermore 
often characterized by additional components: a broad base 
(FWHM$\sim$10000~km~s$^{-1}$), produced by the ionized ejecta; 
and an intermediate 
(FWHM$\sim$1000~km~s$^{-1}$) feature, which originates in the cold-
dense shell (CDS) between the forward and the 
reverse shocks. The kinetic energy of the ejecta is transformed into 
radiation, powering the luminosity of CSM-interacting (CSI) SNe.

\begin{figure}
 \centering
\includegraphics[width=9.5cm]{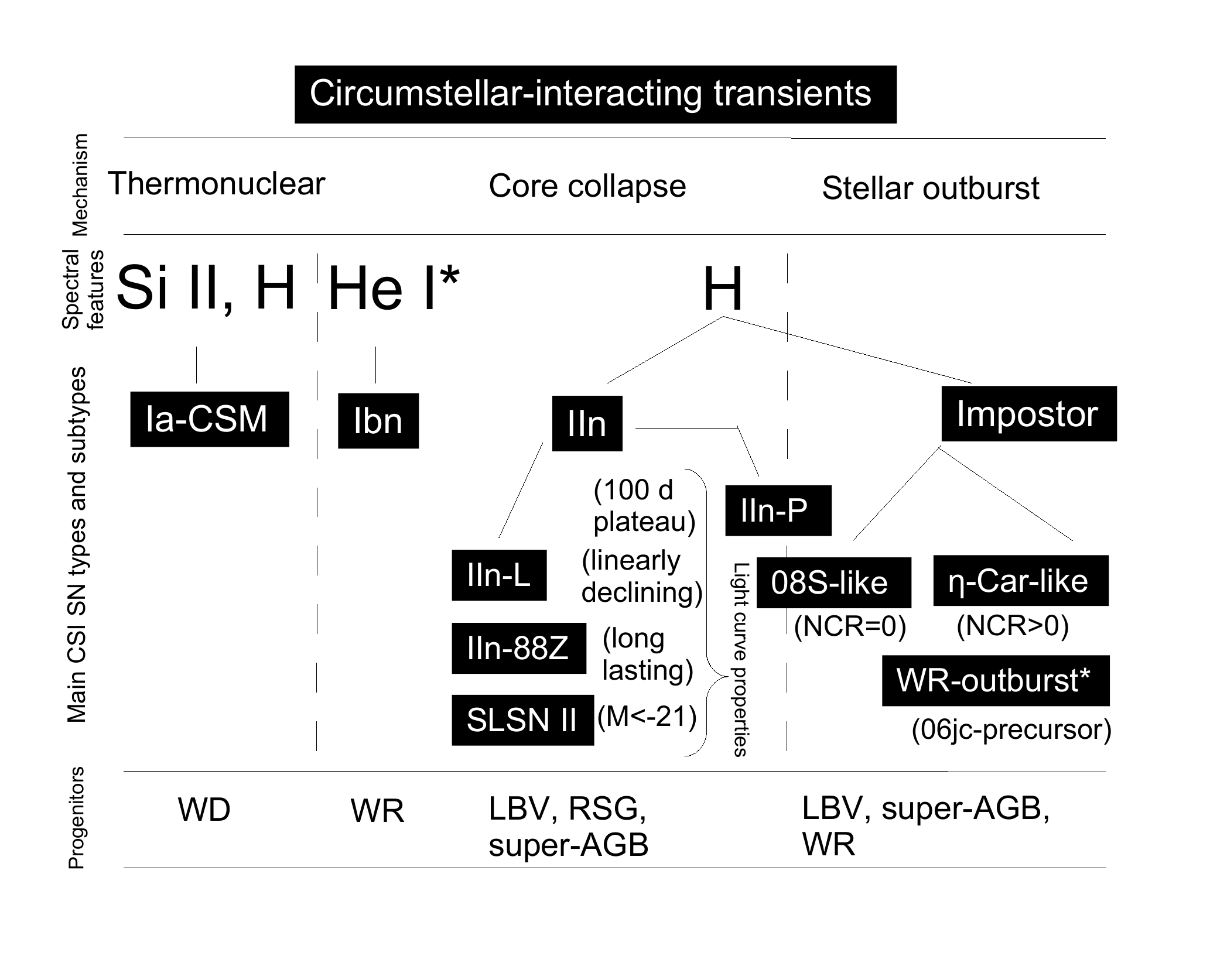} 
  \caption{CSI SN classification scheme. The SN explosion mechanisms, the main elements in the spectra, the light curve properties (only for SN~IIn subtypes), the degree of association (NCR) to \ion{H}{ii} regions (only for SN~IMs) and the possible progenitors are shown in correspondence with their CSI SN subtype. \label{scheme}}
 \end{figure}

SNe~IIn served as the first class of SN identified to be mainly powered by CSM interaction. Their CSM is H-rich, 
as revealed by  their prominent Balmer emission lines. 
Within the SN~IIn family, a wide 
range of observational properties 
is observed \citep{kiewe12,taddia13IIn}.
Their optical light 
curves can decline sharply
by several magnitudes (e.g., 
SN~1994W; \citealp{sollerman98}) or 
settle onto a plateau lasting for years (e.g., SN~1988Z, \citealp{turatto93}; SN~2005ip, \citealp{fox09}, \citealp{smith09_05ip}, \citealp{stritzinger12}, hereafter S12). 

A multiplicity of CSM geometries and densities, as well as a wide range of ejecta masses and kinetic energies \citep[e.g.,][]{moriya14_peakrise} can explain this large variety of observational properties. This multiplicity may also suggest the existence of multiple progenitor channels for SNe~IIn and other CSI~SNe.

 The existence of subclasses within the large SN~IIn group has recently been proposed (\citealp{taddia13IIn,habergham14}, hereafter H14). In the following paragraphs we consider SN~IIn subtypes mainly based on the light curve properties, whereas other CSI SN types are defined based on their spectra.
A scheme summarizing the different CSI~SN types is shown in Fig.~\ref{scheme}.

A group of long-lasting SNe~IIn including SNe~1988Z, 1995N, 
2005ip, 2006jd and 2006qq was described in \citet{taddia13IIn}. These are SNe
whose light curves exhibit a slow decline rate of $\lesssim$0.7~mag~(100~days)$^{-1}$ over a period of time $\gtrsim$ 150 days, or in some cases for several years. Their progenitors have been suggested to be Luminous Blue Variable stars (LBVs, see e.g. SN~2010jl, \citealp{fransson14}). 

The LBV phase is traditionally interpreted as a transitional phase between an O-type star and a Wolf-Rayet (WR) star. According to standard stellar evolution models, a massive star should not end its life in the LBV phase. Recently, \citet{smith15asocial} investigated the association of known LBVs to O-type stars (which spend their short lives clustered in star-forming regions). They found that LBVs are even more isolated than WR stars, and thus conclude that LBVs are not single stars in a transitional phase between O-type and WR stars, but rather mass gainers in binary systems. In this scenario, the LBVs would gain mass from their massive companion stars which would then explode in a SN event that would kick the LBVs themselves far from their birth locations. When H14 and \citet{anderson12} compared the association of SNe~IIn to bright \ion{H}{ii} regions to that of the main CC SN types, it emerged that SNe~IIn display a lower degree of association. As the degree of association to \ion{H}{ii} regions can be interpreted in terms of progenitor mass (see Sect.~\ref{sec:ncr}), this result might suggest a low initial mass for the progenitors of SNe~IIn, which apparently would be in conflict with a massive LBV progenitor scenario. The mechanism proposed by \citet{smith15asocial} reconciles the facts that LBVs are progenitors of at least a fraction of SNe~IIn and at the same time they are found to have a weak positional association to bright \ion{H}{ii} region and O-type stars.

There is indeed evidence that at least some SNe~IIn arise from LBVs. \citet{galyam07} identified a likely LBV in pre-explosion images at the location of SN~IIn~2005gl. SN~2009ip exhibited LBV-like outbursts before the last major event which might be explained with its CC \citep{prieto13,mauerhan13_09ip,smith14_09ip}. For several SNe~IIn from the Palomar Transient Factory
(PTF) there were signs of stellar outbursts before their terminal endpoint \citep{ofek14}. Furthermore, the CSM velocities observed in SNe~IIn are often consistent with those expected for LBV winds ($\sim$100--1000~km~s$^{-1}$), and the large mass-loss rates observed in SNe~IIn are also compatible with 
those of large LBV eruptions, 
ranging between 10$^{-3}$ and 1~M$_{\odot}$~yr$^{-1}$ \citep{kiewe12,taddia13IIn,moriya14}.  
The presence of bumps in the light curves of some SNe~IIn 
can also be understood as the interaction with dense shells produced by episodic mass-loss events (e.g., SN~2006jd, S12).
 In summary, there is thus ample observational evidence from different lines of reasoning that connect LBVs and SNe~IIn.

However, also Red Supergiant stars (RSGs) with superwinds (e.g., VY Canis Majoris) have been suggested to be progenitors of long-lasting SNe~IIn (see e.g., SN~1995N, \citealp{fransson02}, \citealp{smith09_RSG}), since they are characterized by strong mass loss forming the CSM needed to explain the prolonged interaction. 

There are other SNe~IIn which fade faster than the aforementioned 1988Z-like SNe. These have
early decline rates of $\sim$2.9~mag~(100~days)$^{-1}$, and
given the linear decline in their light curves, these events can be labelled Type IIn-L SNe. 

Among the SNe~IIn-L, it is possible to make further subclassifcations based on their spectra. Some SNe~IIn-L show slowly evolving spectra (e.g., SNe~1999el, \citealp{dicarlo02}; 1999eb, \citealp{pastorello02}), with narrow emission lines on top of broad bases lasting for months. 
Other SNe~IIn-L display spectra with a faster evolution, with Type IIn-like spectra at early times and Type IIL-like spectra (broad emission lines, sometimes broad P-Cygni absorption) at later epochs (e.g., SN~1998S, \citealp{fassia00,fassia01}; and SN~1996L, \citealp{benetti06}).  The different spectral evolution of these two subtypes can be traced back to the efficiency of the SN-CSM interaction and to different wind properties. Also for SNe~IIn-L, both LBV \citep[e.g.,][]{kiewe12} and RSG \citep[e.g.,][]{fransson05} progenitors have been considered in the literature. 

We note that both LBVs and RSGs could be responsible for the production of CSI~SNe when they are part of binary systems. 
We have already discussed the mechanism proposed by \citet{smith15asocial} for LBVs. \citet{mackey14} 
explain how interacting supernovae can come from RSGs in binary systems. In their scenario, the binary companion of a RSG can photoionize and confine up to 35\% of the gas lost by the RSG during its life close to the RSG star itself, forming a dense CSM. When the RSG explodes, the SN will appear as a CSI SNe as the SN ejecta will interact with this dense shell.

\citet{mauerhan13} suggested 
the name ``IIn-P" for objects resembling SN~1994W, i.e., SNe with Type~IIn spectra 
showing a $\sim$100~days plateau in the light curve, similar to that of SNe~IIP, followed by a sharp drop and a linear declining tail at low luminosity.
The nature of these events is currently debated; some authors have proposed that 1994W-like SNe arise from the collision of shells ejected by the progenitor star and not from its terminal explosion \citep[e.g.,][]{humphreys12,dessart09}. However, for SN 2009kn a bona-fide core-collapse (CC) origin was favoured by the observations of 
 \citet{kankare12}. Others even suggested SN~1994W was the result of a fall-back SN creating a black hole \citep{sollerman02} and it has also been proposed that SNe~IIn-P might be electron-capture (EC) SNe \citep{nomoto84} coming from the explosion of super asymptotic-giant-branch (AGB) stars \citep[e.g.,][]{chugai04_94W,mauerhan13}. 

An additional member of the family of SNe~IIn are the so-called superluminous SNe II (SLSNe~II). These objects reach peak absolute magnitude of $<-$21~mag  \citep{galyam12}, and SN~2006gy serves as the prototypical example \citep{smith07_06gy}. We note that the mechanism powering these bright transients could be something different from CSM interaction, such as radioactive decay of large amounts of $^{56}$Ni or energy from a magnetar. We do not focus on the host galaxies of SLSNe~II in this paper, with the exception of SN~2003ma, which shows a light-curve shape similar to that of SN~1988Z but brighter by $\sim$2.5~mag \citep{rest11}.

Besides SNe~IIn, in the CSI~SN group we also find the so-called SNe~Ibn, 
or 2006jc-like SNe \citep[e.g., ][]{matheson00,foley07,pastorello07, pastorello08, pastorello08_05la, smith08_06jc, gorbikov13}. These are transients showing He 
emission lines, arising from the SN interaction with a He-rich CSM. 
These SNe likely originate from the explosion of massive Wolf-Rayet (WR) stars. The WR progenitor SN~2006jc was observed to outburst two years before the SN explosion.

Another class of objects arising from CSM interaction is the Type~Ia-CSM 
or 2002ic-like subgroup. These events are interpreted as thermonuclear 
SNe interacting with H-rich CSM 
\citep{hamuy03,aldering06,dilday12,taddia12,silverman13,fox15}, 
although a CC origin has also been proposed 
\citep{benetti06,inserra14}. Their spectra are well represented by the sum of narrow Balmer emission lines and SN~Ia spectra diluted by a blue continuum \citep{leloudas15}.

Finally, a class of transients resembling SNe~IIn is that of the SN impostors (SN~IMs). These events are the result of outbursts from massive stars. Most of the information regarding the observables of SN~IMs has been collected by \citet[][see their table 9]{smith11}. In \citet{smith11} SN~IMs are interpreted as the result of LBV eruptions. Some LBVs may be observed in the S-Doradus variability phase (e.g.
the 2009 optical transient in UGC~2773, \citealp{smith10_impo,smith11}). LBVs in the S-Doradus phase are typically characterized by a variability of 1--2 visual magnitudes, usually interpreted as the result of a temperature variation at constant bolometric luminosity \citep{humphreys94}. 
Occasionally even giant eruptions of LBVs similar to that observed in Eta-Carinae during the 19th century may be observed in external galaxies (e.g. SN~2000ch, \citealp{wagner04,pastorello10}). Their typical luminosities are lower than those of SNe~IIn (M $>-$14), even though some bright events are now suspected to belong to this class (e.g. SN~2009ip, \citealp{pastorello13,fraser13,margutti14}).  Among SN~impostors, objects like SN~2008S have been proposed to be electron-capture SNe from super-AGB stars rather than LBV outbursts (e.g., \citealp{botticella09}; see Sect.~\ref{sec:ncr}).

\subsection{This work}

A possible approach to investigate if the LBVs, 
which produce (at least some) SN~IMs, are the dominant 
progenitor channel for SNe~IIn is to compare the properties of 
the environments for SNe~IIn and SN~IMs. 
In particular the metallicities of the surrounding material can 
be measured via strong emission-line diagnostics. This is the focus of this paper.

If the vast majority of SNe~IIn originate from 
LBVs associated with SN~IMs, then the metallicity distributions of these two 
groups should be similar. Else, if these 
distributions do not match, there could be room for other progenitor channels 
of SNe~IIn. Metallicity measurements of SNe~IIn and SN~IMs based on the host-galaxy absolute magnitudes are provided by H14.  In this paper we provide local metallicity estimates for a large sample of CSI transients, including 
SNe~IIn, Ibn, Ia-CSM and SN~IMs. Our measurements, carried out on data obtained at the Nordic 
Optical Telescope (NOT), are complemented with data available in the literature (e.g., \citealp{kelly12}, hereafter KK12, and H14) and in public archives.

Metallicity measurements at the locations of CSI transients are 
important not only to compare the 
environments of SNe~IIn and SN~IMs, but also to clarify the role of metallicity in the prevalent mass 
losses of these events.
It is well known that, given a certain stellar mass, a higher metallicity drives a larger mass loss
in massive stars, due to stronger line-driven winds \citep[e.g.,][]{kudritzki00}. 
However, the mass-loss rates due to this mechanism, observed in hot stars like Wolf-Rayet (WR), are on the order of 10$^{-5}$~M$_{\odot}$~yr$^{-1}$, which is substantially lower than what is 
required in SNe~IIn.
The massive CSM around SNe~IIn must be produced by larger eruptions, whose underlying mechanism and metallicity dependence
is largely unknown (e.g., gravity waves and super-Eddington winds have been invoked, \citealp{quataert12,smith06}). Since the mass loss of SN~IIn progenitors 
and SN~IMs shapes their CSM and thus the 
appearance of these transients, by looking for correlations between metallicity and observational properties of these transients we aim to constrain 
to what extent 
metallicity is 
an important ingredient. 
In this paper we have collected the observables available in the literature for each SN with measured metallicity at its explosion location.
We also complemented our analysis of the metallicity with that of another important property related to the SN environment, which is the association of SN locations to star-forming (SF) regions. In doing that, we used the results published by H14.

The paper is organized as follows.
In Sect.~\ref{sec:sample} we introduce our SN sample and Sect.~\ref{sec:obs} presents our observations and data reduction procedures. Section~\ref{sec:reduction2} describes how we subtracted the underlying stellar population from each spectrum, how we measured the emission line fluxes and the spectral classification. Section~\ref{sec:oxy} concerns the method to obtain the local metallicity measurements, and Sect.~\ref{sec:results} presents the results on the metallicities of CSI~SNe, which are compared among the different CSI~SN subtypes and with those of other CC~SN classes. In Sect.~\ref{sec:prop} we show the relations between metallicity and observables of CSI~SNe and finally the discussion and conclusions are given in Sect.~\ref{sec:disc} 
and \ref{sec:concl}, respectively.

\section{Sample of CSI transient host galaxies}
\label{sec:sample}

 In Table~\ref{gal} we report the list of 60 transients included in our sample. Thirty-five of them are SNe~IIn, 
six are SNe~Ibn, one is a SN~Ia-CSM, 18 are SN~IMs 
(if we count SN~2009ip also in the SN~IMs, then we have 19 of these transients).
 
With the NOT, long-slit spectra were obtained for the host galaxies of 13 SNe~IIn, 
five SNe~Ibn, one SN~Ia-CSM, and 16 SN~IMs 
(the derived metallicity for SN~2007sv was published in 
 \citealp{tartaglia14}). The host galaxies observed at the NOT are marked with the letter ``o" in the third 
 column of Table~\ref{gal}. With the NOT we also obtained broad-band $R$ and narrow-band H$\alpha$ images for the SNe IIn (except for SN~1995G) and Ibn.

The CSI transients observed at the NOT were chosen 
among those with published spectroscopic and photometric data. These SNe were thoroughly analyzed in the literature, and for most of them estimates of the physical properties of the CSM (e.g., wind velocity, mass-loss rate) were also available. 
This choice was made in order to select 
objects whose observational and physical properties could be related to their host galaxy 
properties, such as the metallicity (see 
Sect.~\ref{sec:prop}). In particular, SN~IMs were chosen among 
those listed by \citet{smith11}. We also decided to observe nearby, 
i.e., resolved, host galaxies (z$<$0.026, see Table~\ref{gal}), to allow for the 
determination of the local metallicity. We discuss the possible biases introduced by this selection in Sect.~\ref{sec:disc}.
 
 We complemented our observed sample with host galaxies whose metallicities (at the center of the galaxy or at the SN position) were already available in the literature (marked with the letter ``l" in the third column of Table~\ref{gal}) or whose spectra were 
 available in public archives (marked with the letter ``a" in the third column of Table~\ref{gal}).
 Eighteen SNe~IIn (excluding SNe~2005ip and 2006jd) have metallicities published in the literature (mainly from KK12 and H14) and four have host-galaxy spectra obtained by the 6dF survey \citep{jones09} and available via NED\footnote{Nasa Extragalactic Database, 
 \href{http://ned.ipac.caltech.edu/}{http://ned.ipac.caltech.edu/}}. 
For two SNe~Ibn their host-galaxy spectra are available in public archives. 
Metallicity measurements were available in the literature for nine SN~IMs 
 (excluding SN~2007sv that we have already published, and including SN~2009ip) and for six of them we also obtained observations at the NOT. Most of the metallicity estimates in the literature for SN~IMs were obtained from P04. All the references are reported in Table~\ref{metal} (4th column).

In summary, our entire sample includes a large fraction of the CSI SNe with published light curves and spectra.
 
Considering the entire sample, on average SN~IMs are located at lower redshift 
($<z_{\rm IM}>$~$=$~0.0020$\pm$0.0004) compared to SNe~IIn ($<z_{\rm IIn}>$~$=
$~0.0276$\pm$0.0092) and Ibn ($<z_{\rm Ibn}>$~$=$~0.0216$\pm$0.0078), since most of 
them were discovered only in nearby galaxies due to their lower luminosity.

\section{Observations and data reduction}
\label{sec:obs}

\begin{figure*}
 \centering$
\begin{array}{cc}
\includegraphics[width=6cm,angle=0]{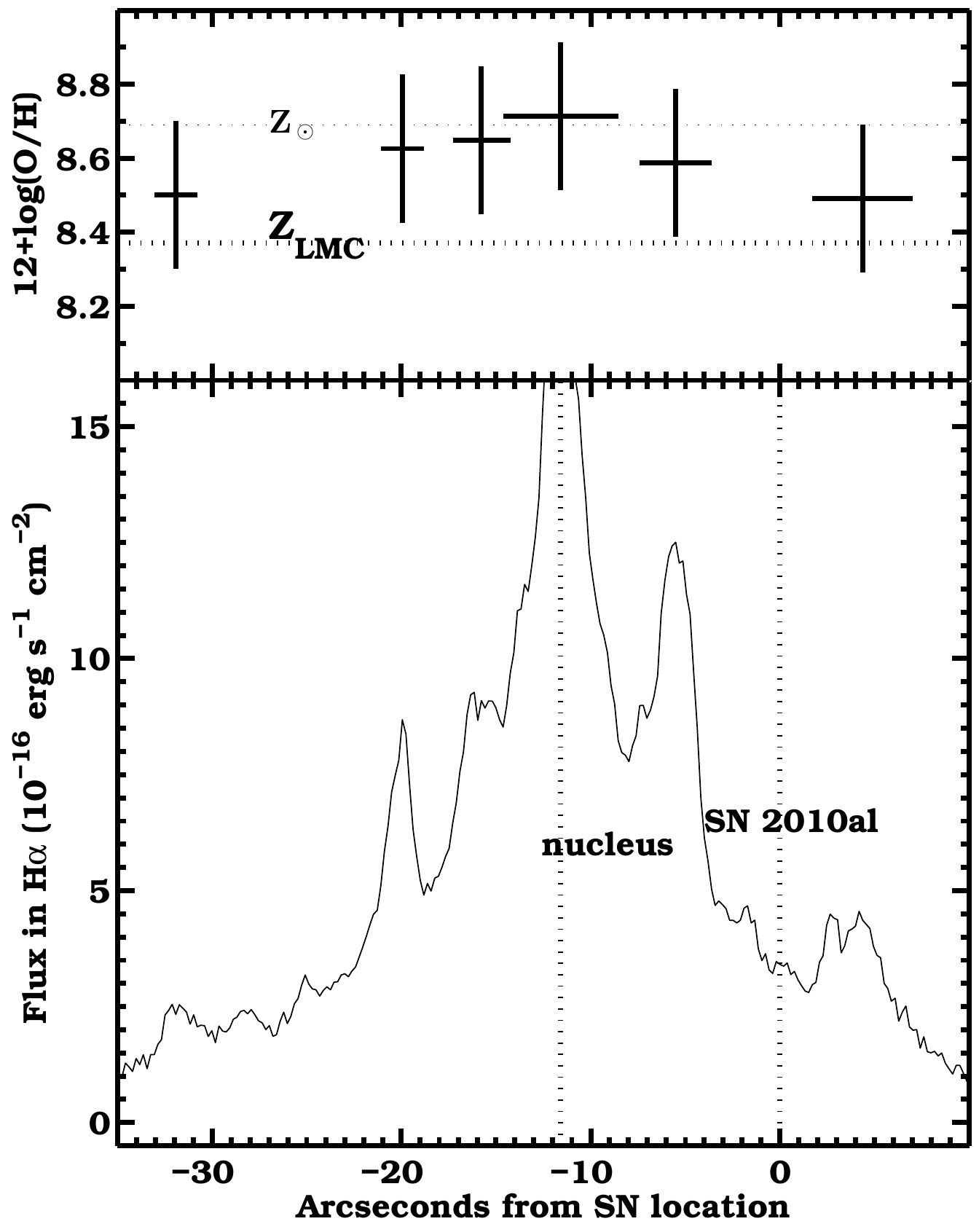}&
\includegraphics[height=10.3cm,angle=90]{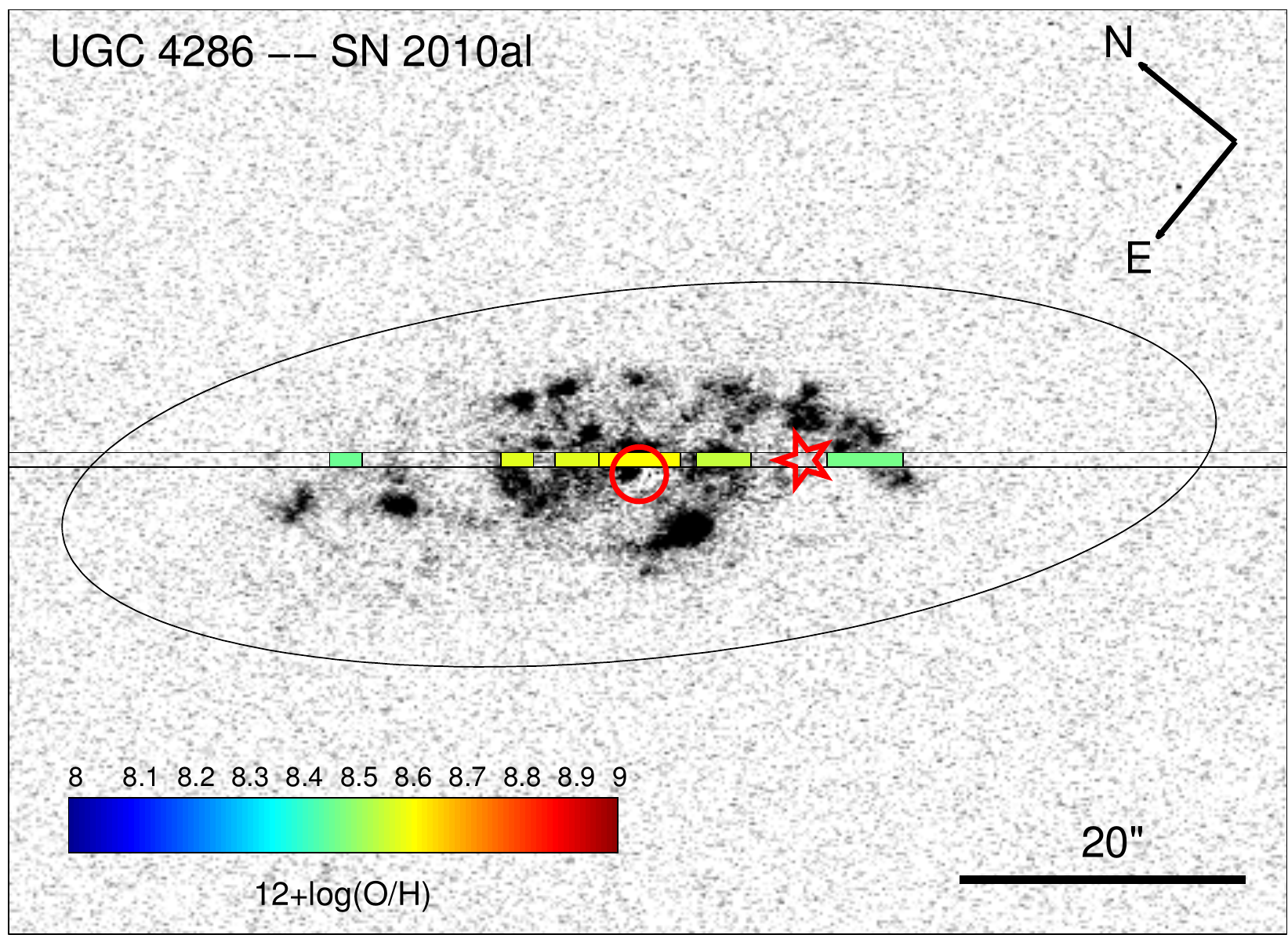} \\
\end{array}$
\includegraphics[width=16.9cm,angle=0]{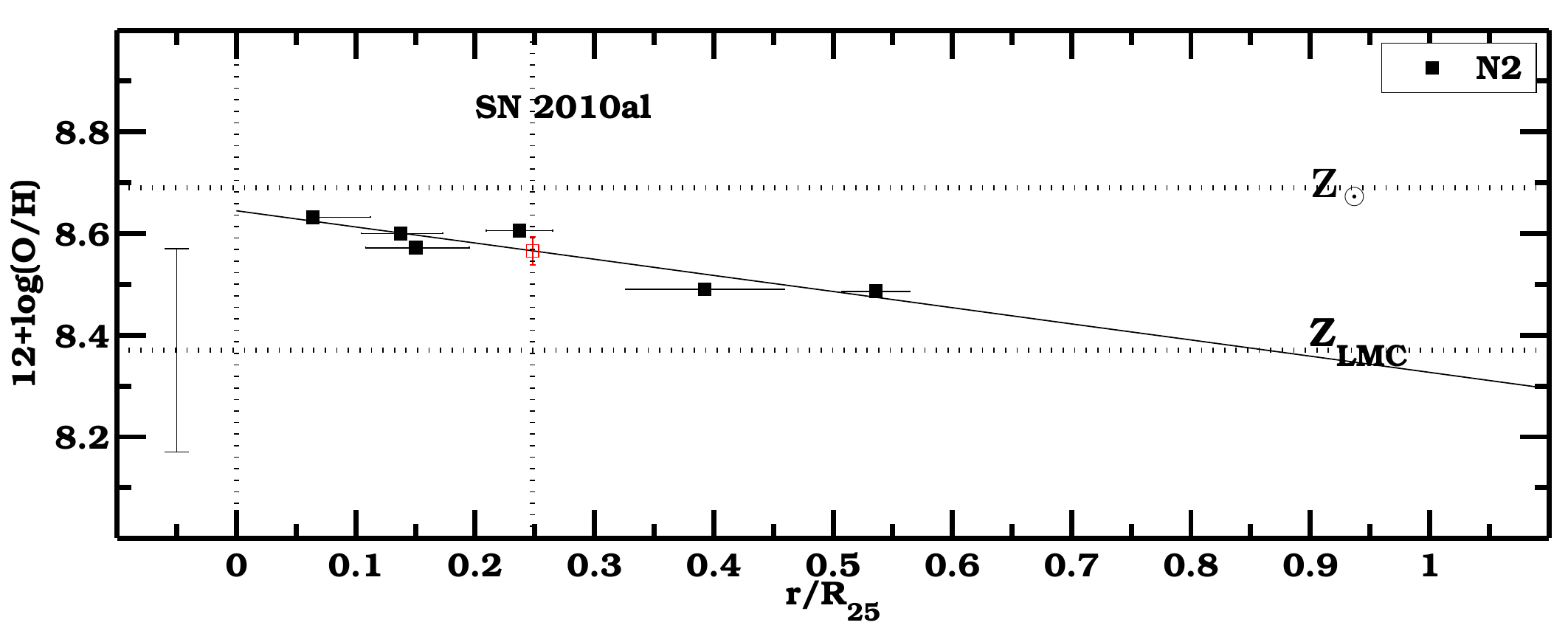}
  \caption{\textit{(Top-right panel)} Continuum-subtracted H$\alpha$ image of UGC~4286, the host galaxy of SN~2010al. The 25th $B$-band magnitude elliptic contour is shown by a black solid line, along with the position of SN~2010al (marked by a red star) and the center of the galaxy (marked by a red circle). The slit position is shown, and a color code is used to present the N2 metallicity measurements at the position of each \ion{H}{ii} region that we inspected. \textit{(Top-left panel)} Flux at the H$\alpha$ wavelength along the slit, shown as a function of the distance from the SN location (marked by a dotted line, like the nucleus position). The N2 measurements are shown at the corresponding positions in the top sub-panel. \textit{(Bottom panel)} Metallicity gradient of UGC~4286. The linear fit on our measurements is shown by a solid line. The interpolated metallicity at the SN distance is marked by a red square and its uncertainty corresponds to the fit error. The error bar ($\pm$0.2) for our N2 measurements is shown aside. The positions of SN and nucleus are marked by vertical dotted lines. The solar metallicity \citep{asplund09} and the LMC metallicity \citep{russell90} are indicated by two horizontal dotted lines\label{sn10al}.}
 \end{figure*} 

\subsection{Nordic Optical Telescope observations and data reduction}
\label{sec:NOTobs}

Observations at the NOT and data reductions were performed following 
the procedures outlined in \citet{taddia13met}, where we studied the 
host galaxies of SN~1987A-like events arising from the explosions of blue 
supergiant stars (BSGs). In this section a brief summary of how our visual-wavelength spectroscopy was obtained and reduced is provided.

Two observational campaigns (P45-004, P49-016) were carried out during 2012 and 2014. Six nights were spent observing the hosts of SNe~IIn, Ibn and Ia-CSM in 2012 (four in April, one of which was lost to bad weather, and two in September).
Four nights were spent observing the host galaxies of SN~IMs (three in April 2014, one in 
September 2014). During April 2014, almost 1.5 of the three nights were lost to bad weather.

In both campaigns our main goal was to determine the metallicity at the exact position of the CSI transients through strong emission-line diagnostics. In order to do that, we obtained long-exposure 
($\gtrsim$~1800~s), long-slit spectra of the Star forming (\ion{H}{ii}) regions within the host galaxies, by simultaneously placing the slit at the SN position and through the galaxy center or through bright \ion{H}{ii} regions near the SN location. 
In order to intercept both the host galaxy center (or bright \ion{H}{ii} regions) and the SN location with the 
slit, we rotated the telescope field by the corresponding angle; then the slit was centered on a pre-determined reference 
star in the field and finally the telescope was offset to point to the SN location. The final pointing 
was checked with a through-slit image. In most cases the slit included a few \ion{H}{ii} regions, allowing 
for a determination of the host galaxy metallicity gradient and hence of the metallicity at the distance of the SN from the host center (see Sect.~\ref{sec:oxy}).

The instrumental setup that was chosen to acquire the host galaxy spectra at the NOT was the same as adopted for the study presented in \citet{taddia13met}, i.e. ALFOSC with grism \#4 (wide wavelength 
range $\sim$3500$-$9000~\AA) and a 1.0$\arcsec$-wide slit (corresponding to the typical seeing on La 
Palma). The obtained spectral resolution is $\sim$16--17~\AA. The exposure times adopted for each spectral observation are listed in Table~\ref{log}.
 
The following procedure was adopted to carry out the spectral reductions. First, the 2D-spectra were bias
subtracted and flat-field corrected in a standard way. 
When available, multiple exposures were then median-combined in order to remove any spikes produced by cosmic rays. 
We extracted and fitted 
with a low-order polynomial the trace of the brightest object in the 
2D-spectrum (either the galaxy 
nucleus, or a bright star, or an \ion{H}{ii} region with a bright continuum). 
The precision of this 
trace was checked by plotting it over the 2D-spectrum. 
We then shifted the same trace in the spatial direction to match the position of each \ion{H}{ii} region visible in the 2D-spectrum, and then extracted a 1D-spectrum for each \ion{H}{ii} region. The extraction regions were chosen by looking at the 
H$\alpha$ flux profile, an example of which is presented in the top-left panel of Fig.~\ref{sn10al}, where 
we also report the width of each spectral extraction. The extracted spectra were wavelength and flux 
calibrated using an arc-lamp spectrum and a spectrophotometric standard star observed the same night, 
respectively. Following \citet{stanishev07}, from each spectrum we removed the second order 
contamination, which characterizes the spectra obtained with grism~\#4. In this study we included all the 
spectra showing at least H$\alpha$, [\ion{N}{ii}]~$\lambda$6584 and H$\beta$ emission lines. We identified the 
location of each corresponding \ion{H}{ii} region by inspecting the acquisition images.

For SNe~IIn, Ibn and Ia-CSM, we also had the chance to observe the host galaxies in a narrow H$\alpha$ 
filter and a broad $R$-band filter, to build H$\alpha$ maps as detailed 
in \citet{taddia13met}. These maps 
further helped us to 
precisely determine the location of each \ion{H}{ii} region that was spectroscopically observed.
An example of a continuum-subtracted H$\alpha$ images is shown in the top-right panel of
Fig.~\ref{sn10al}. Here we indicate the SN position with a red star and the galaxy nucleus with a circle. An 
ellipse marks the 25th $B$-band magnitude contour. Each colored patch within the plotted slit aperture
corresponds to the position of an extracted \ion{H}{ii} region spectrum. We also attempted to use these H$\alpha$ maps to measure the degree of association (NCR, see Sect.~\ref{sec:ncr}) to \ion{H}{ii} regions for each SN. However, since we had to cut the exposure times due to bad weather, we did not reach the desired depth in the H$\alpha$ images. The time lost because of bad weather in the 2014 campaign also prevented us from
obtaining the same photometric observations for the host galaxies of SN~IMs. 
However, we used NCR data from H14 in the discussion about the SN~IIn progenitor scenarios (see Sect.~\ref{sec:disc_progenitor}).
Table~\ref{log} summarize all the observations 
carried out during the two campaigns.

\subsection{Archival data}
\label{sec:archivaldata}

We used archival spectra to complement the dataset for SNe~IIn and Ibn.
All the four spectra retrieved via NED are low-resolution spectra obtained 
at the galaxy center by the 6dF Galaxy Survey \citep{jones09}. 
The spectrum of LSQ12btw was retrieved via the ESO archive and reduced. All these spectra included at least H$\alpha$, [\ion{N}{ii}]~$\lambda$6584 and H$\beta$ emission lines.

\begin{figure}
 \centering
\includegraphics[width=9cm,angle=0]{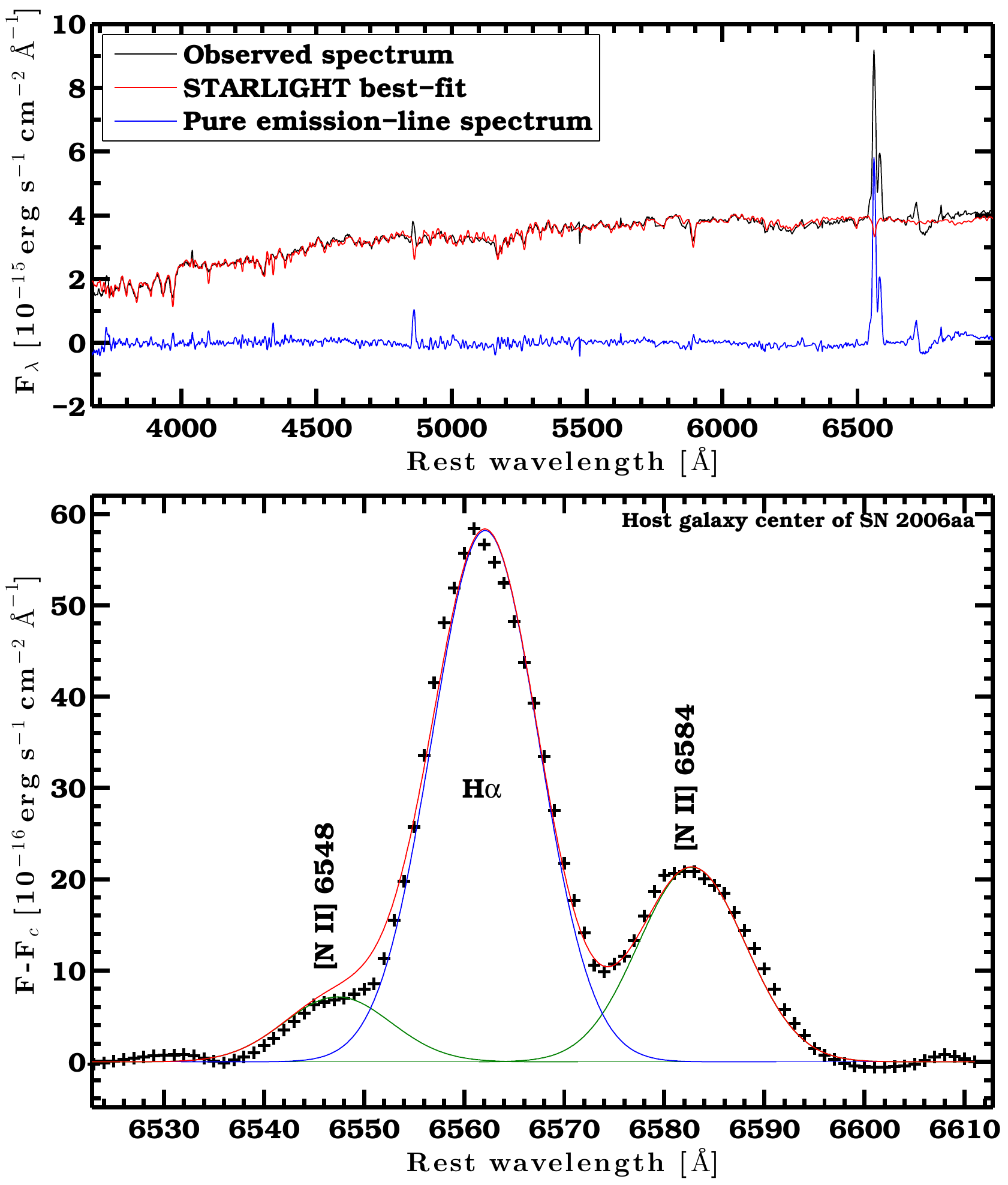}
  \caption{\label{blend} \textit{(Top Panel)} STARLIGHT stellar-population spectrum best-fit to the continuum of the spectrum obtained at the center of the host galaxy of SN~2006aa. The difference between the observed spectrum and the best fit gives the pure emission-line spectrum.
   \textit{(Bottom panel)} The triple-Gaussian fit on the continuum-subtracted H$\alpha$ and [\ion{N}{ii}] lines of the bright \ion{H}{ii} region at 
the center of the host galaxy of SN~2006aa. 
The observed fluxes are represented by black crosses, the best fit in red, the H$\alpha$ component in blue and the [\ion{N}{ii}] components in green.}  
  \end{figure}
  
 \onlfig{4}{ 
  \begin{figure}
 \centering
\includegraphics[width=16cm,angle=0]{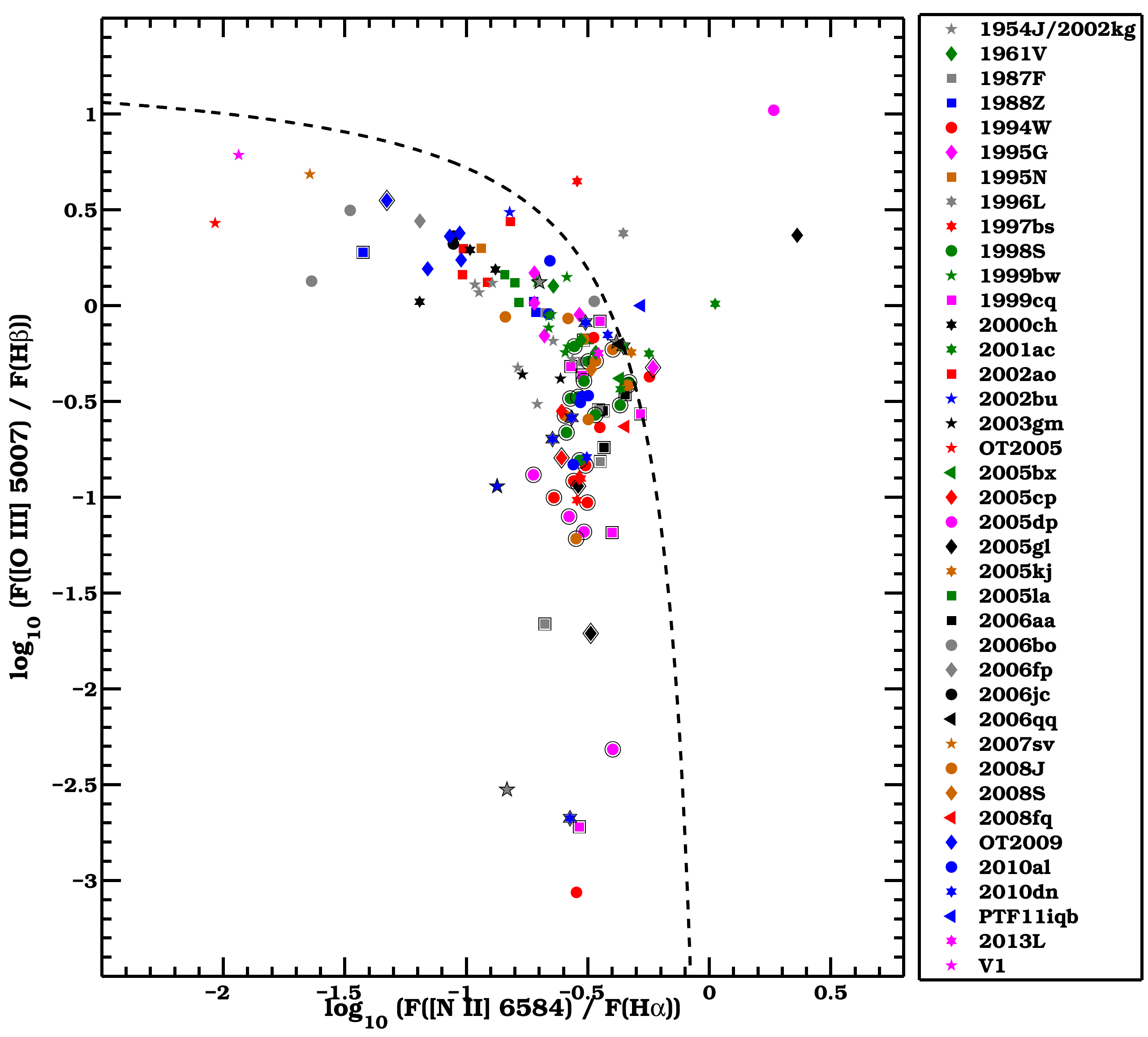}
  \caption{\label{BPT} BPT diagram for each \ion{H}{ii} region that we spectroscopically observed. Above the dashed line \citep[][see Sect.~\ref{sec:reduction2}]{kauffmann03} the line fluxes are AGN contaminated and we rejected those spectra from our metallicity study. Most of the spectra fall in the star-forming part of the BPT diagram, below the dashed line. The symbols marked with black edges indicate that the ratio between [\ion{O}{iii}~$\lambda$5007] and H$\beta$ is an upper limit. All the line ratios are reported in Table~\ref{tab:lineratios}.} 
  \end{figure}}

\section{Stellar population subtraction, line measurements and spectral classification}
\label{sec:reduction2}

In order to accurately measure the emission line fluxes necessary to determine the metallicity, the stellar population component of each spectrum was modelled with the STARLIGHT code \citep{cid05}\footnote{\href{http://astro.ufsc.br/starlight/}{http://astro.ufsc.br/starlight/}} and subtracted from each observed spectrum (corrected for the Galactic extinction). This code linearly combines synthetic spectra from \citet{bc03} and takes into account dust attenuation in order to fit the emission-line free regions of our spectra. The template spectra are stellar population spectra at fifteen different ages between 0.001 and 13 Gyr and three metallicities ($Z~=~$0.2, 1, 2.5~$Z_\odot$). The same templates were used by \citet{leloudas11} in their study of SNe~Ib/c locations.
As an example, the best-fit to the spectral continuum of SN~2006aa host galaxy nucleus is shown in Fig.~\ref{blend} (top-panel), as well as the ``pure" emission line spectrum, as obtained from the observed spectrum subtracted of the best-fit continuum. Once we obtained the pure emission line spectra, we proceeded  to measure the fluxes of [\ion{N}{ii}]~$\lambda6584$ and H$\alpha$, H$\beta$ and [\ion{O}{iii}]~$\lambda$5007.

In order to measure the fluxes of the [\ion{N}{ii}]~$\lambda6584$ and H$\alpha$ emission lines, we followed a fitting procedure similar to that outlined in \citet{taddia13met}. Given the low resolution of our ALFOSC spectra, we had to deblend
[\ion{N}{ii}]~$\lambda6584$,
[\ion{N}{ii}]~$\lambda6548$ and H$\alpha$ through a triple Gaussian fit.
We fixed the width of each Gaussian to be the same, as determined by the
spectral resolution. 
The known wavelength offsets between the centroids of the three
 lines was also fixed. The flux of [\ion{N}{ii}]~$\lambda6548$ was furthermore 
fixed to be 1/3 of that 
 in [\ion{N}{ii}]~$\lambda6584$ \citep[see][]{osterbrock06}. This assumption was needed to allow for a 
 proper fit of this faint line, which could possibly contaminate the flux of H$\alpha$.
Finally, the integrals of the two Gaussians fitted to [\ion{N}{ii}]~$\lambda6584$ and H$\alpha$ provided 
us with the fluxes of these lines (see the bottom panel of Fig.~\ref{blend} for an example of the fitting procedure). 

For each spectrum we measured the H$\beta$ and [\ion{O}{iii}]~$\lambda$5007 line fluxes by fitting them with Gaussians. In several cases we could only place a limit on the flux of [\ion{O}{iii}]~$\lambda$5007.

The ratios of line fluxes such as F([\ion{N}{ii}]~$\lambda6548$) to F(H$\alpha$),  and F([\ion{O}{iii}]~$\lambda$5007) to F(H$\beta$) can be used to determine the metallicity only if the dominant ionizing source for each region are hot massive stars, since strong line diagnostics are based on this condition. To exclude other possible ionizing sources such as shock-excitation or AGN contamination, we classified our spectra based on the BPT diagram \citep{baldwin81}, which shows
log$_{10}$(F([\ion{O}{iii}]~$\lambda$5007)/F(H$\beta$)) versus log$_{10}$(F([\ion{N}{ii}]~$\lambda6548$)/F(H$\alpha$)), which is widely used in order to discriminate the excitation sources of emission line objects.  log$_{10}$(F([\ion{N}{ii}]~$\lambda6584$)/F(H$\alpha$)) is known as N2 and we will use it to determine the metallicity for our SNe (see Sect.~\ref{sec:oxy}).
Gas ionized by different sources occupies different areas across the BPT diagram (e.g., \citealp{kewley01}; \citealp{kehrig12}; \citealp{sanchez14}; \citep{galbany14}; \citep{belfiore15}). 
In Fig.~\ref{BPT} we plot log$_{10}$([\ion{O}{iii}]~$\lambda$5007/H$\beta$) versus N2 for our 
spectra and checked if these points were located within the BPT diagram area 
given by \citet{kauffmann03}, i.e., log$_{10}$([\ion{O}{iii}]~$\lambda$5007/H$\beta$)~$<$~0.61/(N2$-$0.05) $+$ 1.3,  which defines the star-forming galaxies as opposed to galaxies with AGN contamination. The line ratios for each \ion{H}{ii} region are reported in Table~\ref{tab:lineratios}.
 We rejected the spectra 
falling in the other region of the BPT diagram. This occurred for the nuclear spectra 
of the host galaxies of SNe~1994W, 1995G, 2001ac, 2005dp and 2005gl, and for another region of the hosts of SNe~2001ac and 1997bs. 
We also excluded the hosts of SNe 1996L \citep{benetti99}, 2005kj \citep{taddia13IIn} and iPTF11iqb \citep{smith15} from our sample as we only had a nuclear spectrum with [\ion{O}{iii}]~$\lambda$5007/H$\beta$)~$>$~0.61/(N2$-$0.05) + 1.3.

\vspace{0.5cm}

\section{Oxygen abundances at the SN explosion sites}
\label{sec:oxy}

In this section we describe how the metallicity at the SN location was estimated from the observed and archival spectra after stellar population subtraction, line fitting and spectral classification. We also show how we included the metallicity values available in the literature. For the observed spectra we followed the procedure illustrated in \citet{taddia13met}.

\subsection{Local metallicity measurements from observed and archival spectra}

Among the possible emission line diagnostics, we chose to use N2 \citep{pettini04}
for all our metallicity measurements. The oxygen abundance can be obtained from N2 using the following expression presented by 
\citet{pettini04}:\\

12$+$log(O/H)~$=$~9.37$+$2.03$\times$N2$+$1.2$\times$N2$^2+$0.32$\times$N2$^3$.\\

This expression is valid in the range $-$2.5~$<$~N2~$<$~$-$0.3 which corresponds to 7.17~$<$ 12$+$log(O/H)~$<$~8.86.
  This method has a systematic uncertainty of
0.18 which largely dominates over the error from the flux
measurements. Following \citet{thoene09} and \citet{taddia13met}, we adopted 0.2 as the 
total error for each single measurement. We note that \citet{marino13} have recently revised the N2 index using a large dataset of extragalactic \ion{H}{ii} regions with measured T$_{e}$-based abundances. However, we use the calibration by \citet{pettini04} to allow for a direct comparison with other SN types (see Sect.~\ref{sec:othercomp}).

N2 has a larger systematic uncertainty than other methods such as O3N2 \citep{pettini04}. However, N2 presents several advantages that are briefly 
summarized here: 1) for nearby galaxies, the lines needed for this method fall in the part of the CCD with higher efficiency and thus are easier to detect than those needed for the 
other methods (e.g. O3N2 and R23, where lines in the blue or in the near ultraviolet are required).
2) given the fact that H$\alpha$ and [\ion{N}{ii}]~$\lambda6584$ are very close in wavelength, this 
minimize the effects associated with differential slit losses (we did not observe at the parallactic angle as we placed the slit along the SN location -- host-galaxy center direction) and uncertainties on the extinction. 
3) It is well known that there are non-negligible offsets between different line diagnostics 
\citep{kewley08}, and, as N2 was used to estimate the metallicity for many other SNe in the literature (e.g., 
\citealp{thoene09}, \citealp{anderson10}, \citealp{leloudas11}
 \citealp{sanders12}, \citealp{stoll13}, \citealp{kunca13a,kunca13b}, \citealp{taddia13met}),
 this method allows for a direct comparison of CSI transients to other SN types (see Sect.~\ref{sec:othercomp}).
 
We note that there are possible drawbacks when using N2. For instance, the existence of a correlation between the metallicity derived from the N2 parameter and the N/O ratio \citep{perez09} together with the presence of N/O radial gradients across the disks of spiral galaxies (see e.g. P04; \citealp{molla06}) can affect the metallicity measurements obtained with this method. 
The most direct method to determine the oxygen abundances would be via the weak [\ion{O}{iii}]$\lambda$4363 line flux (see e.g., \citealp{izotov06}), but we could not detect this line in our spectra.

\onlfig{5}{\begin{figure}
 \centering$
 \begin{array}{cc}
\includegraphics[width=9cm,angle=0]{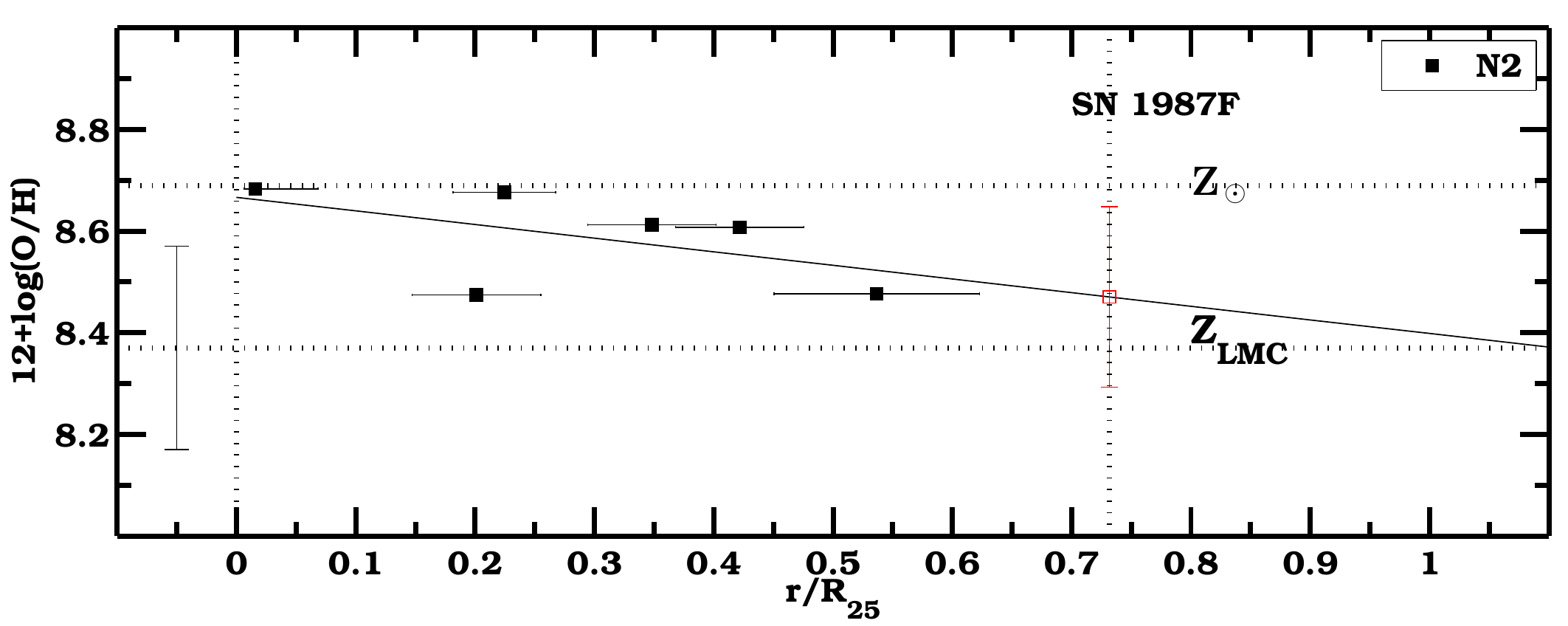}&
\includegraphics[width=9cm,angle=0]{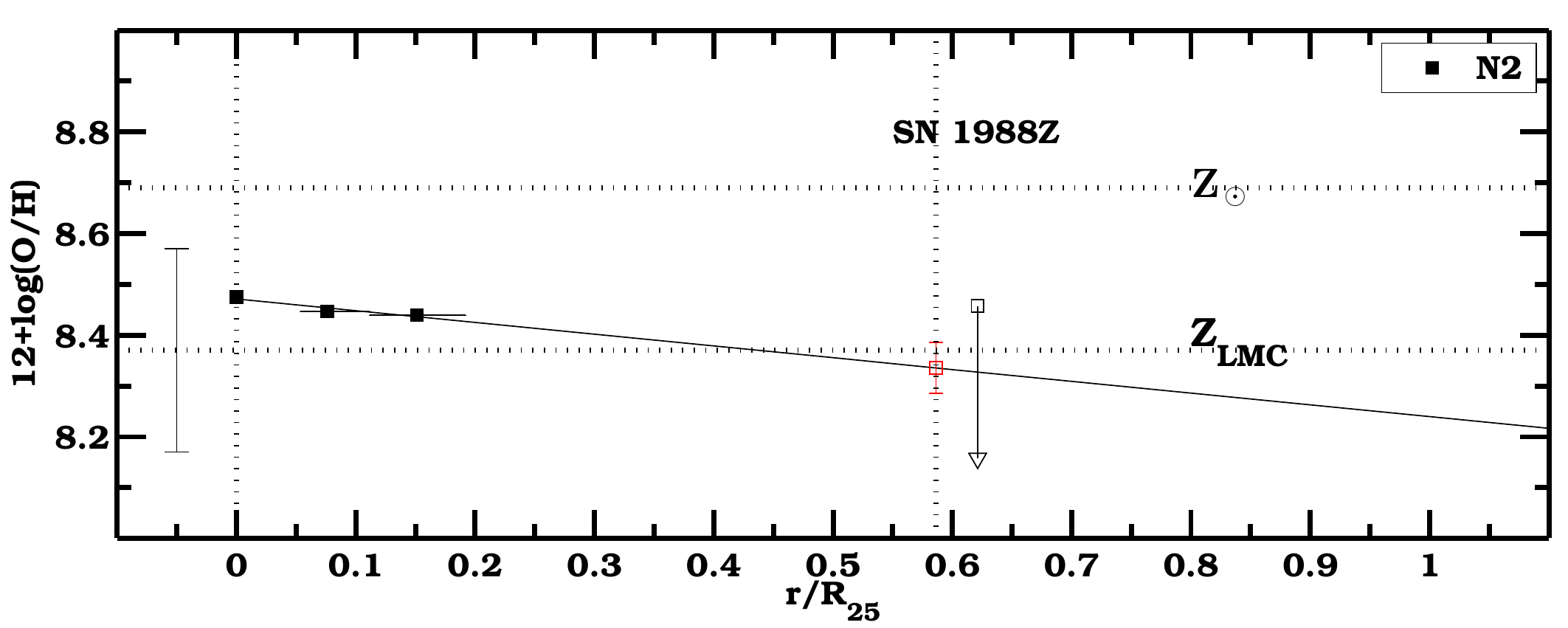}\\
\includegraphics[width=9cm,angle=0]{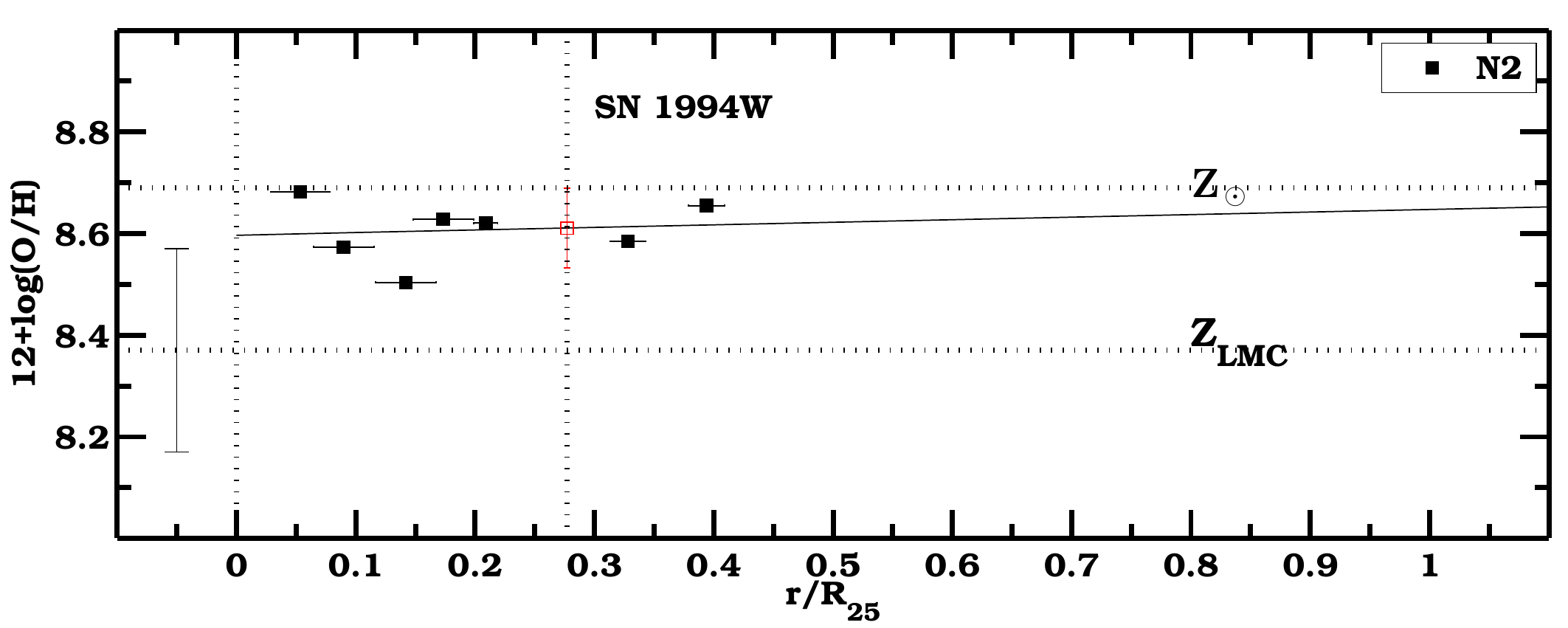}&
\includegraphics[width=9cm,angle=0]{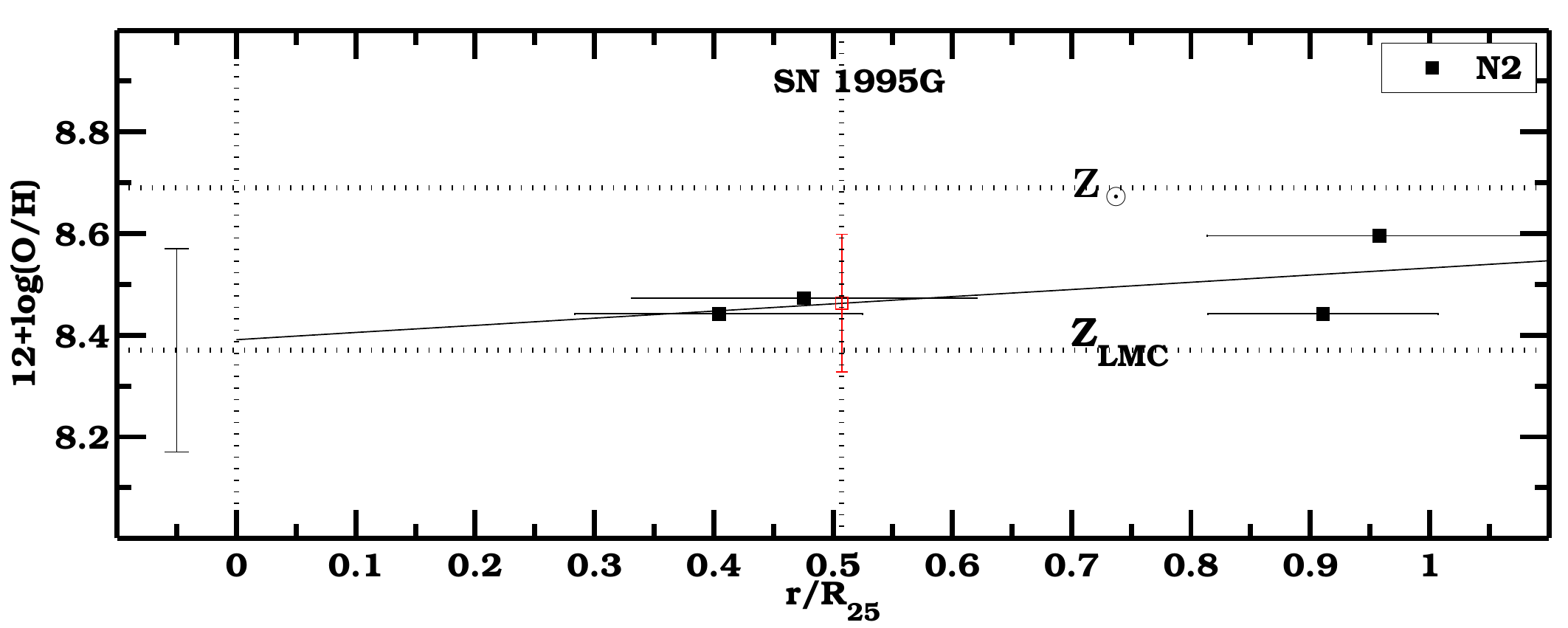}\\
\includegraphics[width=9cm,angle=0]{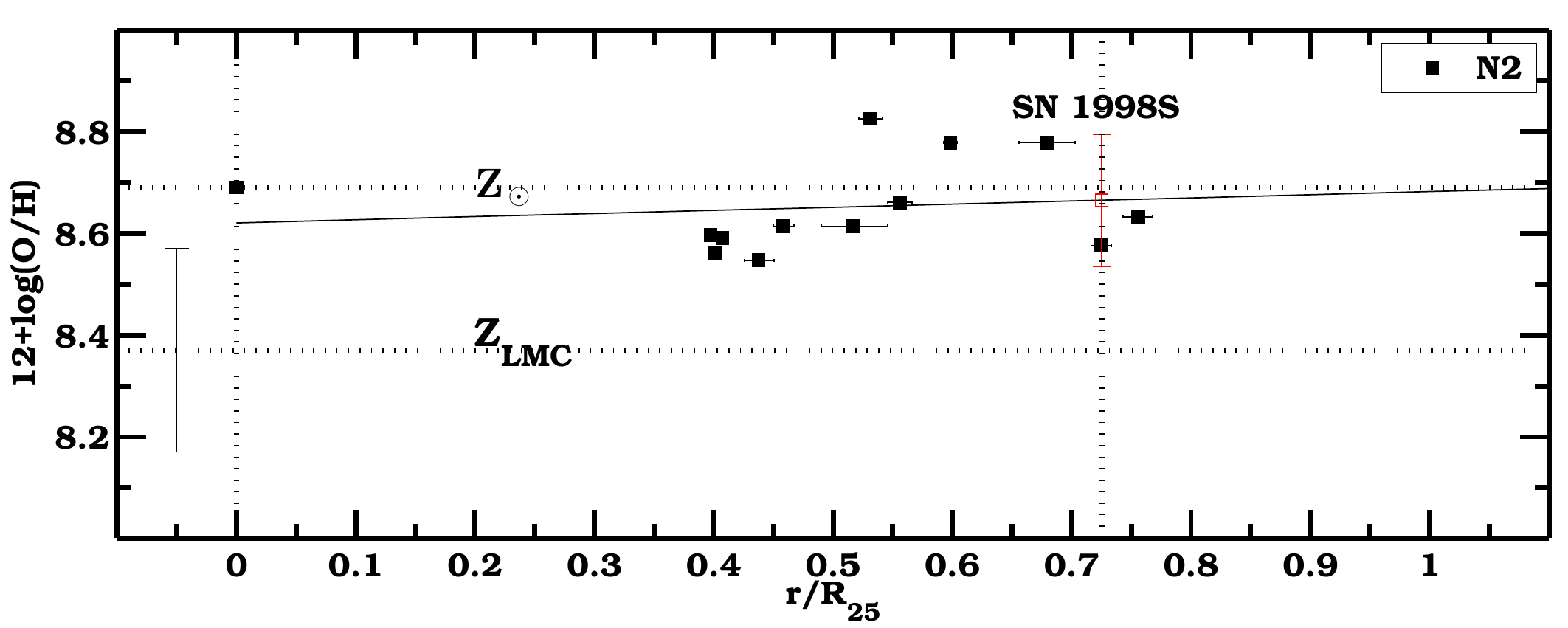}&
\includegraphics[width=9cm,angle=0]{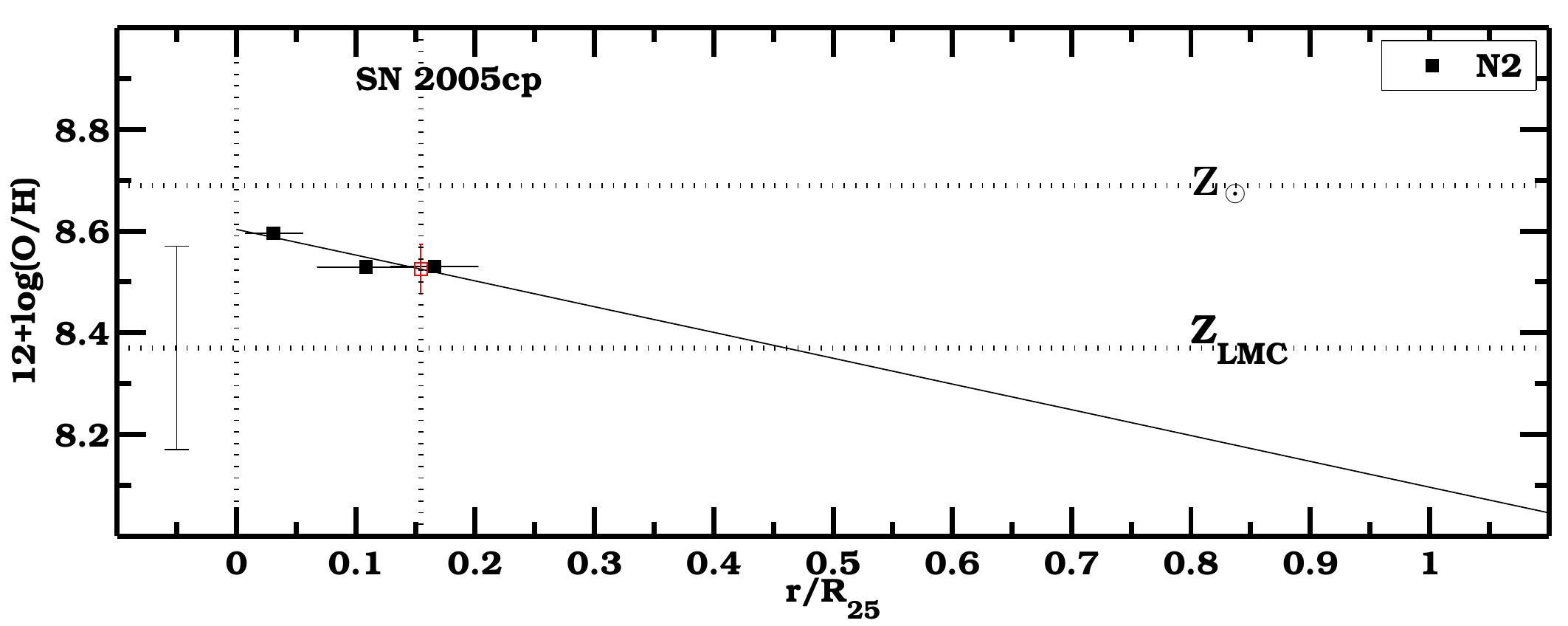}\\
\includegraphics[width=9cm,angle=0]{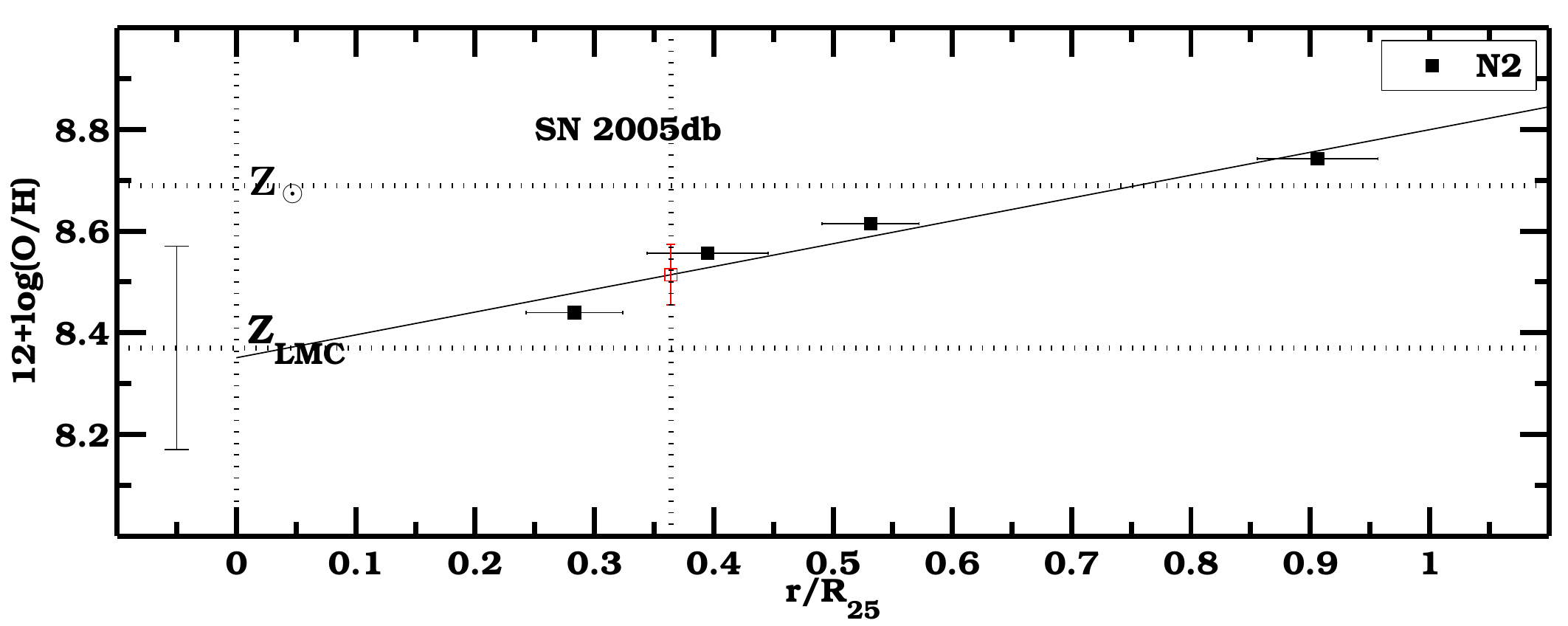}&
\includegraphics[width=9cm,angle=0]{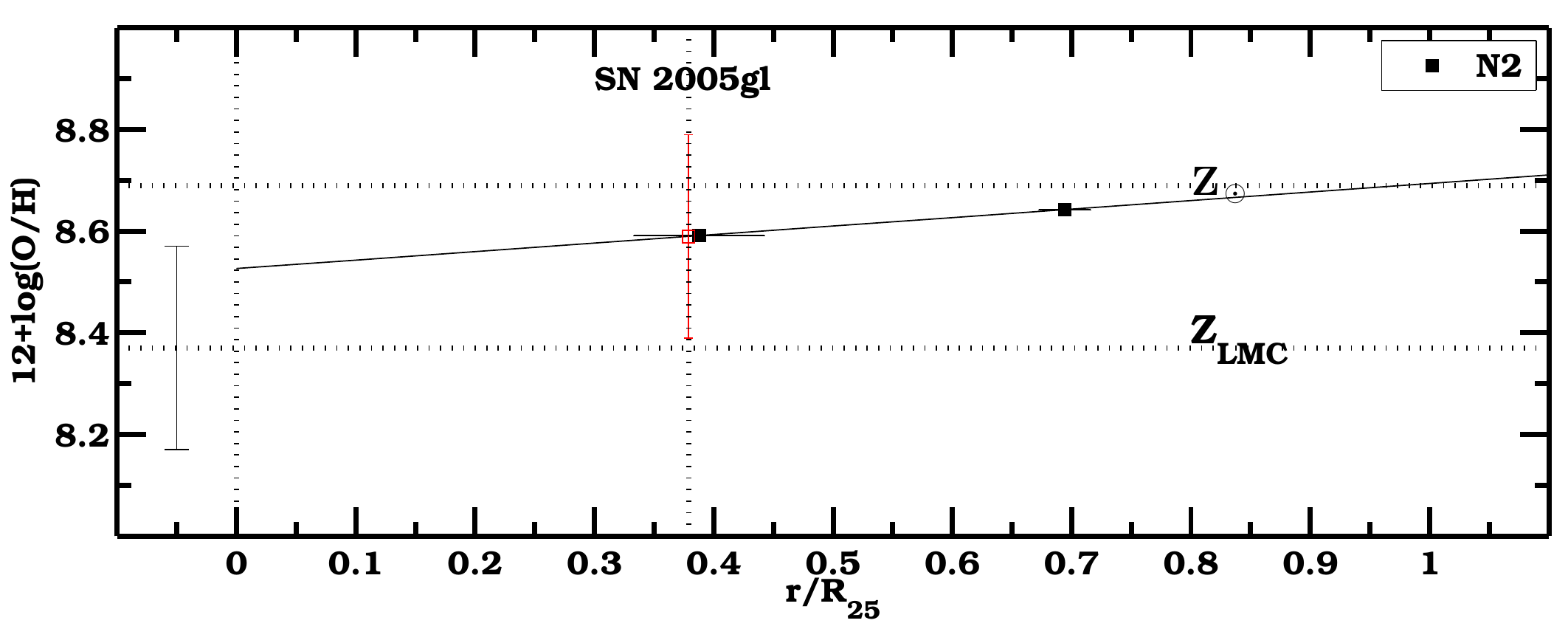}\\
\includegraphics[width=9cm,angle=0]{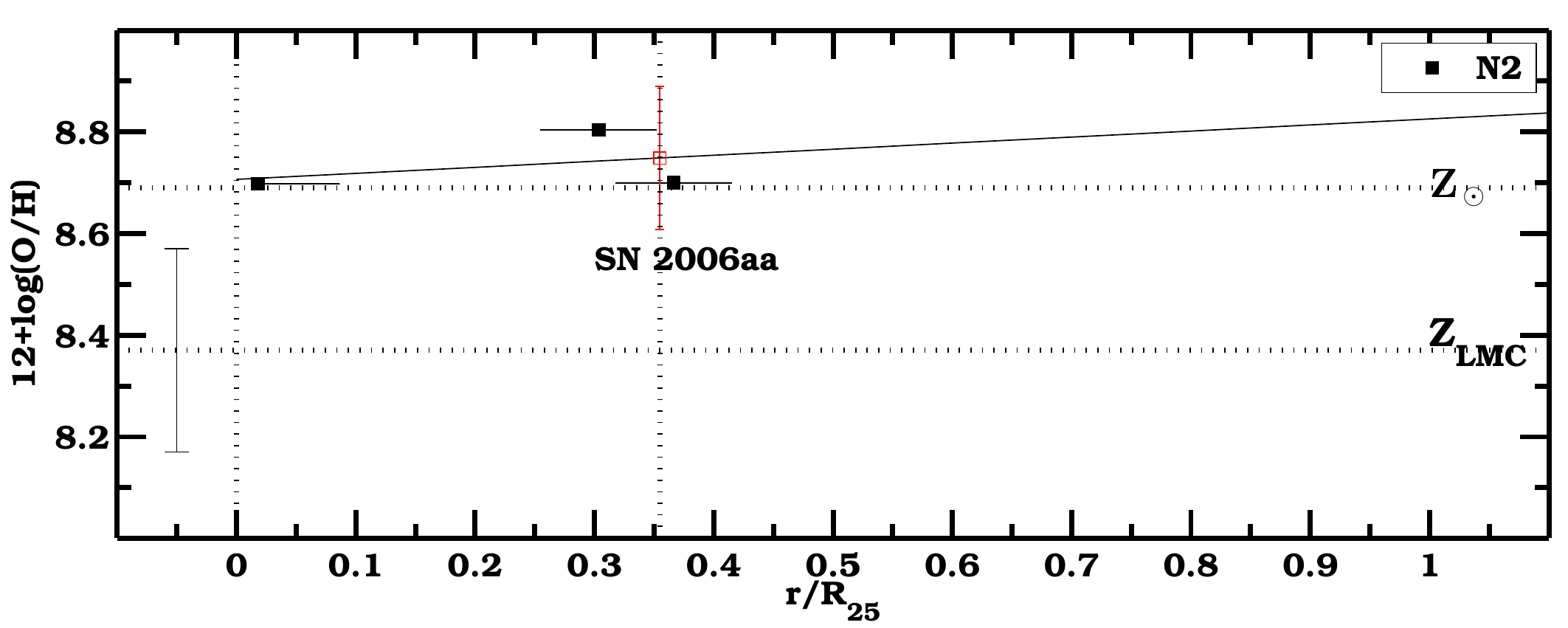}&
\includegraphics[width=9cm,angle=0]{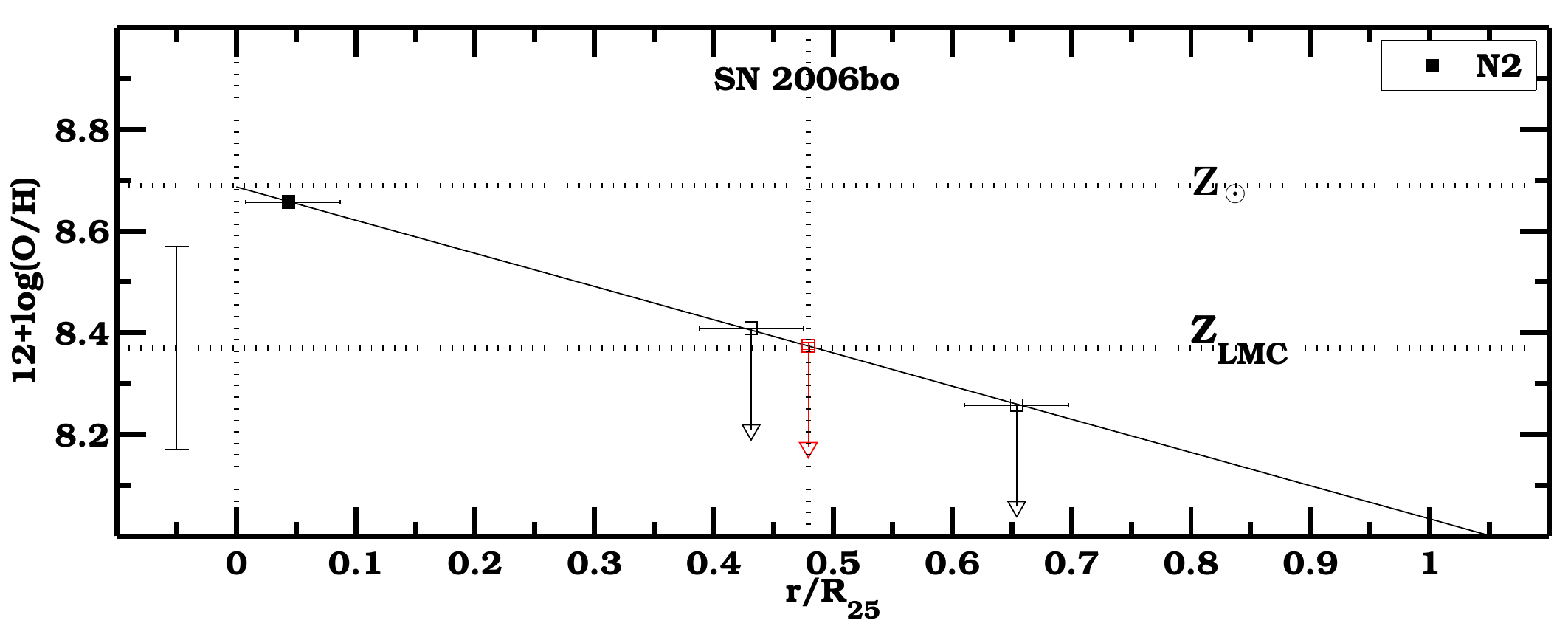}\\
\end{array}$
  \caption{Metallicity gradients of 10 SN~IIn host galaxies observed with the NOT$+$ALFOSC. Symbols and lines as in the bottom panel of Fig.~\ref{sn10al}. Black open squares with arrows correspond to upper limits. The gradients of the hosts of SNe 2006jd and 2005ip are not shown as we could not measure the flux from bright \ion{H}{ii} regions so we used literature data. For the host of SN 1995N, whose gradient is not shown, we only have one measurement and we did not assume any gradient, as it is an irregular and interacting galaxy. \label{allgradIIn}}
 \end{figure}}

\onlfig{6}{\begin{figure}
 \centering$
 \begin{array}{cc}
\includegraphics[width=9cm,angle=0]{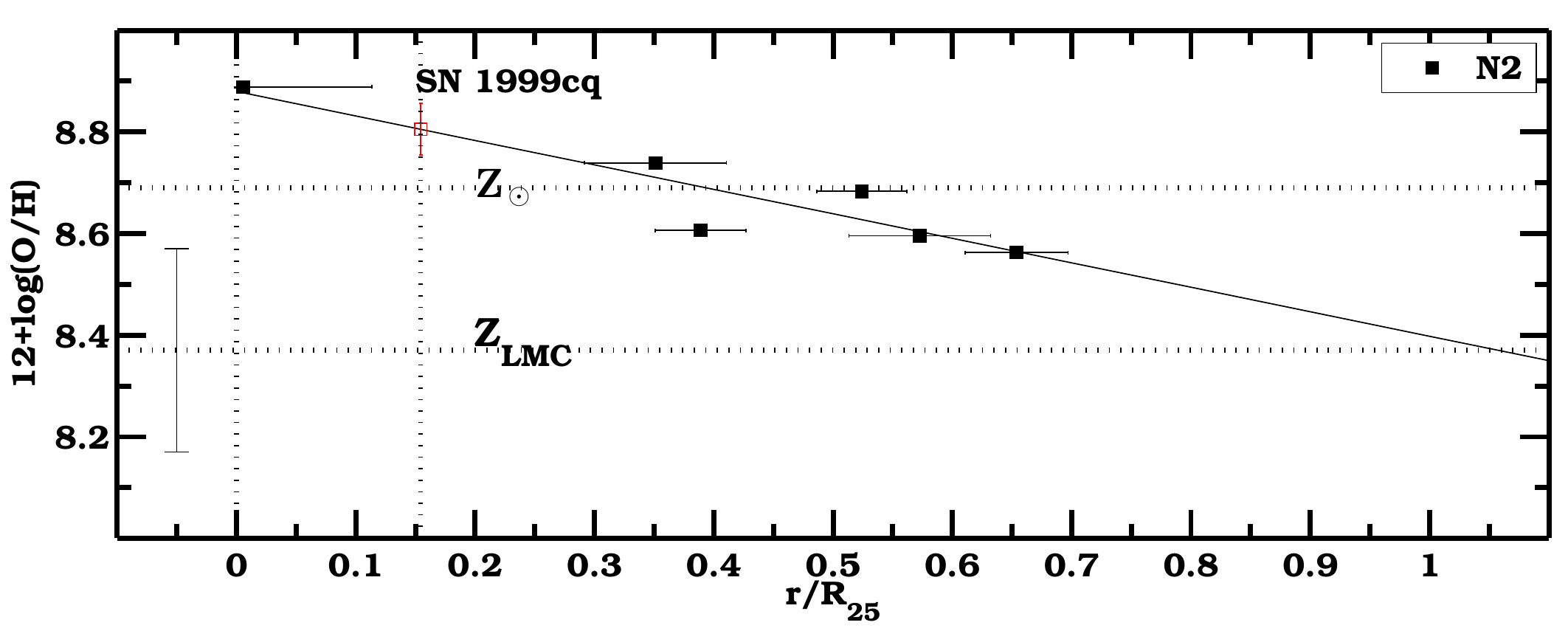}&
\includegraphics[width=9cm,angle=0]{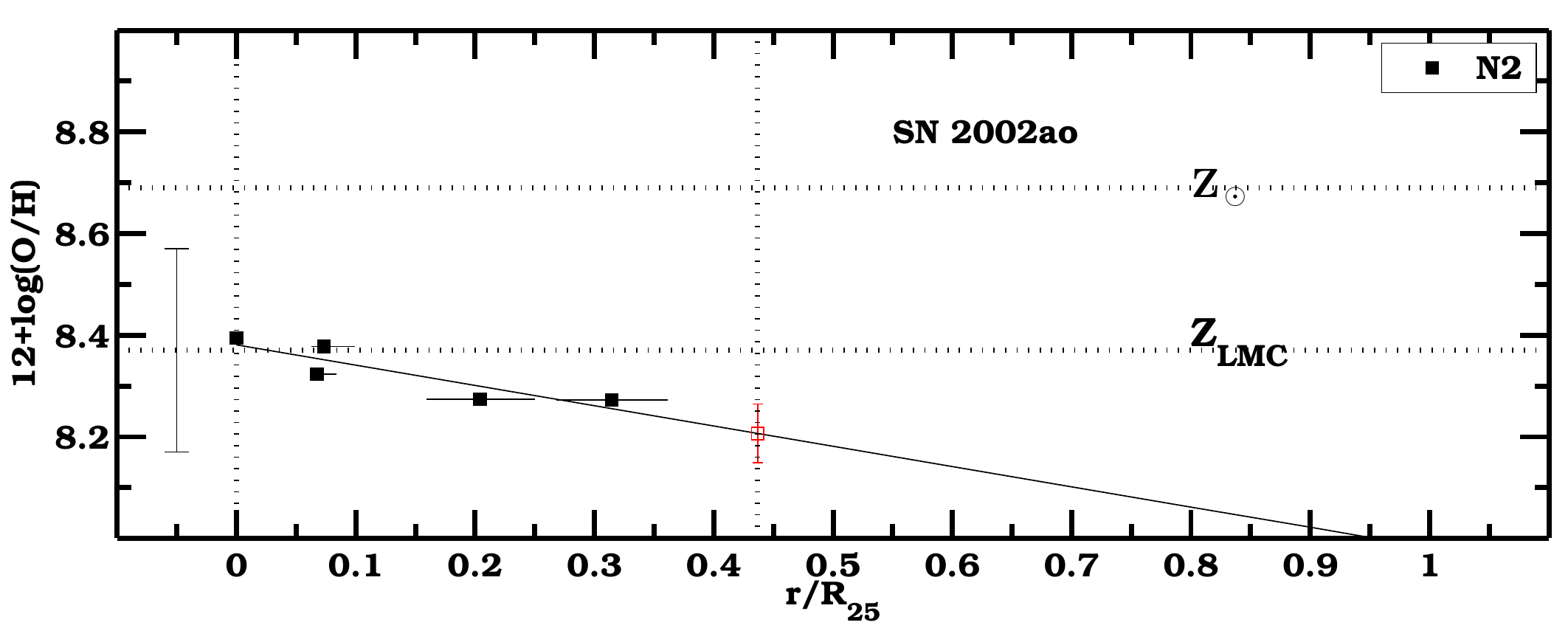}\\
\includegraphics[width=9cm,angle=0]{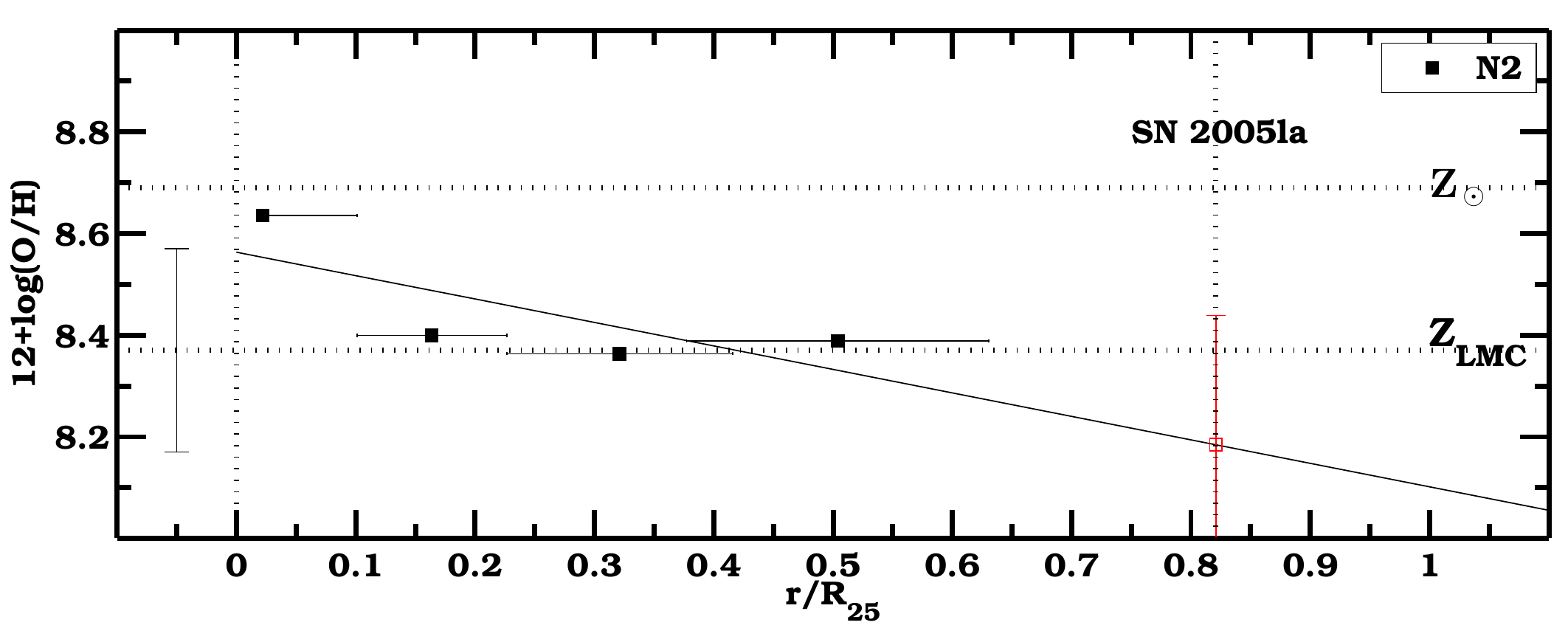}&
\includegraphics[width=9cm,angle=0]{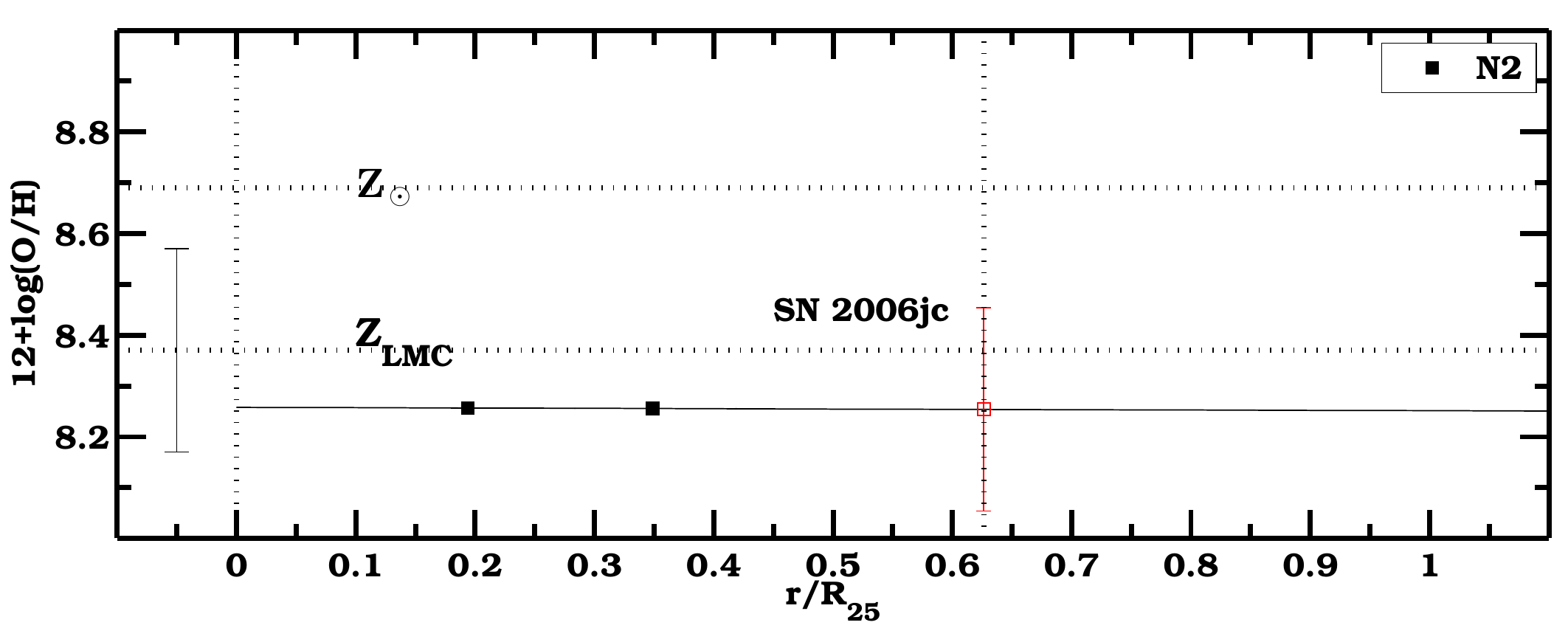}\\
\includegraphics[width=9cm,angle=0]{sn10al_met_grad-eps-converted-to.pdf}&
\includegraphics[width=9cm,angle=0]{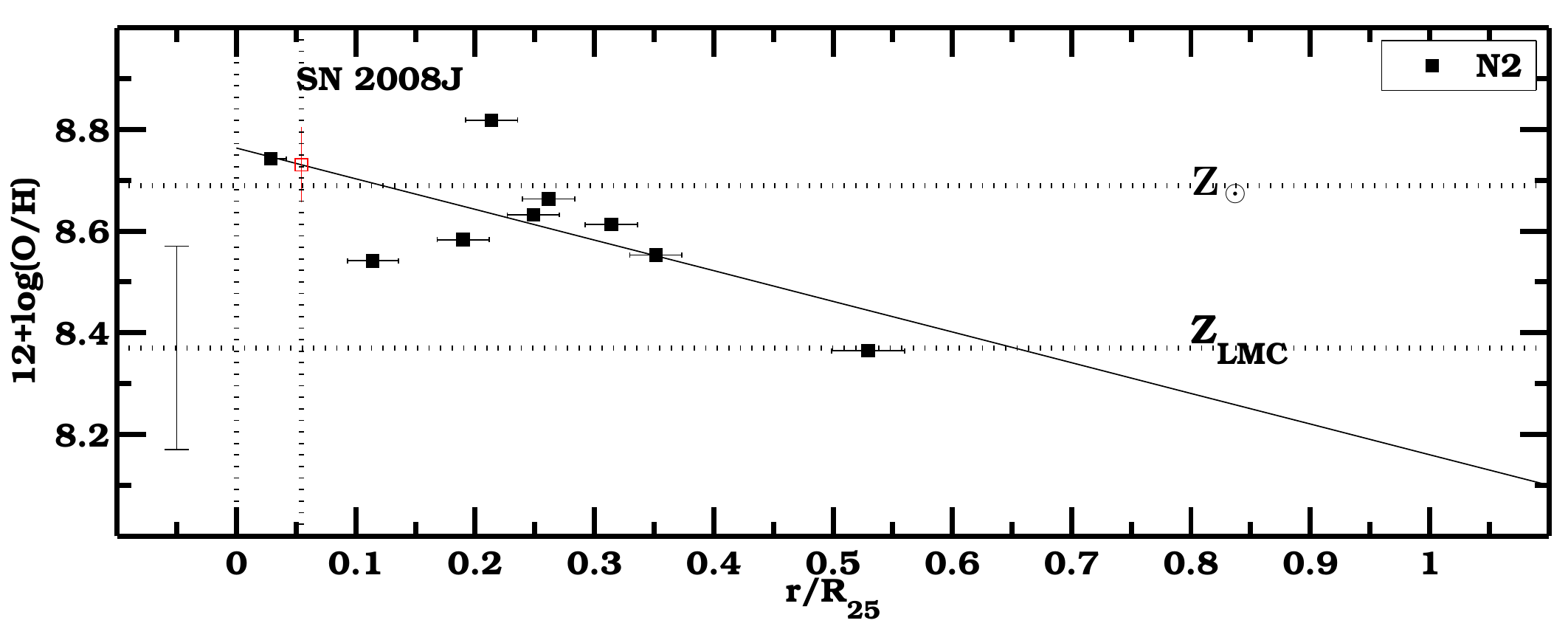}\\
\end{array}$
  \caption{Metallicity gradients of 5 SN~Ibn and 1 SN~Ia-CSM host galaxies observed with the NOT$+$ALFOSC. Symbols and lines as in the bottom panel of Fig.~\ref{sn10al}. \label{allgradIbnIacsm}}
 \end{figure}}

\onlfig{7}{ \begin{figure}
 \centering$
 \begin{array}{cc}
 \includegraphics[width=9cm,angle=0]{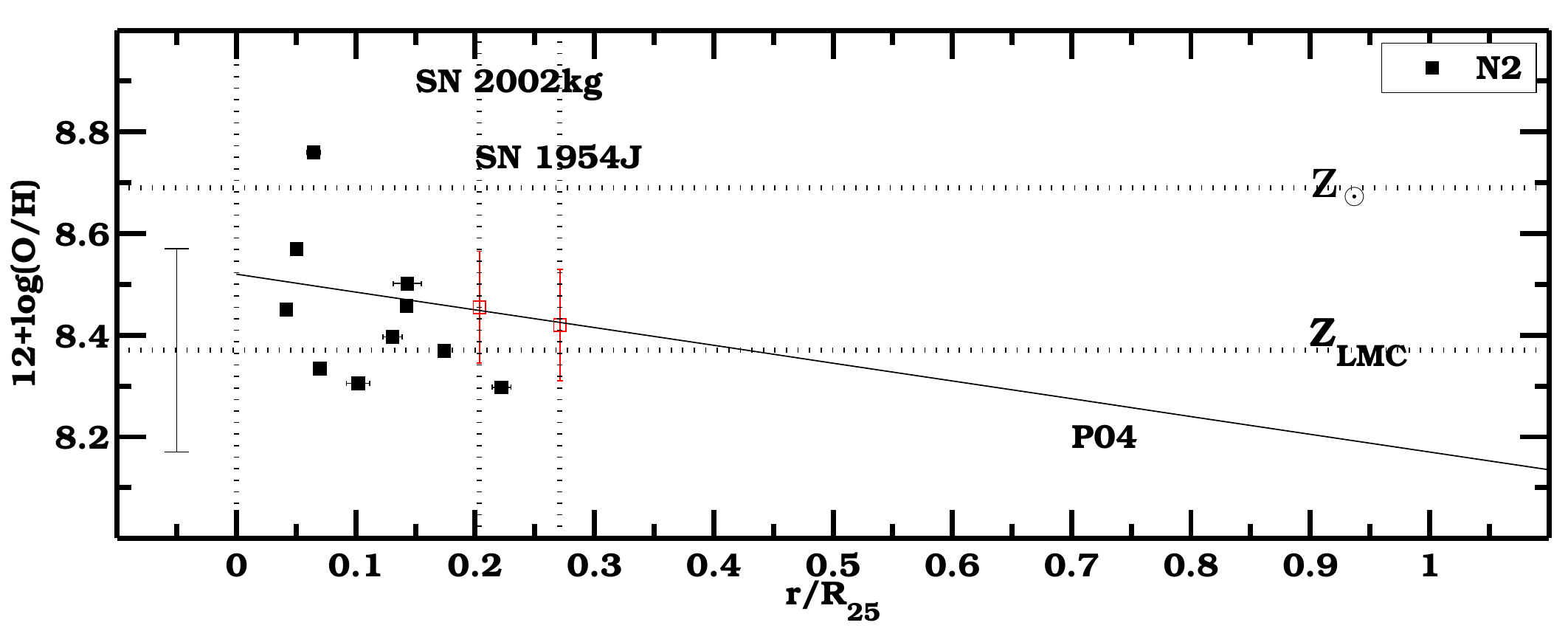}&
 \includegraphics[width=9cm,angle=0]{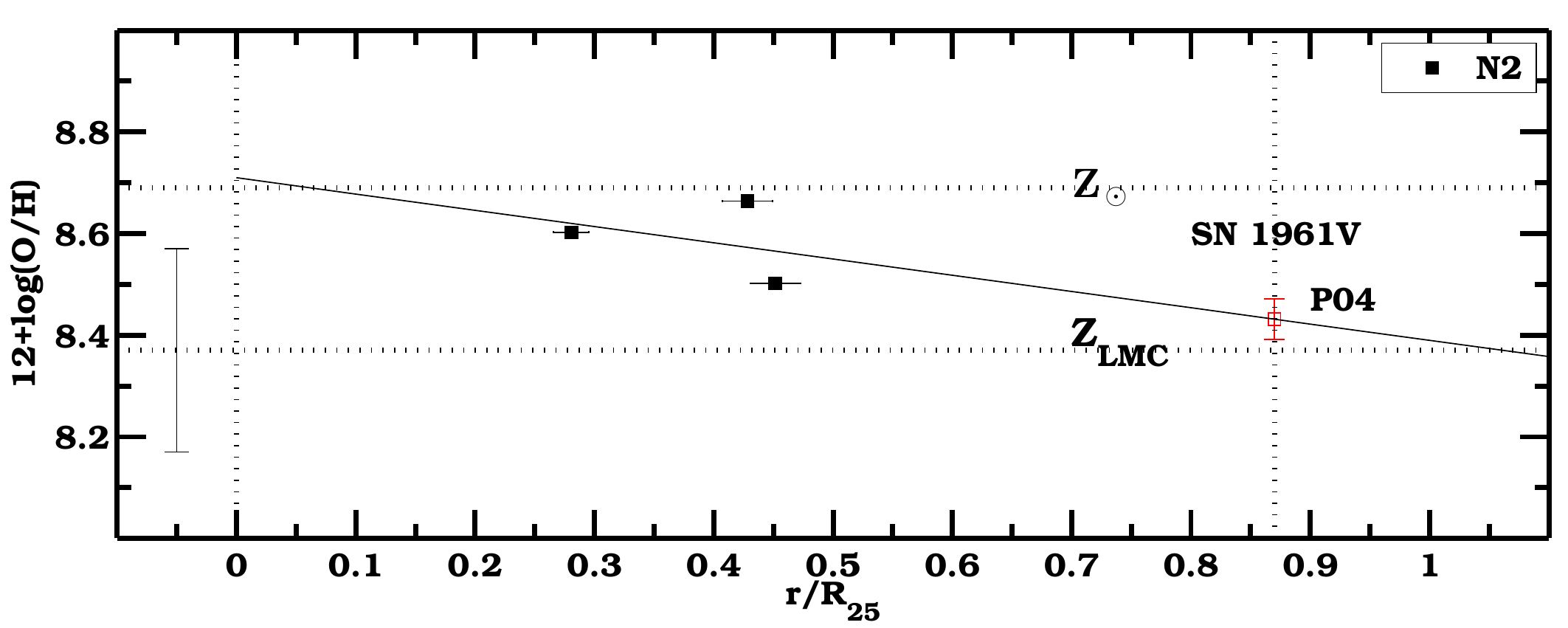}\\
\includegraphics[width=9cm,angle=0]{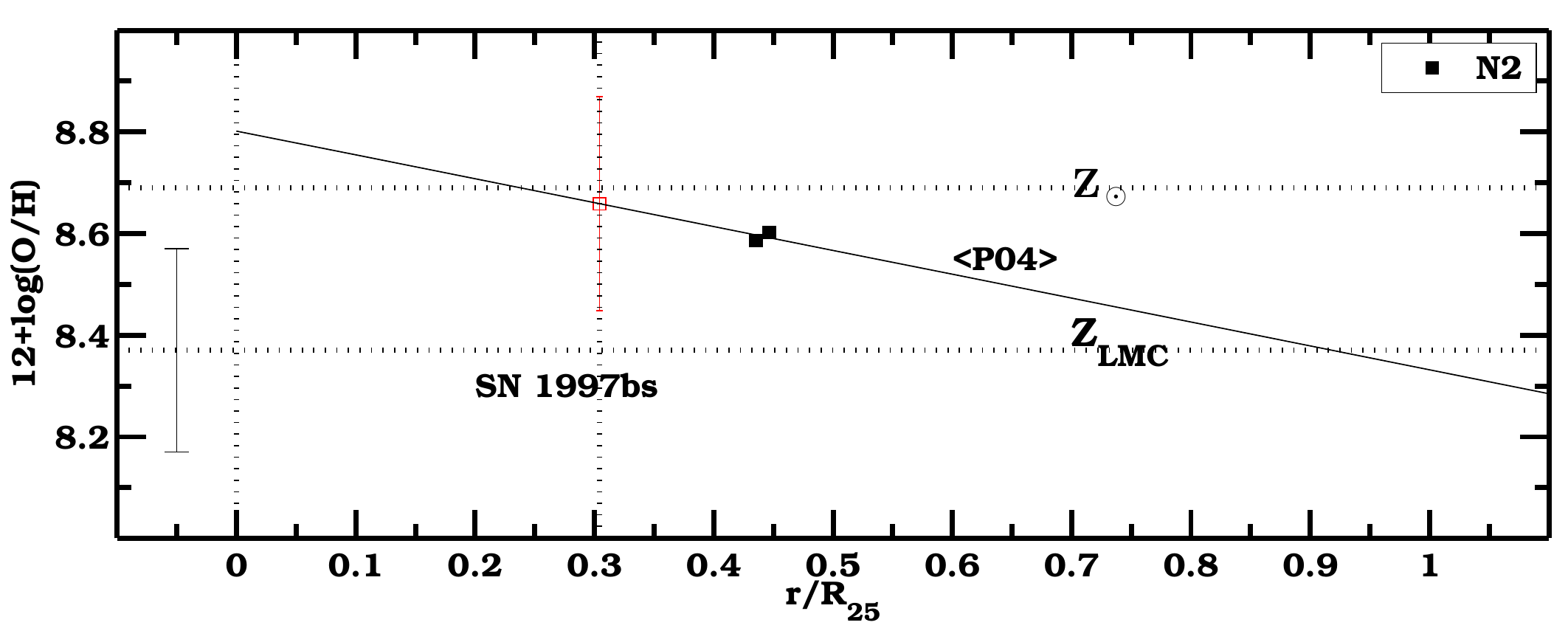}&
\includegraphics[width=9cm,angle=0]{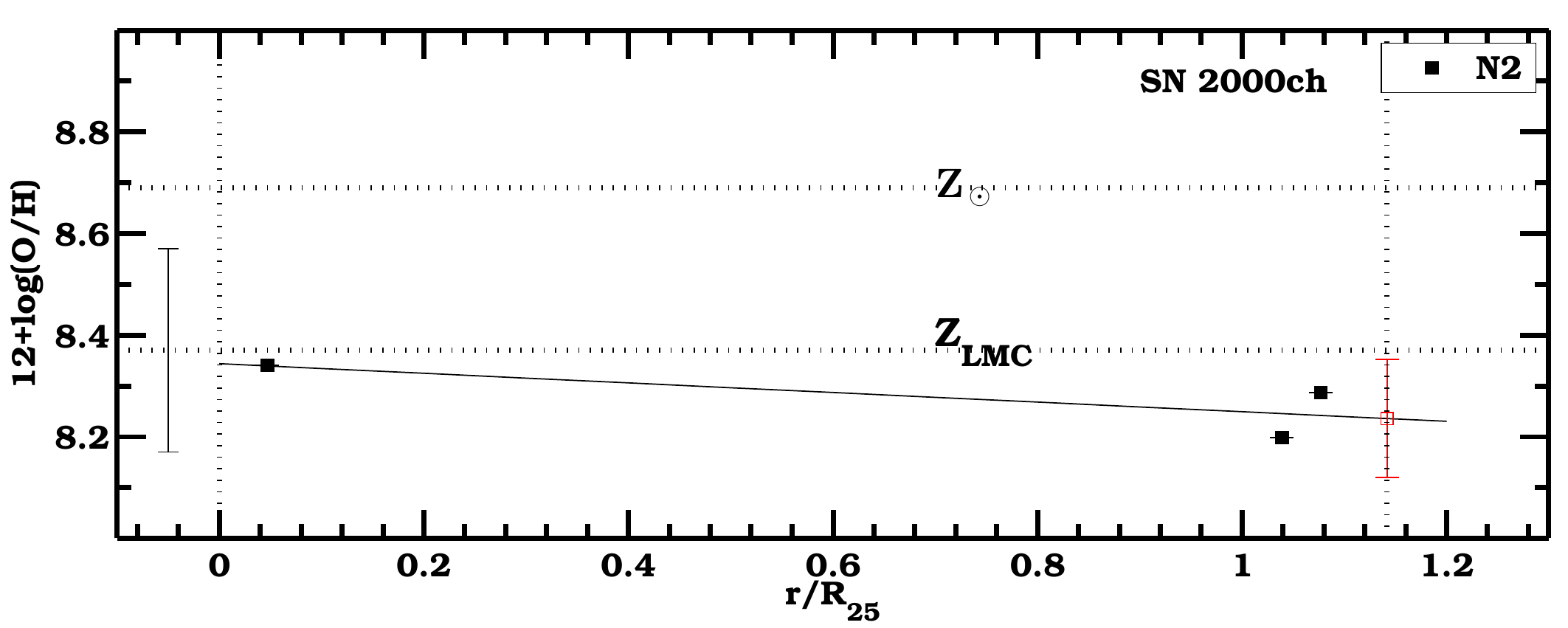}\\
\includegraphics[width=9cm,angle=0]{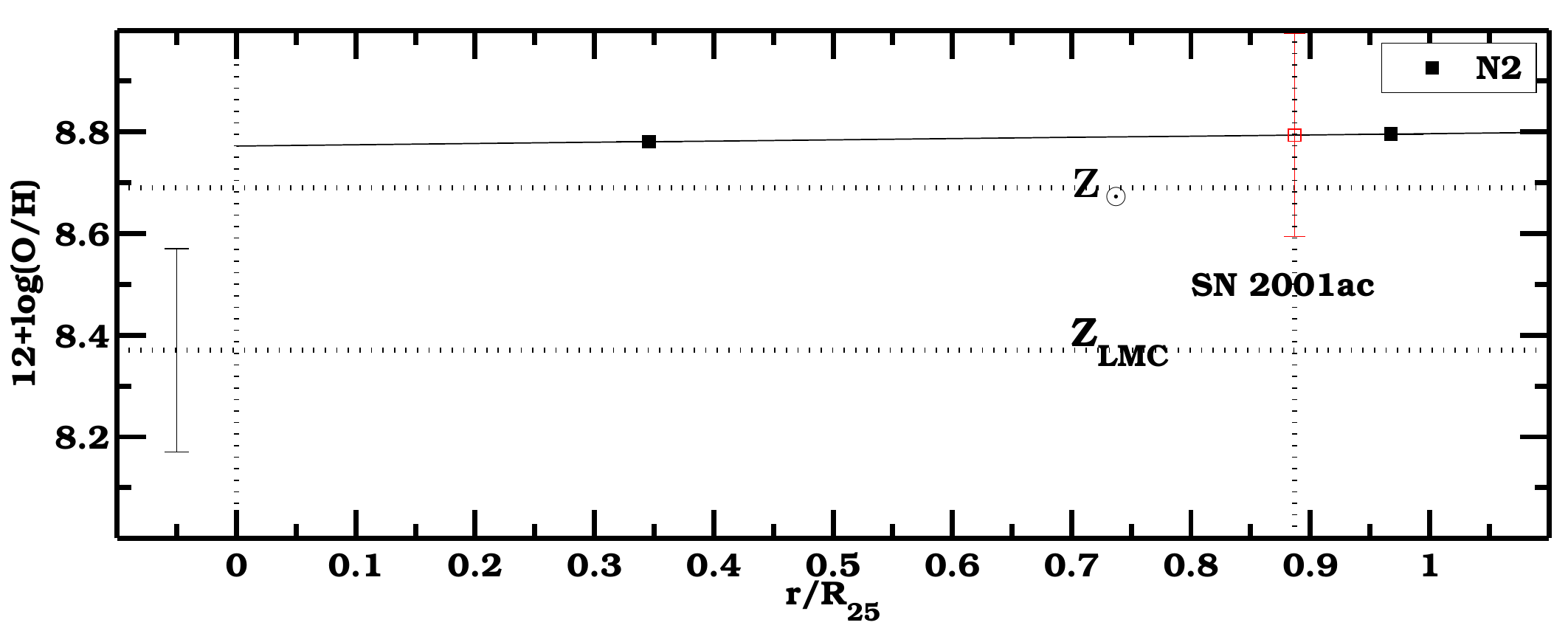}	&
\includegraphics[width=9cm,angle=0]{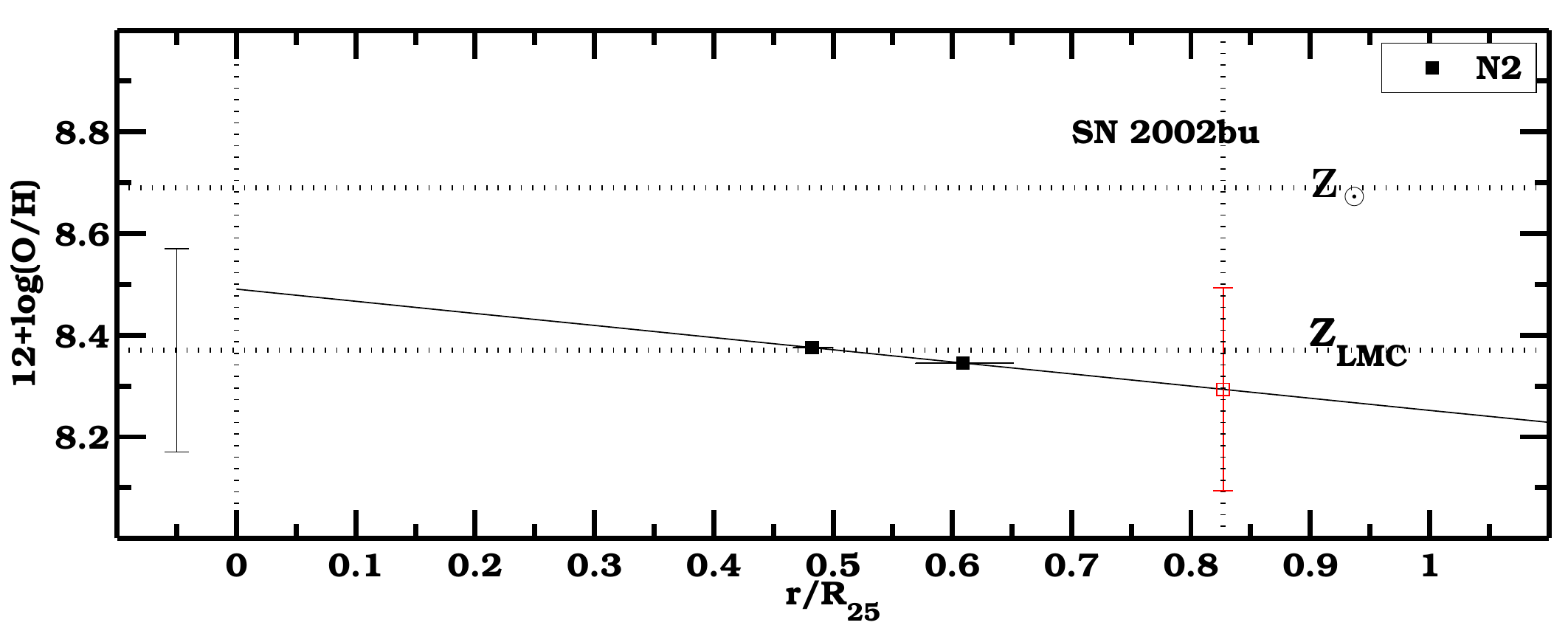}\\
\includegraphics[width=9cm,angle=0]{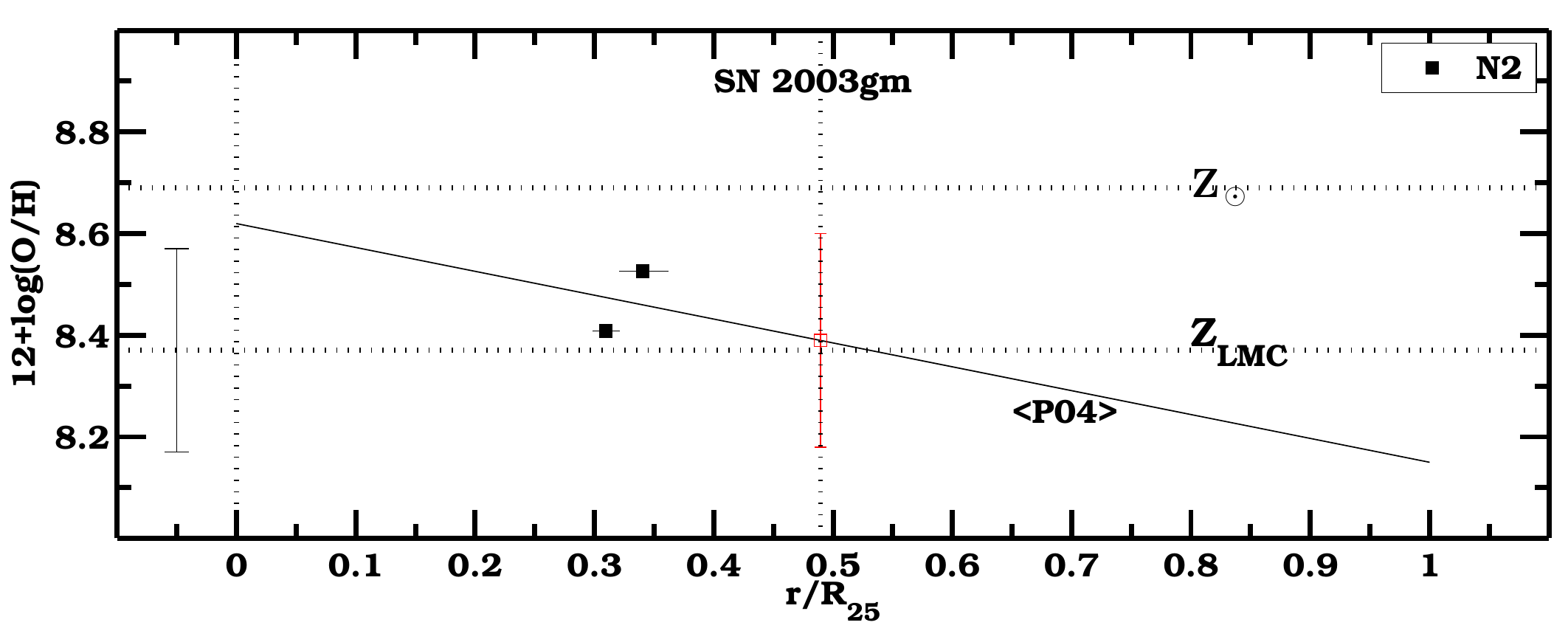}&\includegraphics[width=9cm,angle=0]{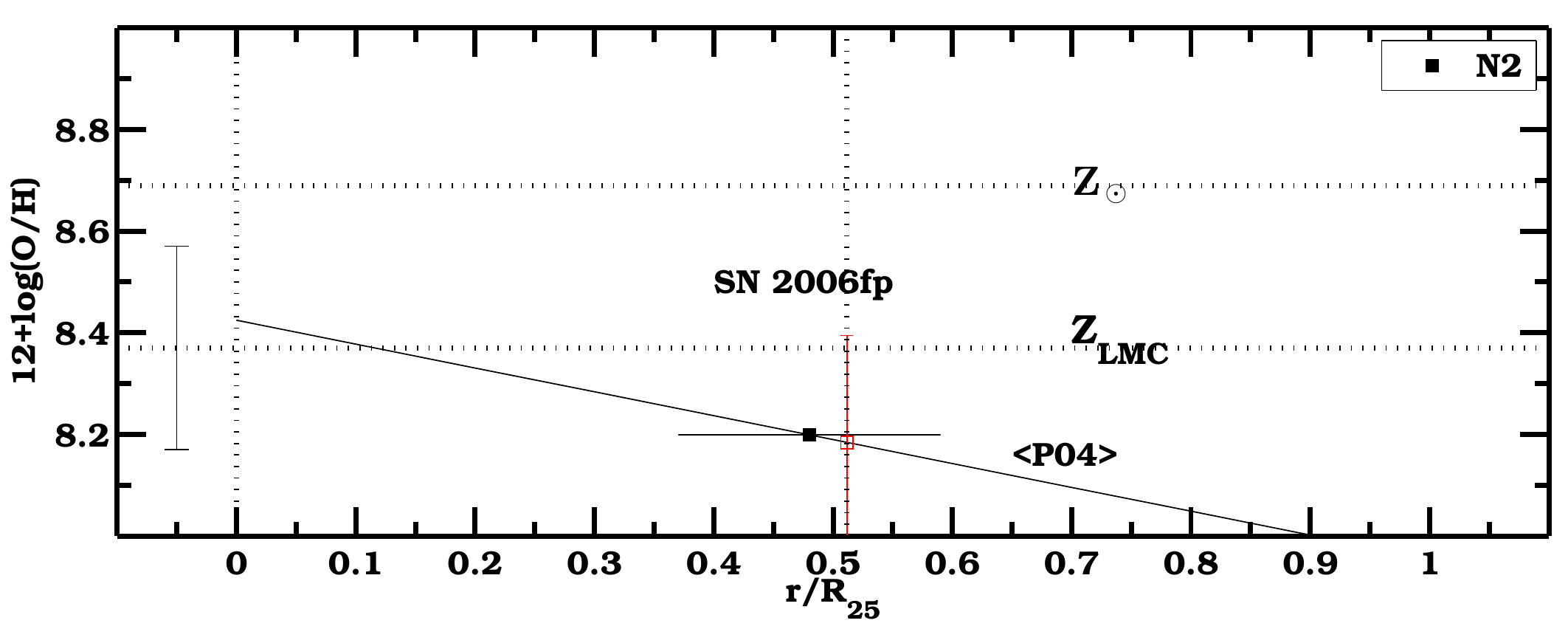}\\
\includegraphics[width=9cm,angle=0]{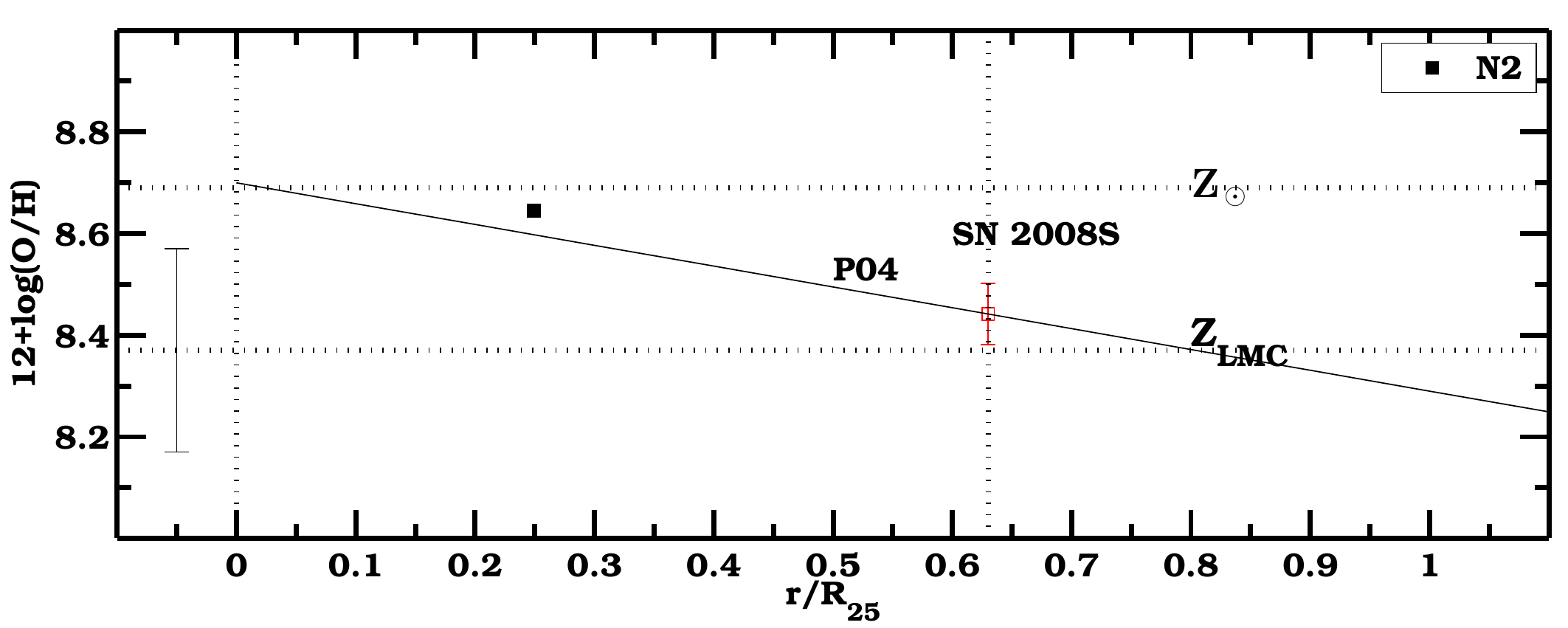}&
\includegraphics[width=9cm,angle=0]{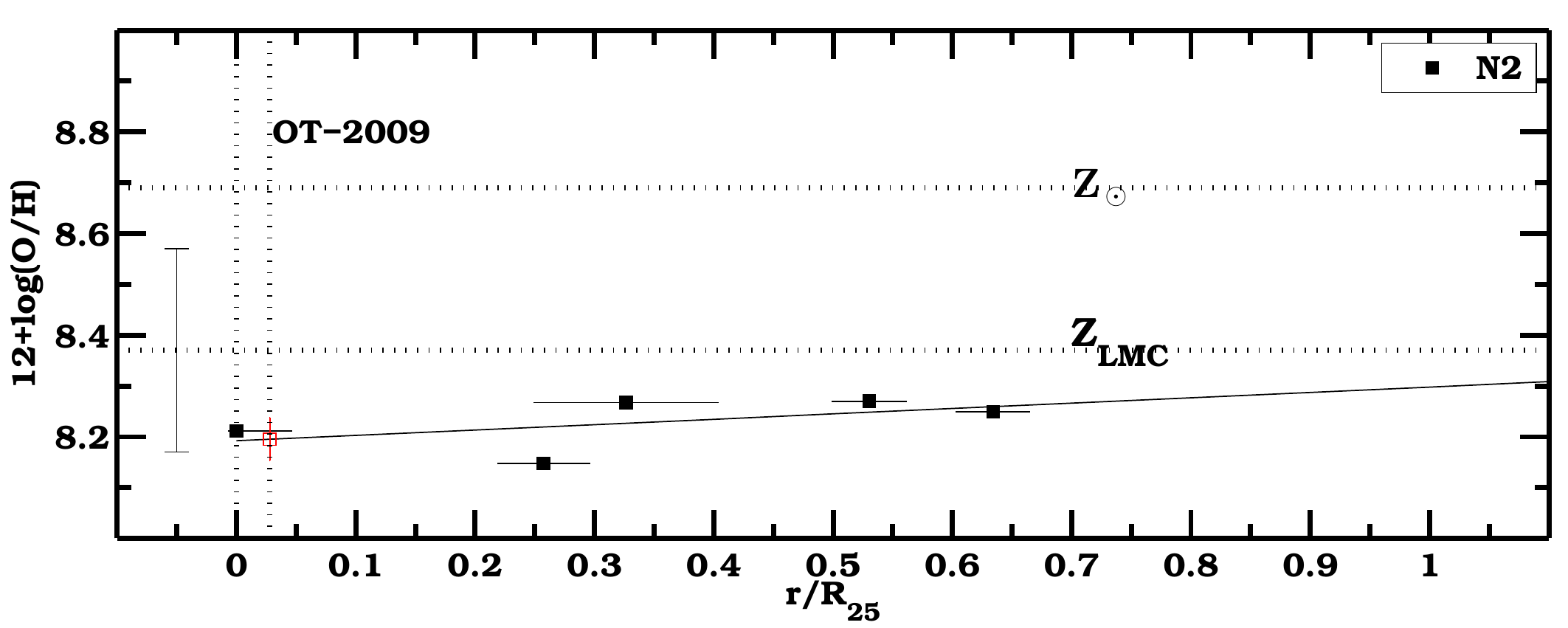}\\
\includegraphics[width=9cm,angle=0]{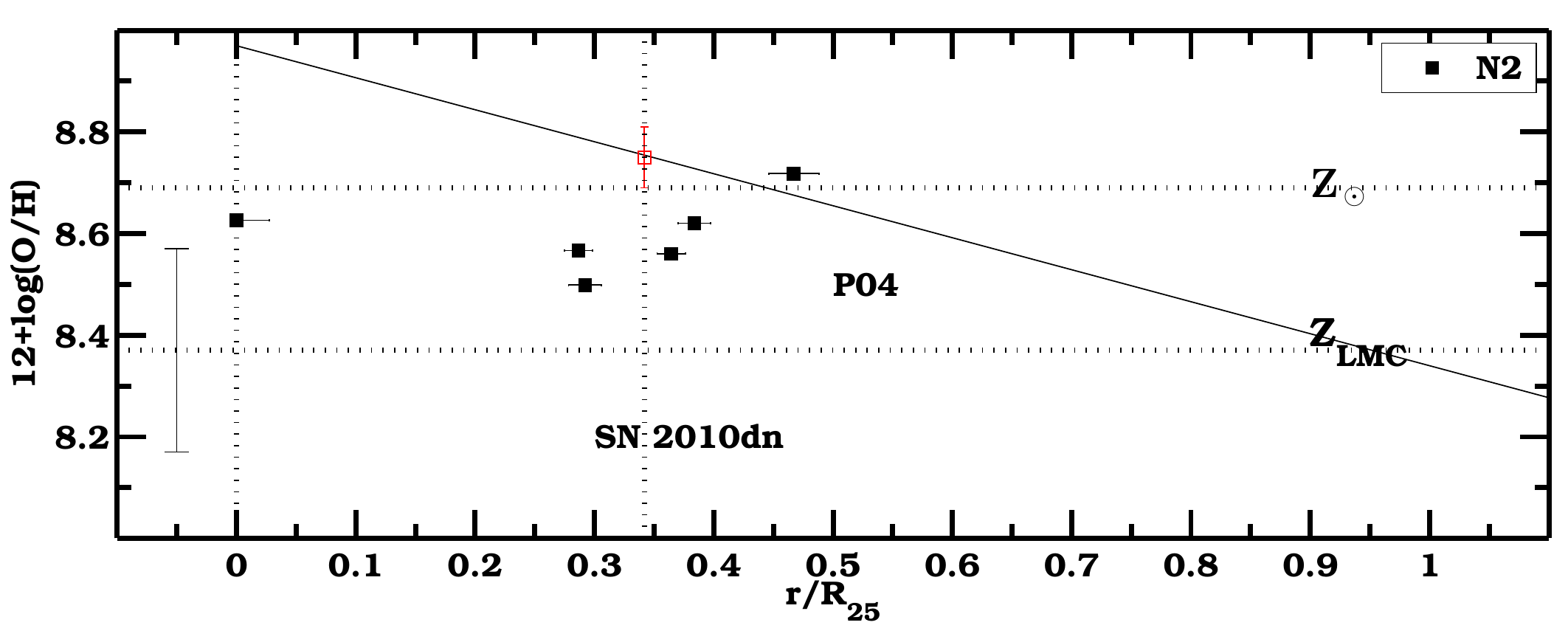}&
\includegraphics[width=9cm,angle=0]{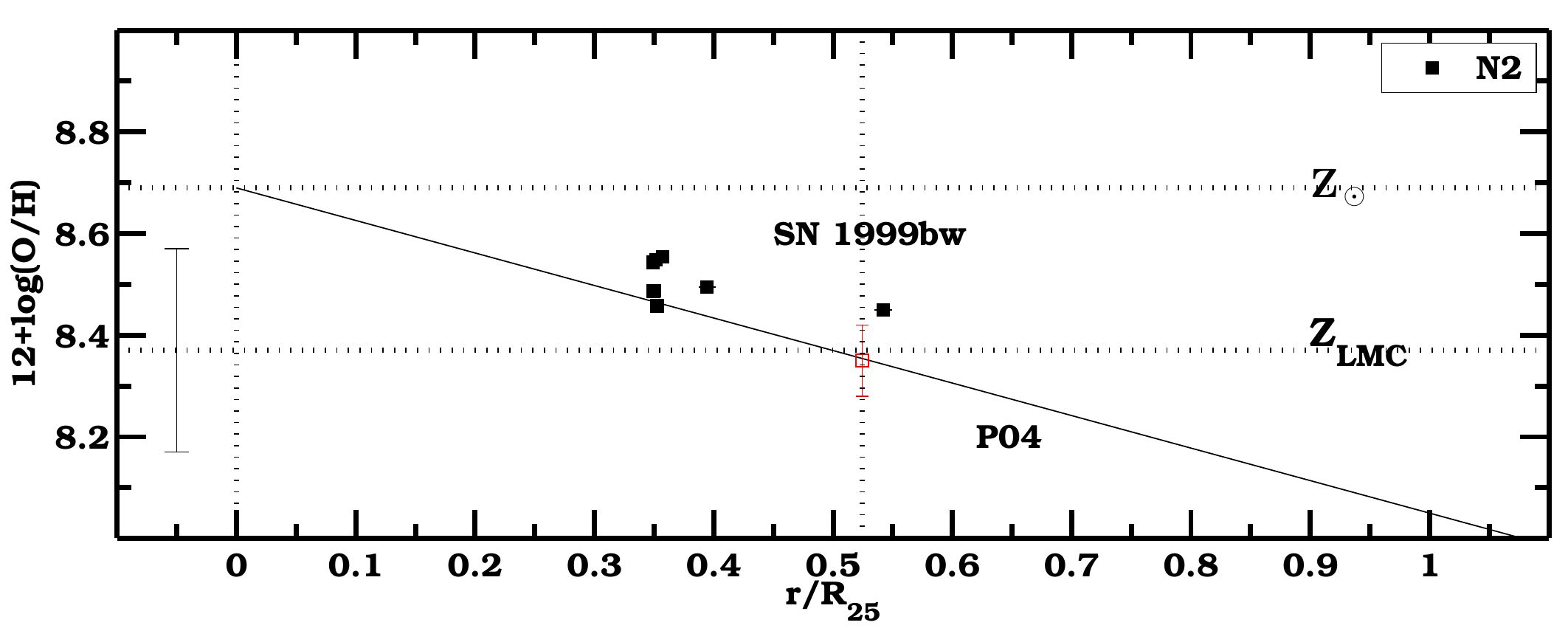}\\
\end{array}$

  \caption{Metallicity gradients of 10 SN~IM host galaxies observed with the NOT$+$ALFOSC. Symbols and lines as in the bottom panel of Fig.~\ref{sn10al}. We label the gradient ``P04" when we use the one measured by \citet{pilyugin04}, and ``$<$P04$>$" when we assumed the average P04 gradient. The data of the hosts of V1 and OT2005 are not shown as for them we do not determined or adopt a gradient, but we only measured the metallicity of one \ion{H}{ii} region close the SN location.\label{allgradSNimpo}}
 \end{figure}}

For each \ion{H}{ii} region with measured N2 metallicity and [\ion{O}{iii}]~$\lambda$5007/H$\beta$)~$<$~0.61/(N2$-$0.05) + 1.3, and for each SN, we computed their de-projected 
and normalized distance from their host-galaxy nuclei, following the method illustrated by \citet{hakobyan09,hakobyan12}.

In order to do that, we established the coordinates of the \ion{H}{ii} region/SN, and collected all the 
necessary information about its host galaxy: nucleus coordinates, major (2 R$_{25}$) and minor (2b) axes, 
position angle (PA) and morphological t-type. 
All these data with the corresponding references are listed in Table~\ref{gal}, which also reports the 
result for the de-projected distance of each SN, r$_{SN}$/R$_{25}$.
We mainly used NED to collect the SN and galaxy coordinates as well as the galaxy dimensions and position 
angle. The Asiago Supernova 
Catalogue\footnote{\href{https://heasarc.gsfc.nasa.gov/W3Browse/all/asiagosn.html}
{https://heasarc.gsfc.nasa.gov/W3Browse/all/asiagosn.html}} (ASC) was used for the t-type, 
SIMBAD\footnote{\href{http://simbad.u-strasbg.fr/simbad/}{http://simbad.u-strasbg.fr/simbad/}} and 
HYPERLEDA\footnote{\href{http://atlas.obs-hp.fr/hyperleda/}{http://atlas.obs-hp.fr/hyperleda/}} was
used when neither NED nor ASC included the aforementioned galaxy data.

\begin{figure*}
 \centering
$ \begin{array}{cc}
\includegraphics[width=9cm]{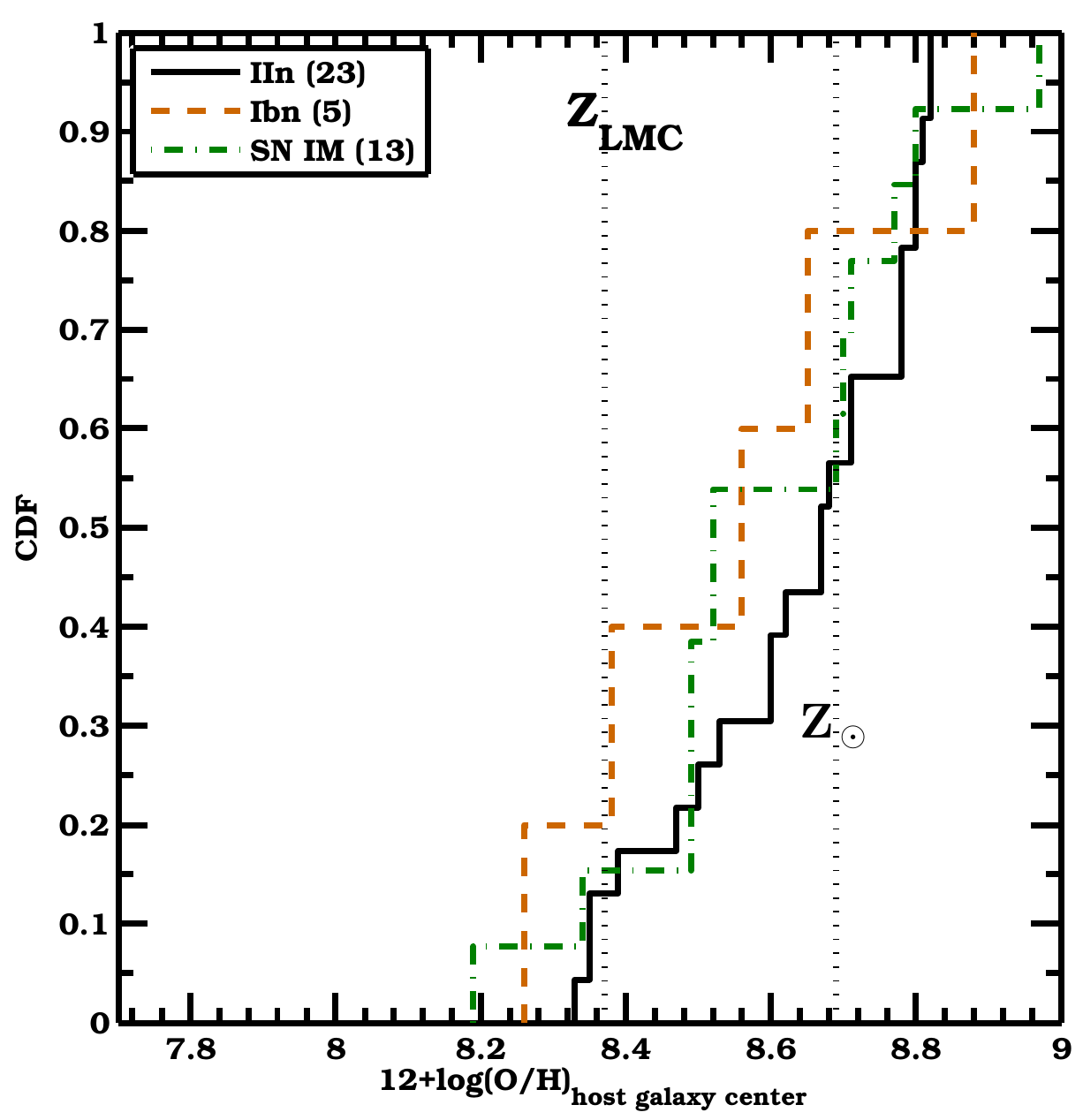}    &
\includegraphics[width=9cm]{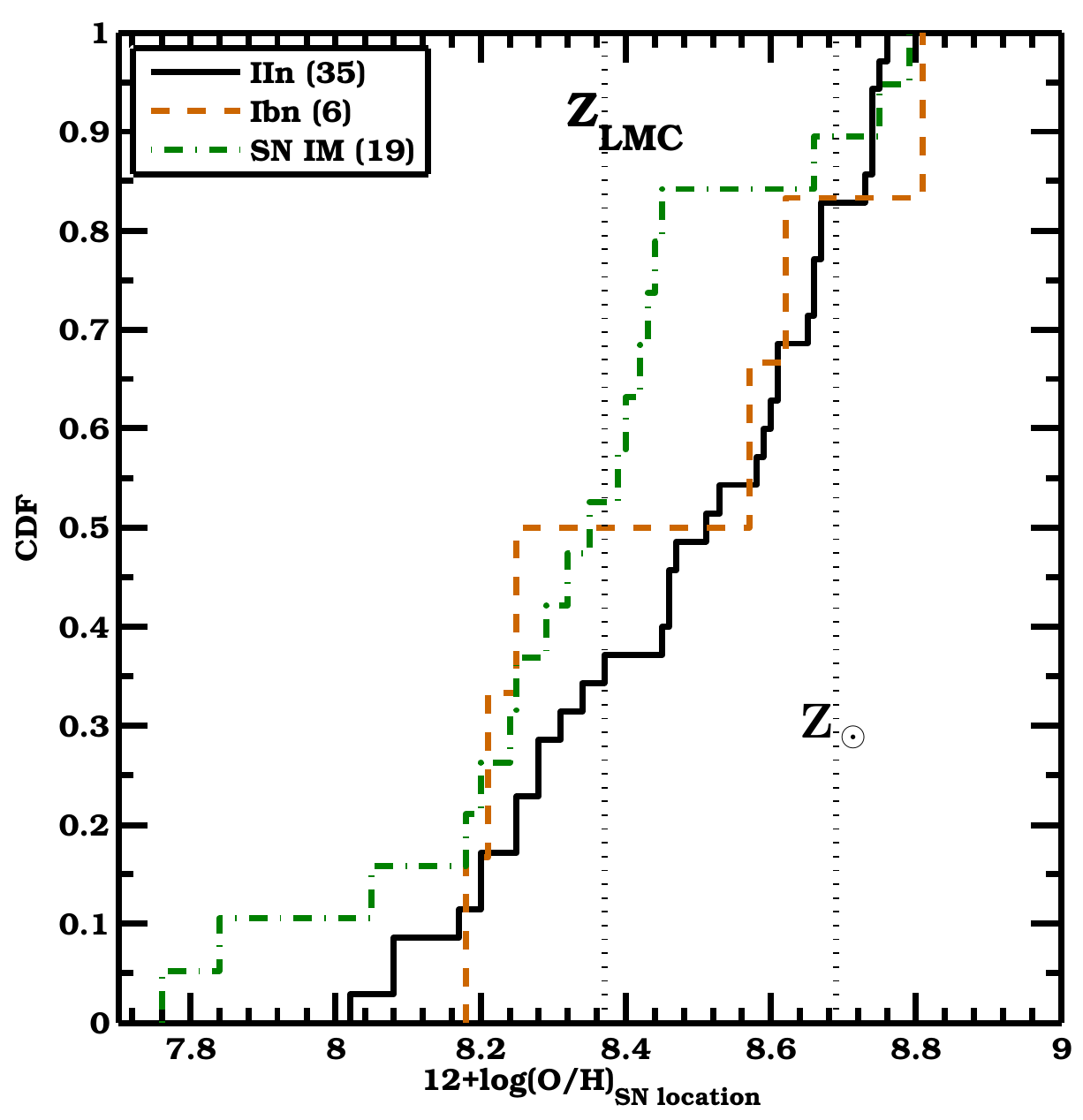}   \\
\includegraphics[width=9cm]{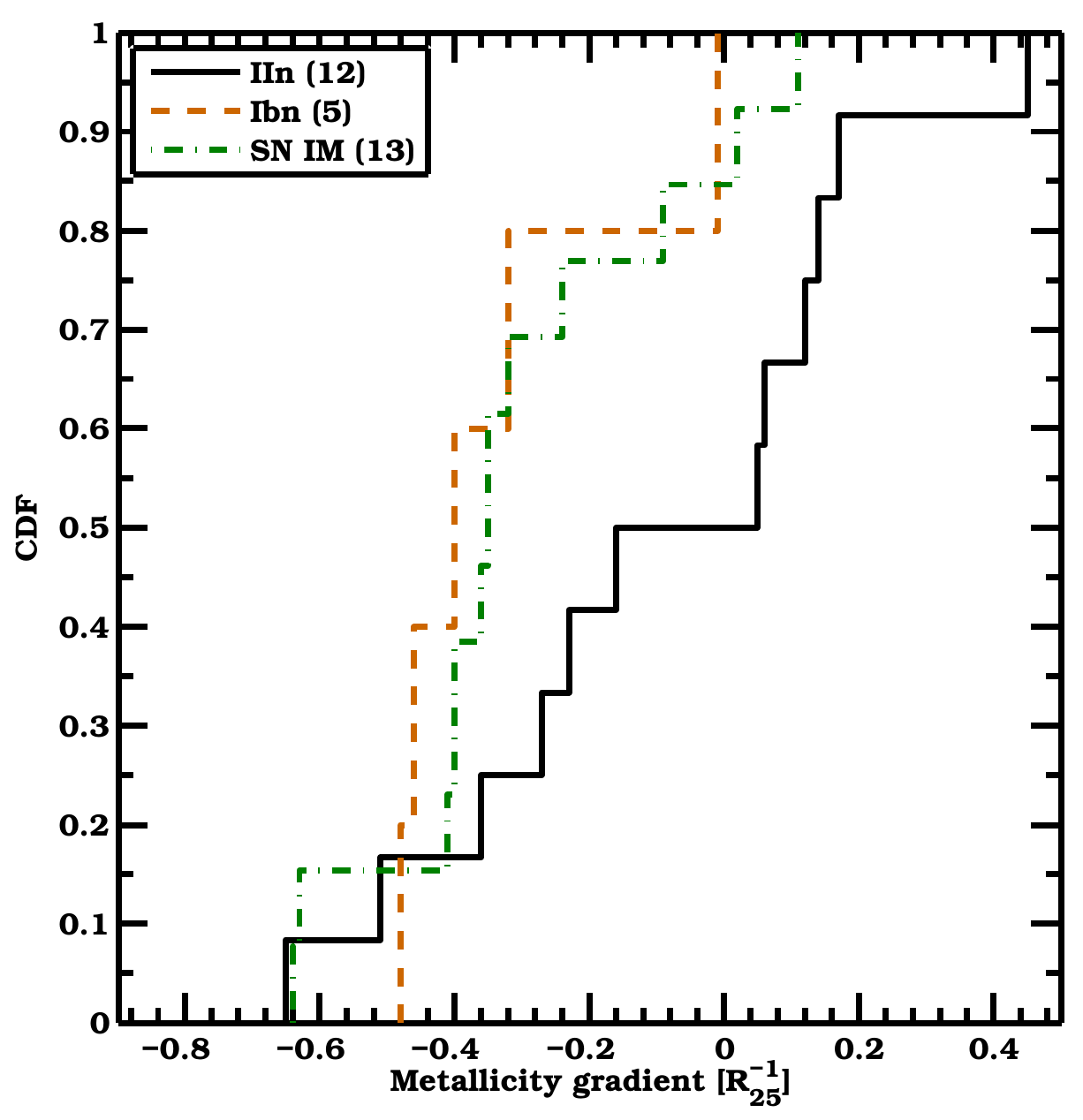}     &
\includegraphics[width=9cm]{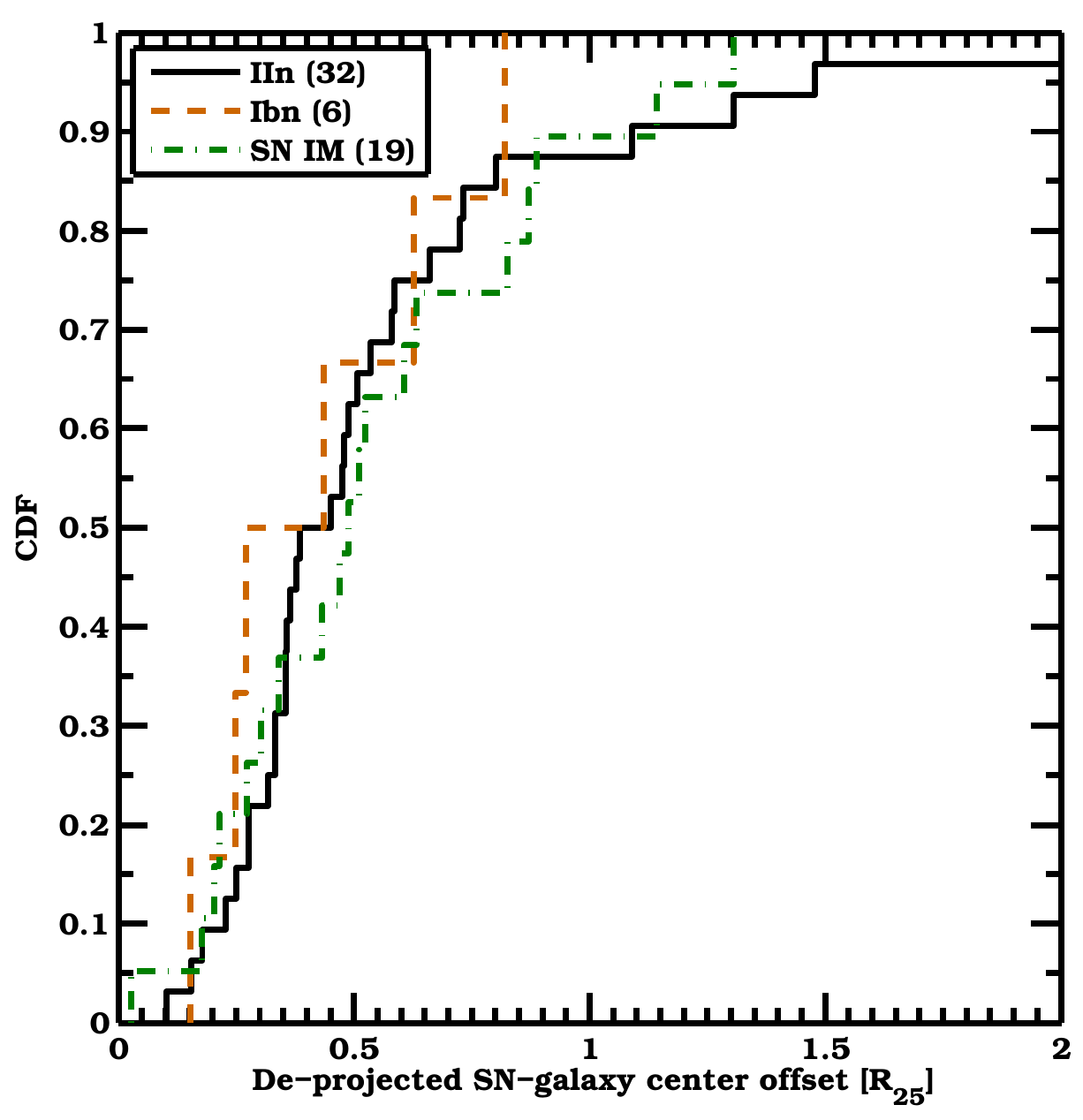}    \\
\end{array}$
  \caption{\textit{(Top-left panel)} Cumulative distribution functions (CDFs) of the 
  central metallicities for SN~IIn, Ibn and SN~IM host galaxies. SN Ibn and SN~IM hosts 
  show slightly lower metallicities than SN~IIn hosts. \textit{(Top-right panel)}  CDFs of 
  the metallicity at the SN location for the same three classes.  SN~IMs and SNe~Ibn are 
  located in lower metallicity environments than those of SNe~IIn. \textit{(Bottom 
  panels)} CDFs of the host metallicity gradients and de-projected SN distances from the 
  host center for SNe~IIn, Ibn, and SN~IMs.\label{cdfmetal}}
 \end{figure*}

\begin{figure*}
 \centering
\includegraphics[width=10cm]{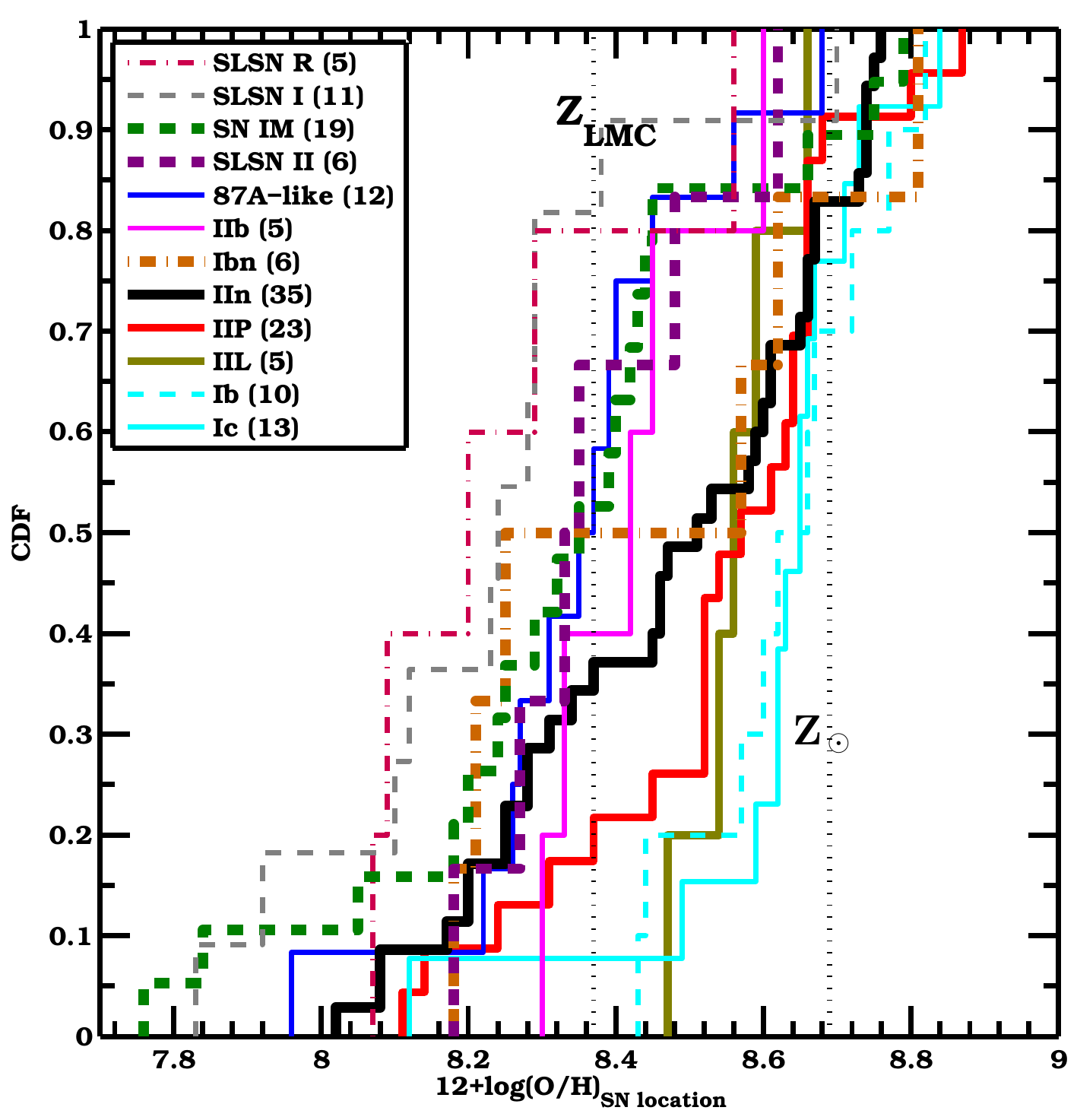} 
  \caption{CDFs of the metallicity at the SN location for CSI transients and for other SN classes from the literature. The legend entries have been ordered by mean metallicity. \label{cdfmetalALL}}
 \end{figure*}

Having obtained the distance from the nucleus for each \ion{H}{ii} region (see the first column of Table~\ref{tab:lineratios}), 
we plotted their metallicities 
versus these distances, and performed a linear chi-square fit to the data.  
An example of such a fit is reported in the bottom panel of 
Fig.~\ref{sn10al}. In most cases, the metallicity was found to 
 decrease as the distance from the center increases (with the exception of nine hosts, where flat or positive gradients were derived). It is well known that there is such a negative metallicity gradient in galaxies (e.g., P04), likely because the star formation rate depends on the local density and therefore more metals have been produced in the denser, central parts of the galaxies. All the data and the fits are shown in Figs.~\ref{allgradIIn}, \ref{allgradIbnIacsm} 
 and \ref{allgradSNimpo}.

The linear fits allowed us to interpolate or extrapolate the metallicity at the computed
 SN distance from the host center (see e.g. the red square in the bottom panel of Fig.~\ref{sn10al}). These 
 values were taken to be the local SN metallicity estimate. We note that these estimates were always 
 found to match the metallicity of the \ion{H}{ii} region closest to the SN.
 In this way, we provided oxygen abundances also for those SNe that were not associated with a bright \ion{H}{ii} region (this is the case of several SN~IMs, see Sect.~\ref{sec:ncr}), and for those SN~IMs or SNe~IIn whose flux was still dominating the emission (e.g., SN~IM 2000ch). 
 All the local and central metallicities, and the metallicity gradients are reported in Table~\ref{metal}.

When more than two \ion{H}{ii} regions were observed, we computed the uncertainty on the 
interpolated/extrapolated SN metallicity as the standard deviation of 1000 interpolation/extrapolations of 
1000 Monte Carlo-simulated linear fits, based on the uncertainty
on the slope and on the central metallicity. These uncertainties
 are those reported in Table~\ref{metal} and do not include the systematic N2 error (see 
 \citealp{taddia13met} for
 a motivation of this choice).
 When only two measurements were obtained, we assumed 0.2 as the uncertainty for the metallicity at the SN position and at the host center.
 
 For most of the galaxies whose spectra were retrieved from online archives, only a single measurement  was possible, often at the host center. In these cases, we adopted a standard metallicity gradient (--0.47~R$_{25}^{-1}$) to extrapolate the metallicity at the SN position. This value corresponds to 
 the average gradient of the large galaxy sample studied by P04. In these cases the error on the SN metallicity included the extrapolation uncertainty, given by the $\sim$0.1~R$_{25}^{-1}$ dispersion of the measured metallicity gradients (see Sect.~\ref{sec:intracomp}). 

\subsection{Local metallicity values from the literature}

In the literature we found several published values for the metallicities of CSI~SN host galaxies.
Some of them were local (e.g. a few from H14), most of them were measurements at the host galaxy center (e.g., all those obtained by KK12, except that of the host of SN 1994Y). KK12 presented O3N2 metallicity estimates based on SDSS spectral measurement performed by an MPA-JHU collaboration and available online at \href{http://www.mpa-garching.mpg.de/SDSS/}{http://www.mpa-garching.mpg.de/SDSS/}. Therefore we retrieved the needed line fluxes for the N2 method for all the SN~IIn hosts presented by KK12 and computed the metallicity measurements in the N2 scale.
H14 and \citet{roming12} present five O3N2 values, so we had to convert them to the N2 scale, using the relation presented by \citet{kewley08}.
In the cases where the metallicity was measured at the host nucleus, we assumed the aforementioned standard gradient to obtain the metallicity at the SN de-projected distance. We note that we could have used the average gradient that we measured in some of our galaxies (see Sect.~\ref{sec:results}), instead of the average gradient from P04. However, the sample of galaxies observed by P04 is larger and representative of many morphological types, so we prefer to adopt their value.
This assumption could potentially affect the results concerning SNe~IIn (for 13 of the 35 SNe~IIn we adopt the P04 average gradient), whereas it is not important for most of the SN~IMs or SN~Ibn, where we do not need to assume any average gradient. However, we will show in the following sections that the SN~IIn results are solid independently of the average gradient assumption.

\section{Metallicity results}
\label{sec:results}
In the following sections we describe the metallicity results for our sample of CSI transients, 
including comparisons among CSI~SN subtypes and to other CC SNe.

\begin{figure*}
 \centering
$ \begin{array}{cc}
 \includegraphics[width=9cm]{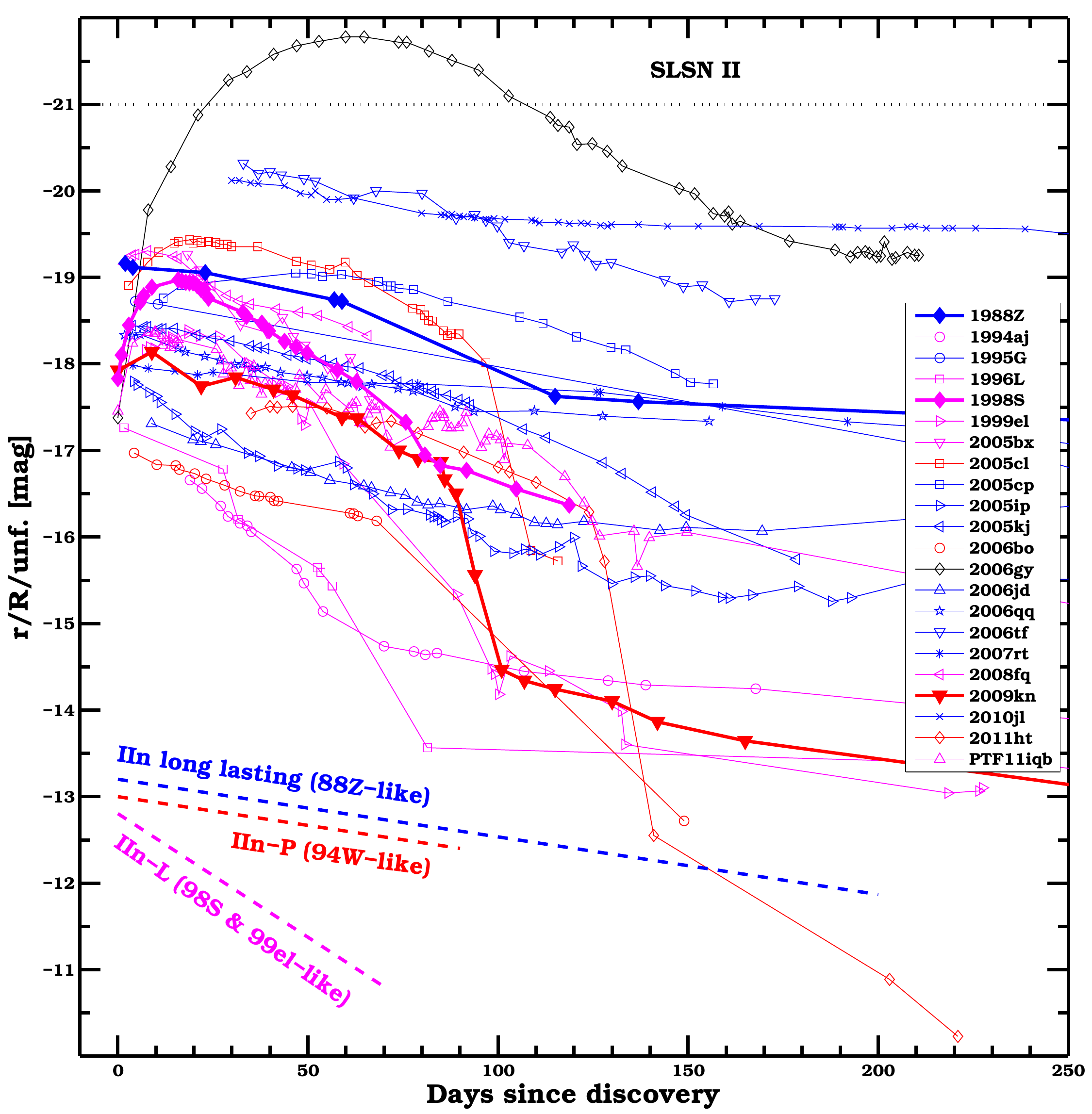} &
\includegraphics[width=9cm]{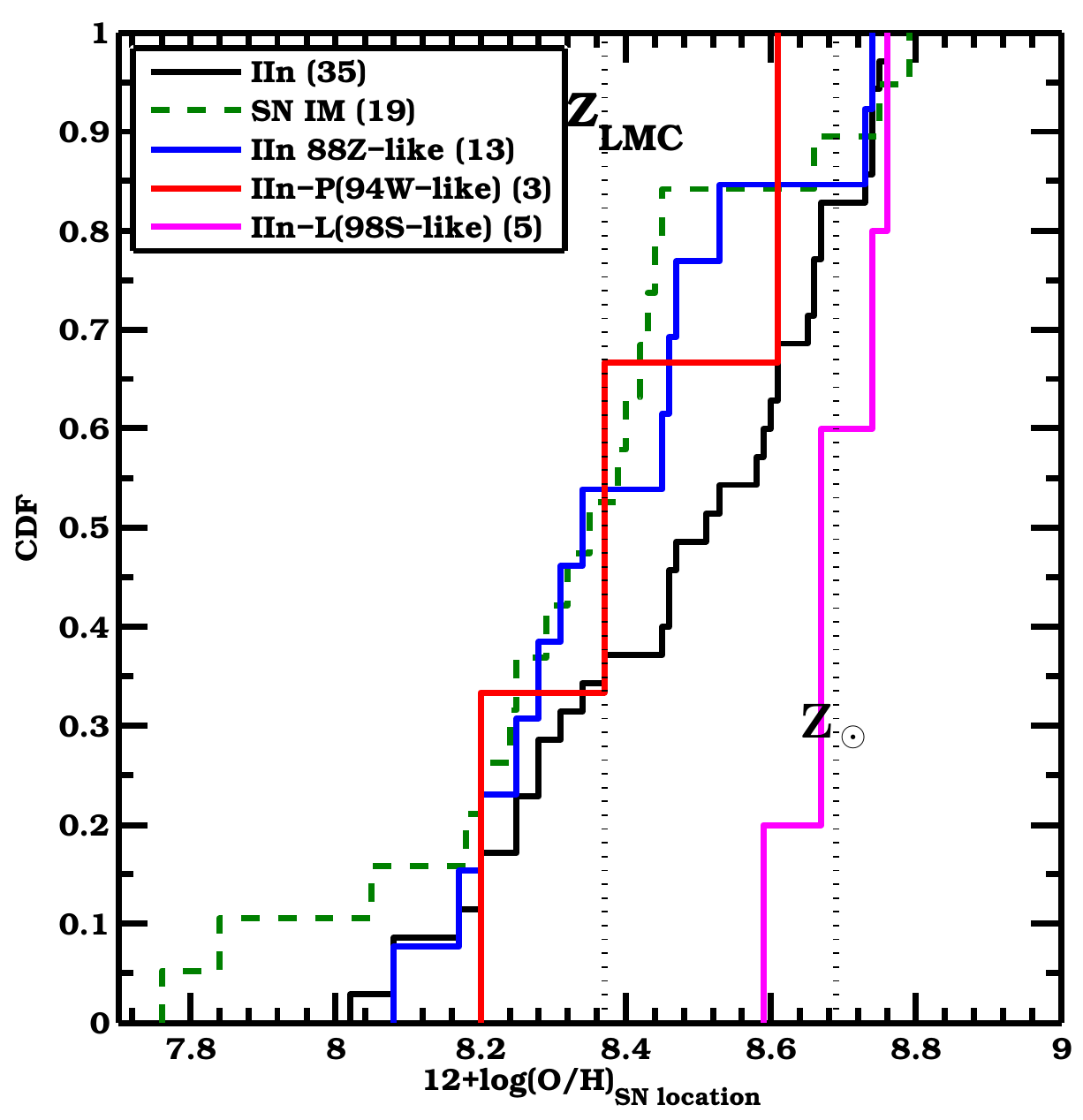}  
 \end{array}$
  \caption{\textit{(Left-hand panel)} SN~IIn light curves from the literature. Three main subclasses are identified (plus the SLSNe) based on the light curve shape. Long-lasting SNe~IIn (1988Z-like; blue) show slow decline ($\sim$0.7~mag~(100~days)$^{-1}$) and sustained luminosity for $\gtrsim$5 months. 
SN~IIn-P (1994W-like; red), show a $\sim$3--4~month plateau followed  by a sharp drop, similar to what is observed in SNe~IIP. SNe~IIn-L (1998S or 1999el-like, depending on the spectral evolution; magenta) show a faster decline ($\sim$2.9~mag~(100~days)$^{-1}$). References for the light curve data and the extinction of SNe with measured metallicity can be found in Table~\ref{tab:phot}. For SNe~1994aj, 1999el, 2005cl, 2006gy, 2006tf and 2009kn the references are \citet{benetti98}, \citet{dicarlo02}, \citet{kiewe12}, \citet{smith07_06gy}, \citet{smith08_06tf}, and \citet{kankare12}, respectively. \textit{(Right-hand panel)} CDFs of the metallicity at the SN location for our CSI transients, including three SN~IIn subclasses (1988Z-like, 1994W-like, 1998S-like). 1998S-like SNe are located at higher metallicity compared to the other subtypes.\label{IInsub}}
 \end{figure*}
 
 \onlfig{11}{\begin{figure}
 \centering
\includegraphics[width=18cm]{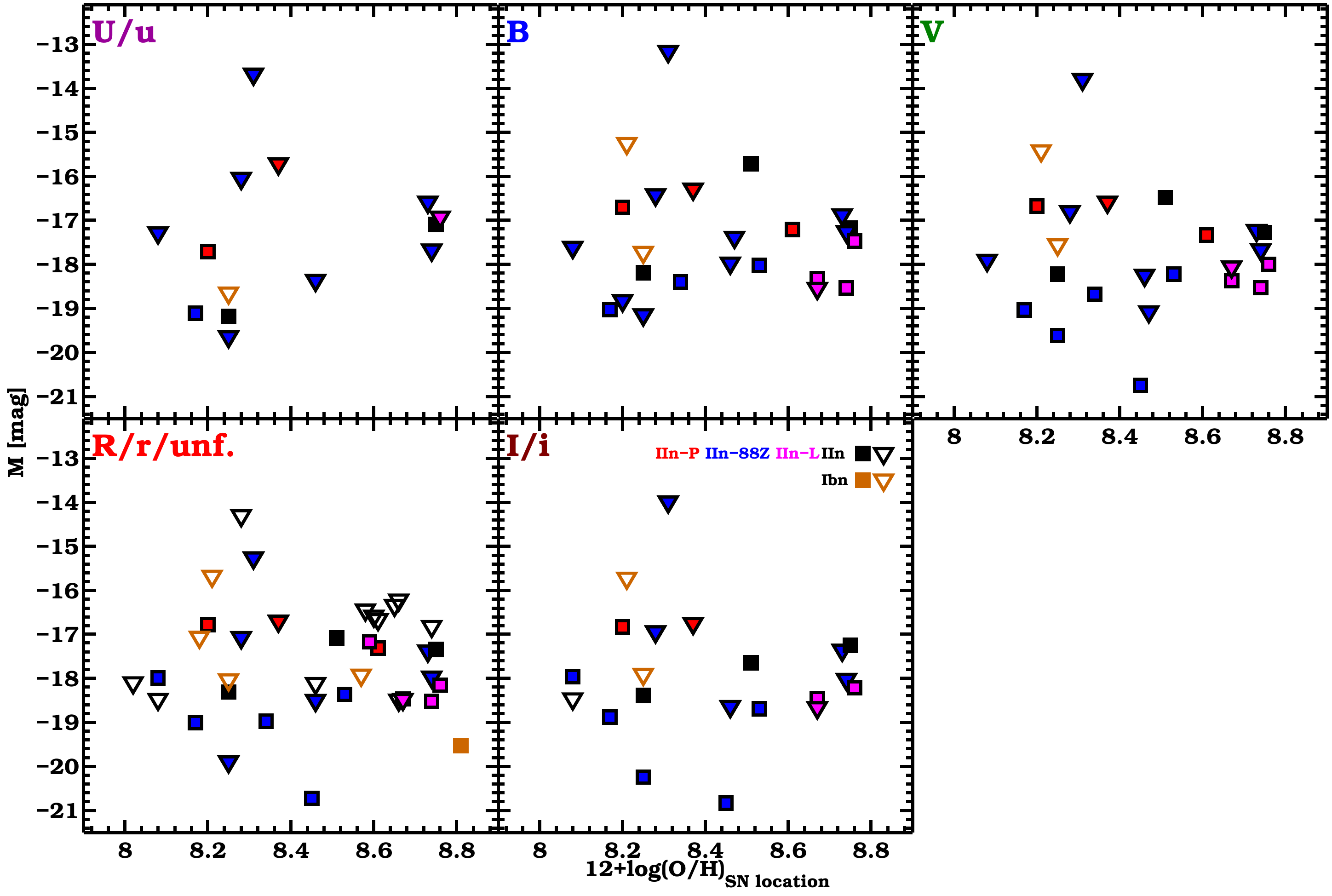} 
  \caption{Absolute peak magnitudes for SNe~IIn and Ibn against the metallicity at the SN position. Triangles are upper limits.\label{Mvsmet}}
 \end{figure}}


\subsection{Metallicity comparison among CSI transients}
\label{sec:intracomp}

In Table~\ref{metal} we report the average values for the metallicity at the host center, at the SN position, 
and for the metallicity gradient. The mean gradients are computed excluding the objects where the 
gradient was assumed to be --0.47~R$_{25}^{-1}$. 

SNe~IIn show slightly higher average oxygen abundance at their locations (8.47$\pm$0.04) than do SN~IMs (8.33$\pm$0.06). The latter values are consistent with the 
metallicity of the Large Magellanic Cloud (LMC has 8.37, \citealp{russell90}), which is sub-solar 
(the Sun has 8.69, \citealp{asplund09}). To understand if this difference is statistically significant, computing the 
mean values is not sufficient. Therefore, we need to compare their cumulative distributions, which are 
shown in the top-right panel of Fig.~\ref{cdfmetal}, through the Kolmogorov-Smirnov (K-S) test. We found 
that the probability that the metallicity distributions of SNe~IIn and SN~IMs are different by chance is only 1\%. If we exclude the CSI SN hosts where we assumed the average P04 metallicity gradient to determine the local oxygen abundance,  this difference still holds (p-value~$=$~1.5\%), with $<$12$+$log(O/H)$>_{IIn}$~$=$~8.48$\pm$0.04 and $<$12$+$log(O/H)$>_{IM}$~$=$~8.31$\pm$0.07. Also if we use the N2 calibration by \citet{marino13}, the K-S test gives a p-value of 1\%.
 To account for the uncertainties of each metallicity measurement (see Table~\ref{metal}), we performed the K-S test between 10$^6$ pairs of Monte Carlo-simulated metallicity distributions of SNe~IIn and SN~IMs. The 68\% of the resulting p-values were found to be $\leq$0.17, suggesting that the difference is real.
  We discuss this result and its implications for the SN~IIn progenitor scenario in 
Sect.~\ref{sec:disc}. 
SNe~Ibn show an average metallicity (8.44$\pm$0.11) similar to that of SNe~IIn
and a distribution that appears in between those of SNe~IIn and SN~IMs, although here our sample is limited to only 6 SNe~Ibn and there is no statistically-significant difference with the other two distributions (p-values$>$56\%).

The average oxygen abundances at the host center for SNe~IIn and SN~IMs are 8.63$\pm$0.03 
and 8.59$\pm$0.06, respectively. These distributions (top-left panel of Fig.~\ref{cdfmetal}) 
turned out to be similar, with p-value$=$47\%. SN~Ibn hosts exhibit a slightly lower mean central metallicity than SNe~IIn (8.55$\pm$0.11, but with p-value$=$54\%). Metallicities at the center are expected to be higher than those at the SN explosion site due to the negative metallicity gradient typically observed in the galaxies (e.g., P04).

The mean gradient for the host galaxies of SNe~IIn, Ibn and SN~IMs is $-$0.10$\pm$0.09~R$_{25}^{-1}$, 
$-$0.33$\pm$0.09~R$_{25}^{-1}$ and $-$0.31$\pm$0.06~R$_{25}^{-1}$, respectively. These 
absolute values are slightly lower than the average gradient ($-$0.47~R$_{25}^{-1}$) of the galaxies 
studied by P04 and adopted for the SNe in our sample with a single metallicity estimate at the galaxy 
center. 
The gradient distribution of SN~IIn hosts shows higher values than those of SN~IMs and SNe~Ibn (bottom-left panel of Fig.~\ref{cdfmetal}), however the K$-$S 
tests result in p-values~$>$~12\%.

SNe~IIn, Ibn and SN~IMs show similar average distances from the host center, with 0.64$\pm$0.15~R$_{25}$, 0.43$\pm$0.10~R$_{25}$ and 0.54$\pm$0.08~R$_{25}$, respectively. Their 
distributions are almost identical (p-values$>$49\%, see the bottom-right panel of Fig.~\ref{cdfmetal}).

We note that if we had assumed our average metallicity gradient instead of that from P04 for the hosts with a single measurement at the center, then the metallicity distribution of SNe~IIn (which is the only one that would be affected by a  different choice of the average metallicity gradient)
would have been at slightly higher metallicity, making the difference with SN~IMs even more significant.

\subsection{Metallicity comparison to other SN types}
\label{sec:othercomp}

Figure~\ref{cdfmetalALL} shows the cumulative distributions of the metallicity at the SN location for our 
CSI transients, along with those of other SN types previously published in 
\citeauthor{taddia13met} (2013b, see figure~15 and references in the text) and \citet{leloudas14}. All the metallicities included in the plot are local and obtained 
with the N2 method in order to enable a reliable comparison among the different SN classes. It is evident 
that the SN~IMs have relatively low metallicities along with SN~1987A-like events, 
SNe~IIb and SLSN~II. When we compare the metallicity distribution of SN~IMs to that of SNe~IIP, their difference 
has high statistical significance (p-value~$=$~0.09\%; even higher if the comparison is to SNe~Ibc). SNe~IIn 
show higher metallicities than SN~IMs, but still lower than those of SNe~IIP (p-value$=$29\%) and SNe~Ibc (p-value$=$2.5\%, highly significant). All the SN types show higher metallicities compared to SLSNe~I and SLSNe~R.

\section{Host galaxy properties and CSI transient observables}
\label{sec:prop}
In the following sections we investigate any potential correlations between the 
observables of CSI transients and the properties of their environment, 
in particular the 
metallicity. This exercise aims to establish to what extent metallicity plays an important role in shaping the appearance of these events.

\subsection{Metallicity and SN properties}

\subsubsection{Metallicity and SN~IIn subtypes}

 \onlfig{12}{\begin{figure}
 \centering
$ \begin{array}{cc}
\includegraphics[width=9cm,angle=0]{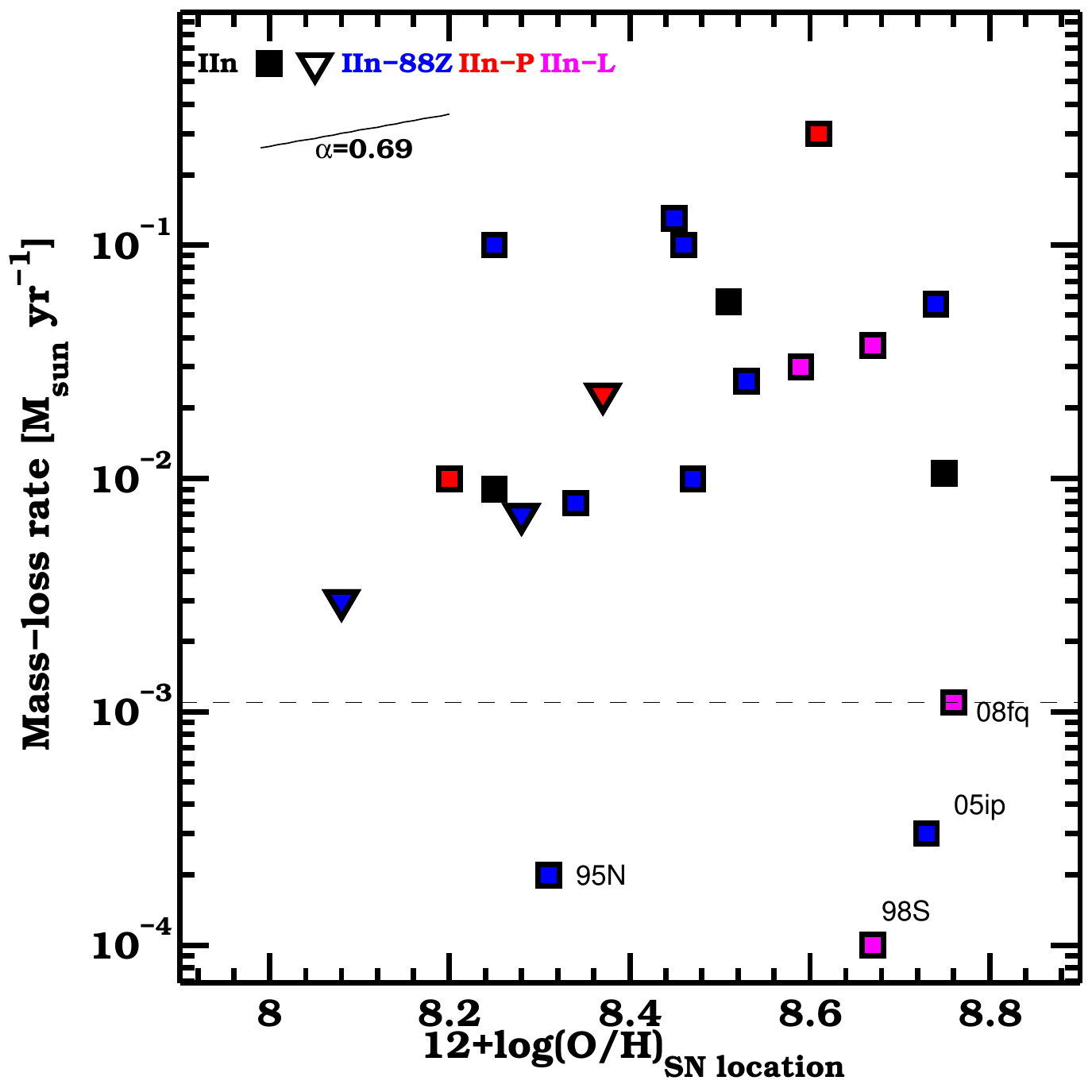} &
\includegraphics[width=9cm,angle=0]{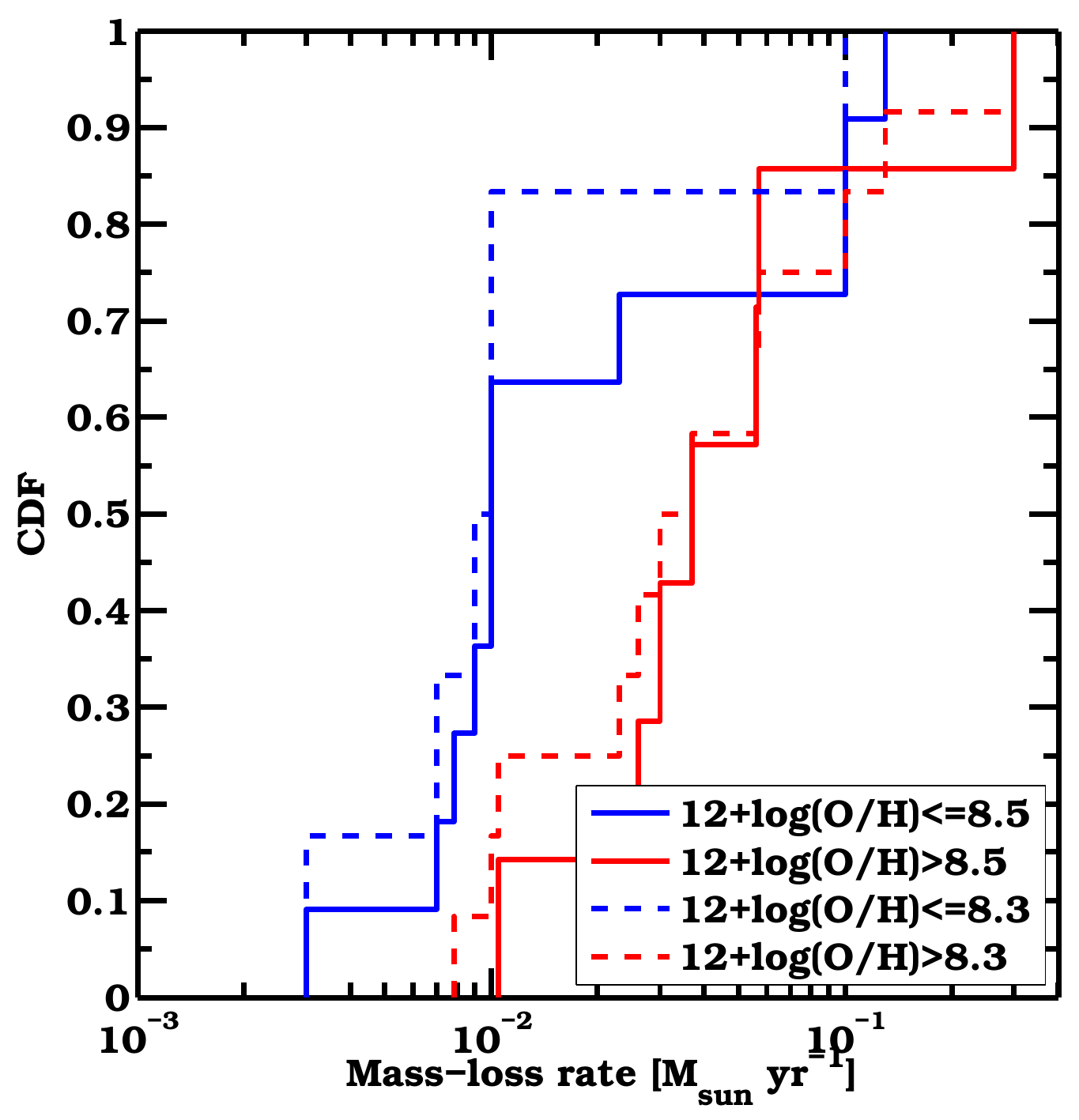} \\
\includegraphics[width=9cm,angle=0]{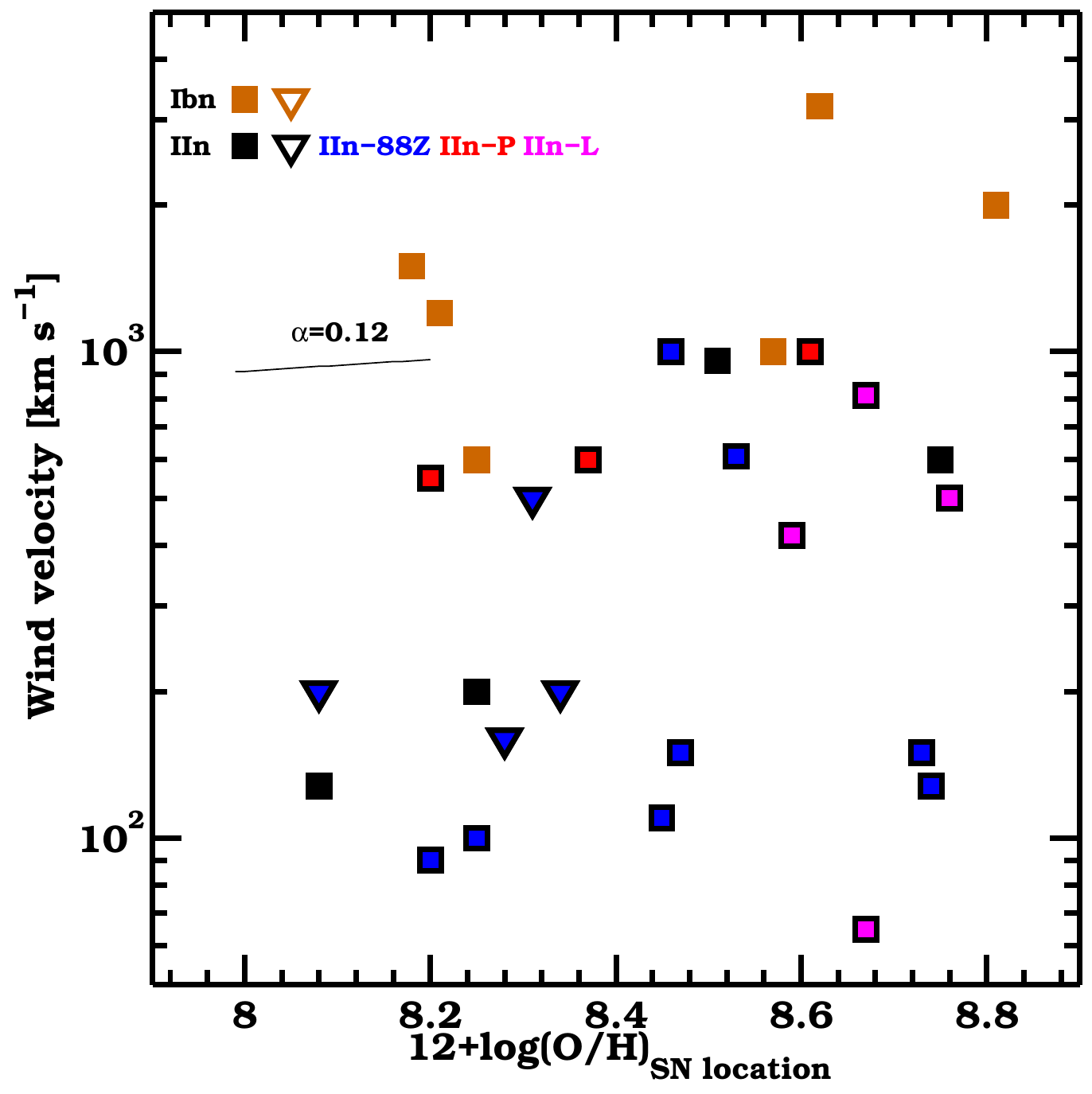} &
\includegraphics[width=9cm,angle=0]{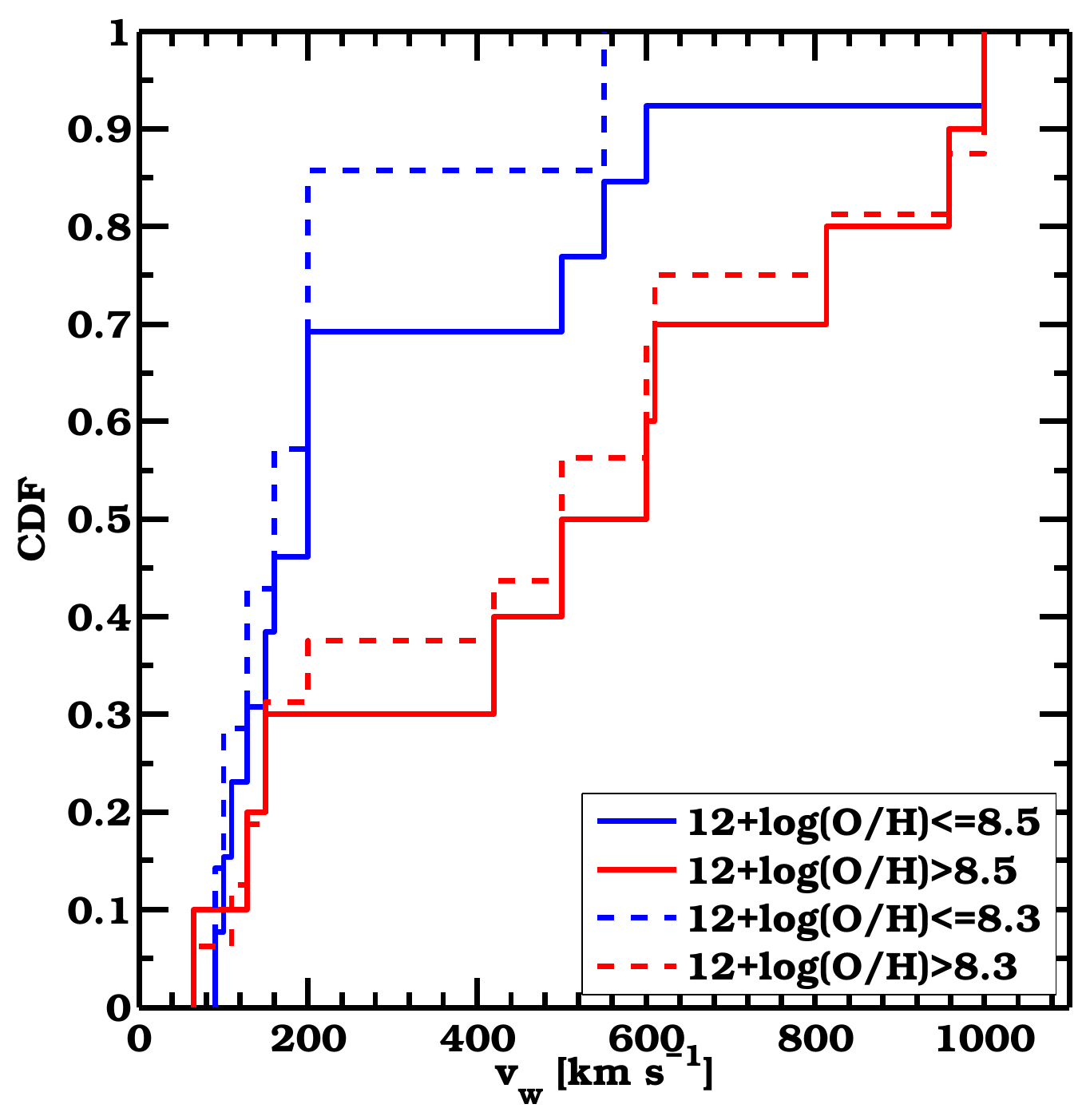}\\
\end{array} $
  \caption{\textit{(Top-left panel)} Mass-loss rates of SNe~IIn versus local metallicity measurements. Triangles are upper limits. Above 10$^{-3}$~M$_{\odot}$~yr$^{-1}$, the mass loss rates appear to be higher at higher metallicity.
The power-law (PL) dependence on metallicity ($\alpha~$=$~$~0.69 \citep[e.g.,][]{vink11}) of the line-driven mass-loss rates is shown with a black segment and is consistent with the data.  SNe~IIn-L (98S-like) typically show lower mass-loss rates than long-lasting SN~IIn (88Z-like).
\textit{(Top-right panel)} Mass-loss rate CDFs for metal-poor and metal-rich SNe~IIn, shown to better highlight that at higher metallicities the mass loss rates tend to be higher. Two different metallicity cuts are shown, at log(O/H)$+$12~$=$~8.3 and 8.5. 
\textit{(Bottom-left panel)} Wind velocities of SNe~IIn and SNe~Ibn versus local metallicities. Triangles are upper limits. The PL dependence on metallicity ($\alpha~$=$~$~0.12 \citep[e.g.,][]{kud02}) of the line-driven wind velocities rates is shown with a black segment, and is consistent with the data. 
SNe~IIn-L (98S-like) typically show higher wind velocities than long-lasting SN~IIn (88Z-like).
\textit{(Bottom-right panel)} Wind-velocity CDFs for metal-poor and metal-rich SNe~IIn, shown to better highlight that at higher metallicities the wind velocities tend to be slightly higher. Two different metallicity cuts are shown, at log(O/H)$+$12~$=$~8.3 and 8.5. \label{masslossvsmet}}
 \end{figure}}
 

In the introduction we described three SN~IIn subgroups, mainly based on the light curve shapes, i.e., 1988Z-like (or long-lasting SNe~IIn), 1994W-like (or SNe~IIn-P), and 1998S-like (or SNe~IIn-L with fast spectral evolution). A collection
of SN~IIn light curves from the literature, including most of the light curves of SNe~IIn with measured metallicity, is shown in Fig.~\ref{IInsub} (left-hand panel), where we display the different subtypes with different colors. 
In our sample we selected the events belonging to the different SN~IIn subclasses (see the second columns of Tables~\ref{tab:phot}--\ref{tab:spec}) to look for possible trends in the metallicity. In Fig.~\ref{IInsub} (right-hand panel) we plot their metallicity CDF. It was possible to subclassify only 21 of the 35 SNe~IIn in our sample, i.e. those with sufficient photometric coverage in the literature.
We found that our SN~IIn-L (we only have 1998S-like SNe) are typically located at solar metallicity ($<$12$+$log(O/H)$>_{IIn-L}$~$=$~8.69$\pm$0.03).
Long-lasting SNe~IIn are found at lower metallicity ($<$12$+$log(O/H)$>_{IIn-88Z}$~$=$~8.38$\pm$0.06), similarly to SN~IMs. Among them, the object showing the largest oxygen abundance is SN~2005ip.
The difference between SN~1998S-like events and long-lasting SNe~IIn is statistically significant (p-value~$=$~0.004) and would be even larger if we had adopted our measured average metallicity gradient instead of that from P04, since we assumed the P04 gradient for the hosts of three (out of five) SNe~IIn-L. It would also be statistically significant if we exclude the hosts where we assumed the P04 gradient from both groups.
1994W-like SNe exhibit a metallicity distribution similar to that of 1988Z-like SNe   ($<$12$+$log(O/H)$>_{IIn-P}$~$=$~8.39$\pm$0.12). In this group we include only SNe~1994W, 2006bo and 2011ht, which do show a plateau followed by a sharp decay.

\subsubsection{Metallicity and SN~IIn/Ibn magnitude at peak}

It has been noticed in the literature that some bright SNe~IIn exploded in low-metallicity environments
\citep{stoll11}. In this paper, we have the opportunity to test if such a trend can be confirmed for a larger 
sample of SNe~IIn. For each CSI~SN we collected their peak apparent magnitudes in five optical bands ($U$/$u$, $B$, $V$, $R$/$r$/unfiltered, $I$/$i$) when available in the literature. We also collected 
the visual Galactic extinction for each SN ($A_{V}(MW)$, from
\citealp{schlafly11}) and, when available in the literature, an estimate of the host extinction ($A_{V}(h)$).

Peak 
magnitudes and extinctions are summarized in Table~\ref{tab:phot}, with references provided.
We computed the distance from the redshift (see Table~\ref{gal}), assuming 
H$_0$~$=$~73.8~km~s$^{-1}$~Mpc$^{-1}$ \citep{riess11}. 
Using the distance and the extinction estimate, from the apparent peak magnitude of each filter we 
computed the absolute peak magnitude ($M_{u/U,B,V,R/r/unf.,I/i}$), which are reported in Table~\ref{tab:abs} and plotted against the local 
metallicity estimates in Fig.~\ref{Mvsmet}. When the absolute peak magnitude is only an upper limit, it is 
marked with a triangle. This figure does not show any clear correlation or trend between metallicity and luminosity at peak. When we group our events in different metallicity bins (i.e., log(O/H)$+$12~$\leq$~8.2, 8.2~$<$log(O/H)$+$12~$\geq$~8.4,  8.4~$<$~log(O/H)$+$12~$\geq$~8.6, log(O/H)$+$12$>$~8.6) and compare their absolute magnitude distributions via K$-$S tests, no statistically significant difference is found. We thus see no evidence that normal SNe~IIn exhibit brighter absolute magnitudes at lower metallicities. However, we note that SLSNe~II (M$_{r/R}$~$<$~--21), which might also be powered by CSI, are typically found at lower N2 metallicities (8.18~$<$~log(O/H)$+$12~$>$~8.62) than those of our normal SN~IIn sample \citep{leloudas14}.

\subsubsection{Metallicity and dust emission}

\citet{fox11} observed a large sample of SNe~IIn at late epochs in the mid-infrared (MIR) with Spitzer.
Among their SNe~IIn, 17 events are included in our sample and six of them show dust emission (including SN~2008J that was later re-typed as a Ia-CSM; \citealp{taddia12}).
We compare the metallicity distribution of those with dust emission against those without dust emission, and do not find any statistically significant difference (see Fig.~\ref{cdfdust}, left-hand panel), despite the fact that one could expect the formation of more dust at higher metallicity. However, we notice that the detection of dust emission is likely more affected by other factors, such as the luminosity of the SN itself. Since the optical SN radiation is reprocessed and emitted in the MIR, a luminous SN can reveal the presence of dust better than a faint SN. However, a too high radiation field may also destroy the circumstellar dust. 
Indeed we see a statistically-significant difference in $r/R/unf.$-band peak absolute magnitudes between the SNe with and without dust emission (p-value~$=$~5\%), with the dust-rich SNe being brighter by 1.3~mag on average. (see Fig.~\ref{cdfdust}, right-hand panel). Here we include the objects with only a limit on the peak magnitude in the computation of the CDFs.

\onlfig{13}{\begin{figure}
 \centering
\includegraphics[width=16cm]{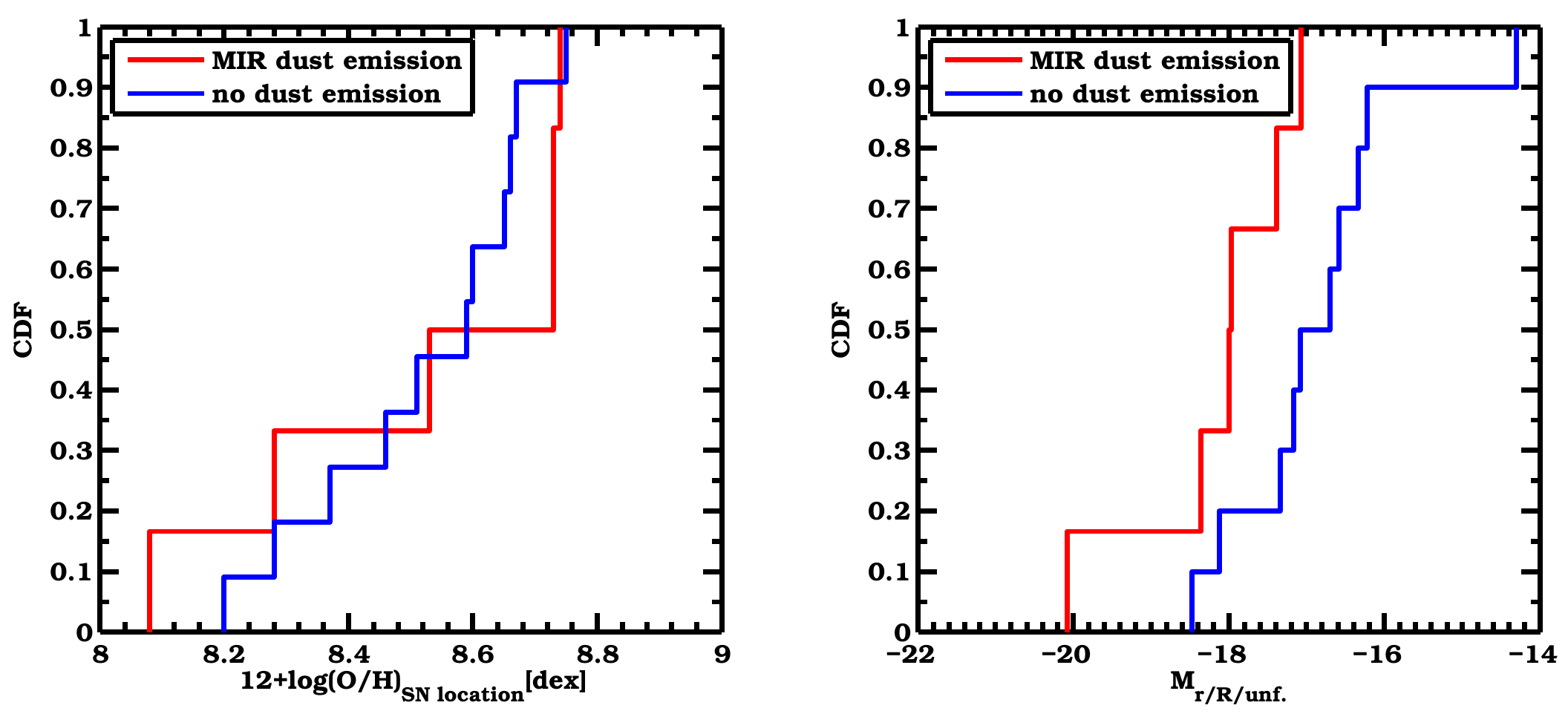} 
  \caption{\textit{(Left-hand panel)} Metallicity CDFs for CSI~SNe showing and not showing MIR dust emission at late epochs \citep{fox11}. These CDFs do not show any statistically significant difference. \textit{(Right-hand panel)} CDFs of the peak absolute $r/R/unf.$ band magnitudes for the same CSI~SNe showing and not showing MIR dust emission at late epochs. These CDFs show that SNe with MIR dust emission tend to be brighter at peak.\label{cdfdust}}
 \end{figure}}

\subsubsection{Metallicity and SN~IIn/Ibn CSM properties}
For several of our SNe~IIn and Ibn, we collected data about their CSM/progenitor wind velocities (v$_w$, as 
measured from the blue-velocity-at-zero-intensity BVZI, FWHM or P-Cygni absorption minima of their narrow lines), and their mass-loss rates ($\dot{M}$, estimated from spectral analysis, e.g. \citealp{kiewe12} or via light curve modelling, e.g. \citealp{moriya14}). We list these quantities in Table~\ref{tab:spec}, with their references. In Fig.~\ref{masslossvsmet} (left-hand panels) we plot $\dot{M}$ and v$_w$ against the local metallicity measurements.
It is important to stress that the mass-loss rates are sometimes obtained by measuring the shock velocity from the broad component of the emission lines. This might be misleading in some cases, as the broadening is sometimes produced by Thomson scattering in a dense CSM \citep{fransson14}.

If we consider only mass-loss rates $\gtrsim$10$^{-3}$~M$_{\odot}$~yr$^{-1}$, a trend 
of higher $\dot{M}$ with higher metallicities is visible in the top-left panel of Fig.~\ref{masslossvsmet}.
This behavior is also what is expected in hot stars, where line-driven winds drive the mass loss, but this is typically for much lower mass-loss rates.
The metallicity dependence of the mass-loss rates due to line-driven winds is a power-law (PL) with $\alpha~$=$~$~0.69 \citep[e.g.,][]{vink11}. 
We show this PL with a black segment in Fig.~\ref{masslossvsmet}, and we can see that it is roughly consistent with the observed in the data. 
When we compare the mass-loss rate CDFs of metal-poor and metal-rich SNe~IIn, the difference between the two populations is found to be statistically significant (p-value~$=$~0.030--0.034, depending on the metallicity cut, see the top-right panel of Fig.~\ref{masslossvsmet}). We found five SNe with relatively low mass-loss rate ($\dot{M}~\lesssim~$10$^{-3}$~M$_{\odot}$~yr$^{-1}$) that seem to deviate from the aforementioned trend. These are SNe~1998S and 2008fq, which belong to the SN~IIn-L subclass, and SNe~2005ip and 1995N, two long-lasting (i.e., 1988Z-like) events. We note that all the SNe~IIL whose metallicity measurements are available in the literature on the N2 scale \citep{anderson10} show relatively high metallicity (log(O/H)$+$12$\sim$8.6), similar to those of SNe~IIn-L. Furthermore, the SNe~IIL class is characterized by relatively weak CSM interaction and relatively low mass-loss rates up to 10$^{-4}$~M$_{\odot}$~yr$^{-1}$ (e.g., SN~1979C and SN~1980K, \citealp{lund88}), again resembling SNe~IIn-L, which typically show lower mass-loss rates than SN~1988Z-like events (see the top-left panel of Fig.~\ref{masslossvsmet}). In this sense, SNe~IIn-L might be considered as close relatives to SNe~IIL (SN~1998S was sometimes defined as a SN~IIL, \citealp{chugai08}), having only slightly stronger CSM interaction, as suggested by the persistent emission lines in their early phases.

In the bottom-left panel of Fig.~\ref{masslossvsmet}, there are evidence that SNe~IIn at higher metallicity tend to have larger wind velocities than those located in low metallicity environments. However, when we compare the wind velocity CDFs of metal-poor and metal-rich SNe~IIn (see the bottom-right panel of Fig.~\ref{masslossvsmet}), this difference is only marginally significant (p-value~$=$~0.14--0.27). SNe~Ibn show higher wind velocities than SNe~IIn, regardless the metallicity. We note that, in line-driven winds, the metallicity dependence of the wind velocity is weak and can be described by a PL with $\alpha$~$=$~0.12 \citep{kud02}. We plot this PL with a black segment in Fig.~\ref{masslossvsmet}, and we can see that is roughly consistent with the observed trend. 
SNe~IIn-L (98S-like) typically show higher wind velocities than long-lasting SN~IIn (88Z-like).

We discuss these results and their implications on the mass-loss mechanism in Sect.~\ref{sec:disc_implimassloss}.

\subsection{Metallicity and SN~IM properties}

\onlfig{14}{\begin{figure}
 \centering
\includegraphics[width=8.3cm,angle=0]{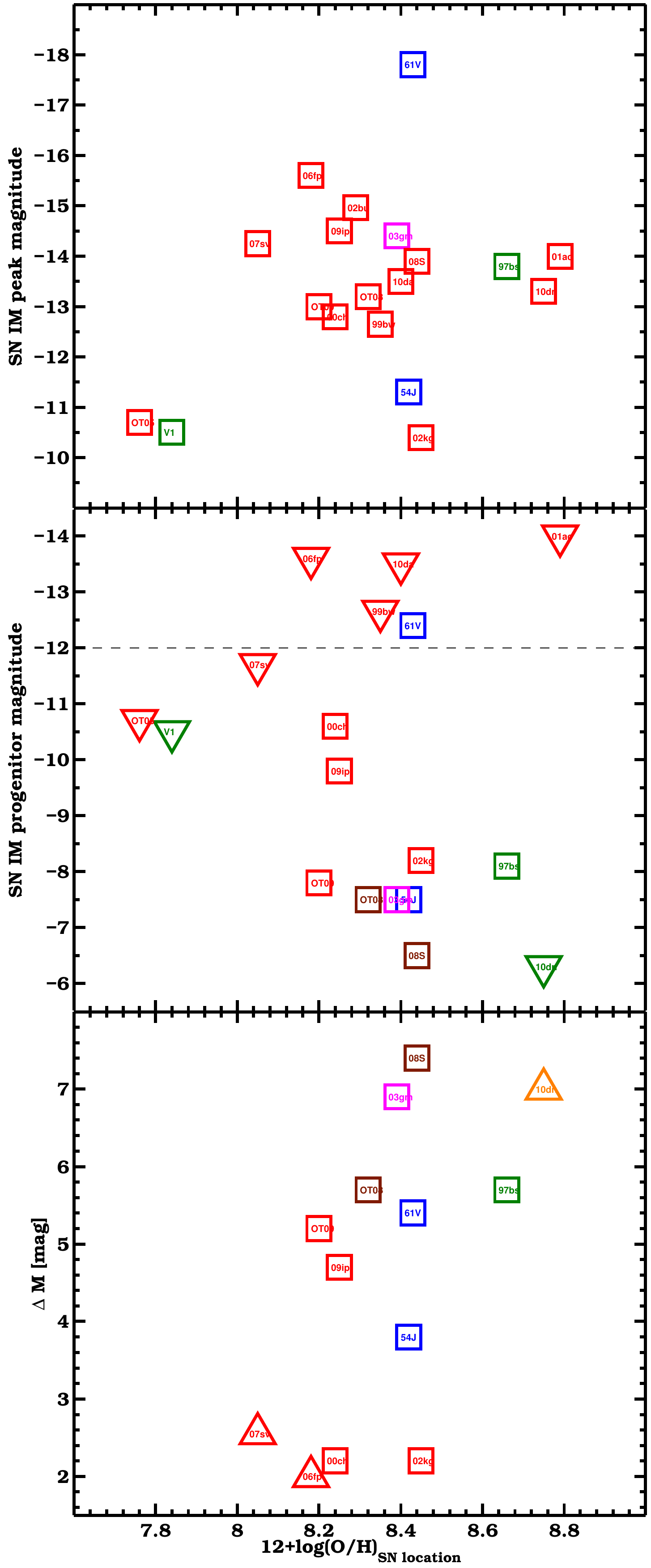} 
  \caption{\textit{(Top panel)} SN~IM peak absolute magnitudes (from \citealp{smith11}) versus local metallicity. \textit{(Central panel)} Absolute magnitude of SN~IM progenitors (from \citealp{smith11}) versus local metallicities. Triangles are lower limits. With the exception of SN~1961V and of those events where we have limits poorer than $-$12~mag, at lower metallicity the luminosity seems to be higher. \textit{(Bottom panel)} The difference between each peak outburst and the corresponding progenitor magnitude versus metallicity is shown. Triangles are lower limits. The data suggest larger $\Delta$M at larger metallicity. 
Red corresponds to $R/unf.$ bands, blue to $B$ band, green to $V$ band, magenta to $I$ band, brown to NIR, orange to $R/V$.\label{metVSimpoprop}}
 \end{figure}} 

The comprehensive paper by \citet{smith11} lists the absolute magnitudes of the outburst peak(s), the absolute magnitude of each SN~IM progenitor before the outburst, the expansion velocity during the outburst (V$_{exp}$, from spectral analysis), the 
characteristic decay time (t$_{1.5}$, i.e. the time the SN~IM spent to fade by 1.5~mag from peak), the H$\alpha$ equivalent width (EW), the presence of multiple peaks and sharp dips in the light curves. \citet{smith11} noted that there are no clear correlations among these observables. 

On the other hand, we have found some indications of possible trends with respect to the metallicity at the SN~IM location.
 In Fig.~\ref{metVSimpoprop} (central panel), the SN~IM progenitor absolute magnitude 
 before the outburst is plotted versus the metallicity. Squares correspond to detections, triangles to upper limits. The different colors correspond to the different pass-bands (see caption). It is evident how (in the optical) most of the fainter SN~IM progenitors are located at higher metallicity, and the brighter ones at lower metallicity, with the exception of SN~1961V (and of those events where we only have poor limits, $>-$12~mag). Note that \citet{smith11} suggest a CC origin for this event. On the other hand, \citet{vandyk12} suggests that the 1961V event was the outburst of a LBV progenitor star as it is still detectable in recent HST images. 
  
 In the bottom panel of Fig.~\ref{metVSimpoprop}, the difference in magnitude 
 ($\Delta$M) between the progenitor of the outburst and the peak of the eruptive event is 
 shown versus the metallicity. The higher the metallicity at the location of the transient, 
the larger seems to be the difference between the peak luminosity during 
the outburst and the luminosity of the progenitor. 
 
As both the SN~IM progenitor magnitudes and $\Delta$M show a possible trend with the metallicity, it follows that these two observables also show a trend, with lower $\Delta$M values for the more luminous SN~IM progenitor.
 
The large variations in optical magnitudes ($\Delta$M~$>$~2) observed in these SN~IMs
 are probably due to actual variations in the bolometric luminosities, rather than 
being an effect of variations in temperature at constant bolometric luminosity as in most of 
the Galactic LBVs, which are characterized by smaller $\Delta$M \citep[see e.g.,][and 
reference therein]{clark09}. 
However, we also note that $\Delta$M might be affected, or even produced, 
by the formation of dust in the CSM surrounding the quiescent SN~IM progenitor \citep{kochanek14}.

The top panel of Fig.~\ref{metVSimpoprop} reveals that there is no clear trend between peak outburst luminosity and metallicity. The same is true for expansion velocity, H$\alpha$ EW and t$_{1.5}$. Furthermore, we found no difference in the metallicity distributions of SN~IMs with and without multi-peaks in the light curves, or between SN~IMs with and without sharp dips in the light curves.

\subsection{Association to SF regions and CSI transient properties\label{sec:ncr}}

The degree of association of a SN to star forming (SF) regions in a galaxy can be interpreted in terms of SN progenitor zero age main sequence mass \citep{james06,anderson08,anderson12}, i.e., the stronger the association, the larger the progenitor mass.
A way to quantify the degree of association is to measure the NCR index, i.e.
the normalized cumulative rank pixel value. This number is obtained by ranking the pixels in the H$\alpha$ continuum-subtracted host-galaxy image, based on their counts. Then, the CDF of the pixel counts is built and the NCR value is the CDF value of the pixel corresponding to the SN position.

\begin{figure}
 \centering
\includegraphics[width=9cm,angle=0]{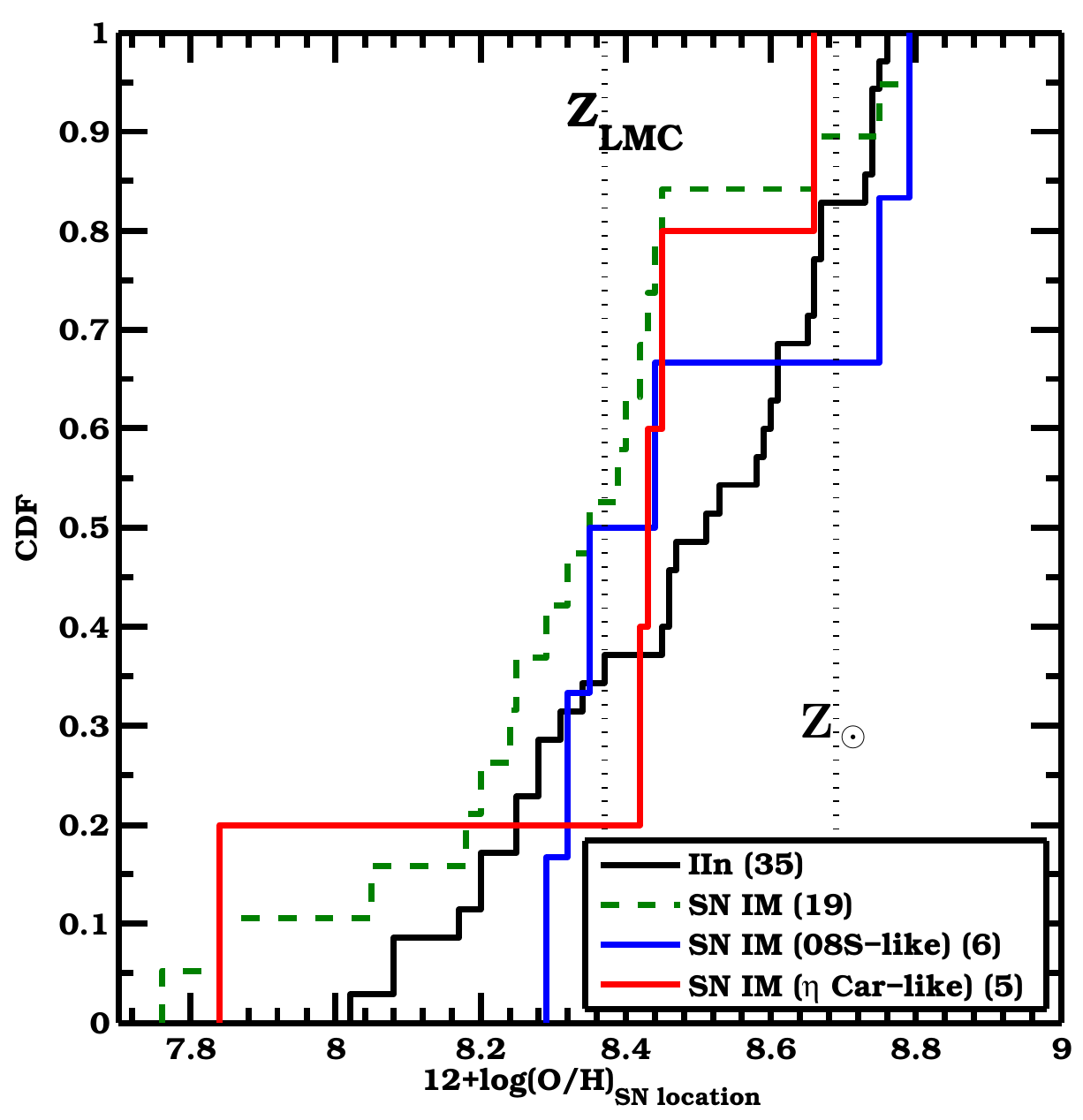} 
  \caption{CDFs of the metallicity at the SN location for SN~IM subtypes, mainly based on NCR values.\label{cdf_impo_subtype}}
 \end{figure}
 
 H14 presented a comprehensive study of the association to SFRs for CSI transients, where they introduced a subclassification of the SN~IMs: SN~2008S-like and $\eta$ Carinae-like. The first ones (SNe~2008S, 1999bw, 2001ac, 2002bu, 2010dn, 2008-OT)
do not show any association to bright \ion{H}{ii} regions (NCR$=$0) and are often enshrouded in dusty environments. The second category (SNe~1954J, 1961V, 1997bs, 2002kg and V1) 
are associated with SF regions (NCR$>$0). Also because of the weak association to \ion{H}{ii} regions, the first class of SN~IMs has been suggested to be related to 8--10~$M_{\odot}$ super-AGBs \citep{botticella09,pumo09,wanajo09,prieto10,kochanek11} rather than to more massive LBV progenitors \citep[however, see][]{smith15asocial}, which are more likely to produce $\eta$ Carinae-like IMs.
 When comparing the metallicity distributions of these two populations, we do not find any statistically significant difference (see Fig.~\ref{cdf_impo_subtype}).  However the numbers are small and the uncertainties are large. The same is true for SNe~IIn and Ibn.
If we plot the SN~IM properties from \citet{smith11} (see previous section) against the NCR values from H14, we do not find any clear trends or correlations. Furthermore, the NCR does not exhibit any clear trends with peak absolute magnitudes, mass-loss rate or CSM/wind velocity in SNe~IIn. In the last column of Table~\ref{metal} we list all the available NCR values from H14.
 
We note that \citet{crowther13} concluded that the observed association between CC~SNe and \ion{H}{ii} regions provides only weak constraints upon the progenitor masses.
 
\section{Discussion}
\label{sec:disc}
\subsection{Implications for the SN~IIn progenitor scenario}
\label{sec:disc_progenitor}
\begin{figure*}
 \centering
\includegraphics[width=16cm]{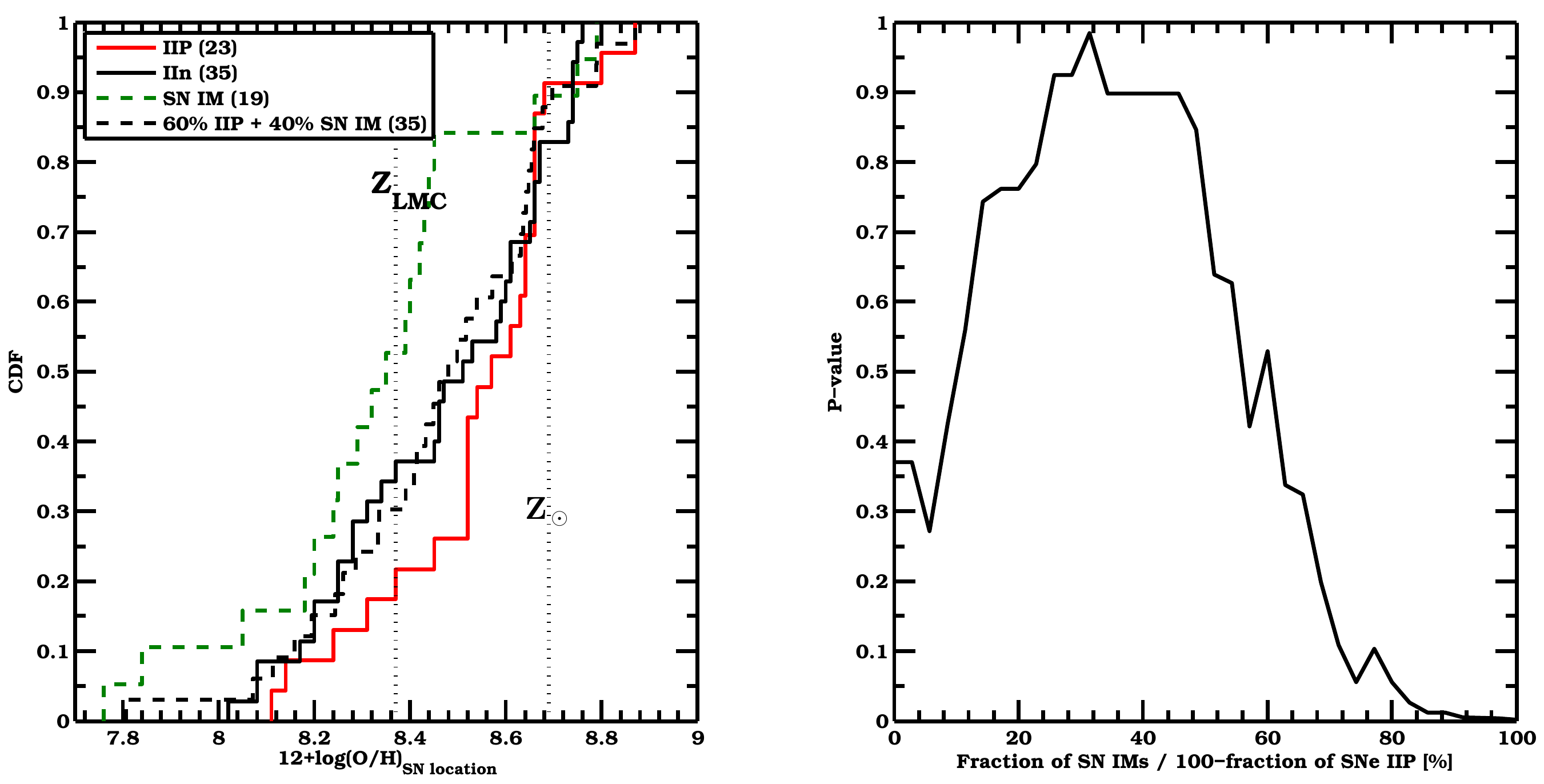}
\includegraphics[width=16cm]{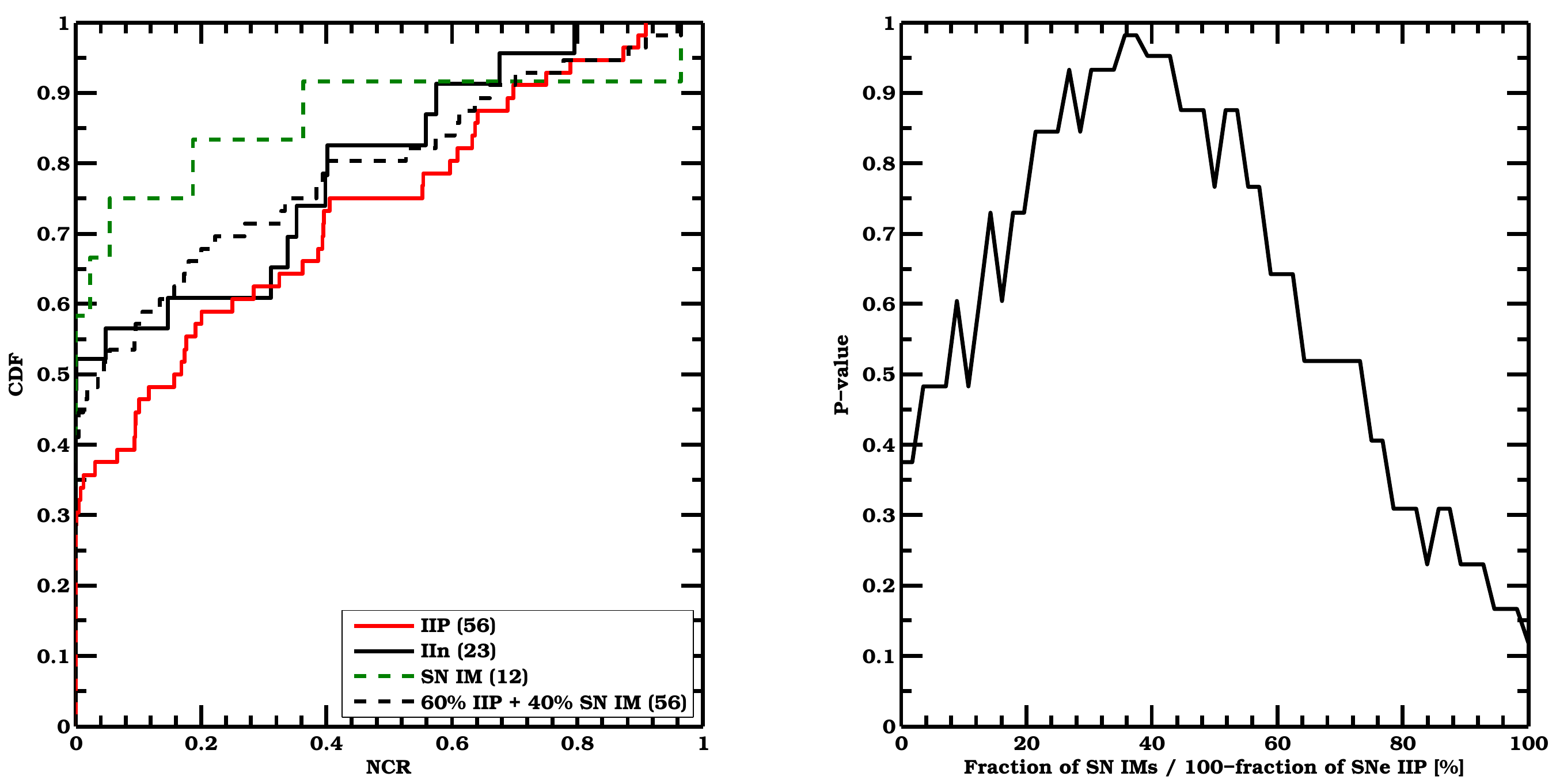} 
  \caption{ \textit{(Top-left panel)} Metallicity CDFs of SNe~IIn, SN~IMs, SNe~IIP and of a simulated SN progenitor population constructed of SN~IM/LBVs and SN~IIP/RSGs, which fit the observed SN~IIn metallicity CDF. \textit{(Top-right panel)} p-value of the K-S test between the metallicity distributions of SNe~IIn and the simulated progenitor population as a function of the fraction of LBVs in the simulated progenitor population. \textit{(Bottom-left panel)} NCR CDFs of SNe~IIn, SN~IMs, SNe~IIP and of a simulated SN progenitor population made of SN~IM/LBVs and SN~IIP/RSGs, which fit the observed SN~IIn NCR CDF. Data from H14 and \citet{anderson12}. \textit{(Bottom-right panel)} p-value of the K-S test between the NCR distributions of SNe~IIn and the simulated progenitor population as a function of the fraction of SN~IM/LBVs in the simulated progenitor population. \label{cdf_progIIn}}
 \end{figure*}

As described in the introduction, there are several indications that LBVs, which likely produce SN~IMs, 
are also progenitors of at least a fraction of SNe~IIn. 
On the other hand, the large variety of SNe~IIn may suggest the existence of multiple precursor channels. 
The latter is also suggested by our
result on the different metallicity distributions of SNe~IIn and SN~IMs. Since we know that at least some SNe~IIn are produced by the terminal explosion of LBVs (see e.g., \citealp{pastorello07}, \citealp{galyam07}, \citealp{ofek14}), and assuming that most of the SN~IMs are produced by LBVs, then the metallicity distribution of SNe~IIn can be considered to be a combination of
the metallicity distribution of SN~IMs with that of one or more alternative progenitor populations. 
These other progenitors must have higher metallicity than that of SN~IMs, which would explain why SNe~IIn 
are found at higher metallicity compared to SN~IMs.

RSG stars with superwinds like VY Canis Majoris have been invoked as possible progenitors of some SNe~IIn 
\citep[e.g.,][]{fransson02,smith09_RSG}. If all the SNe~IIn which are not produced by LBVs are actually 
produced by RSGs, and if we assume that the metallicity distribution of SNe~IIP (which do have RSG progenitors, \citealp{smartt09}) and that of RSGs producing SNe~IIn are the same, 
then we can combine the metallicity 
distributions of SN~IMs and SNe~IIP to reproduce the observed metallicity distribution of SNe~IIn.
This exercise will allow us to roughly estimate what fractions of LBVs and RSGs that are precursors of SNe~IIn, in a simple two-progenitor populations scenario. This is 
done by maximizing the p-value of the K-S test between the SN~IIn metallicity distribution and that of 
the combined distributions of SN~IMs/LBVs$+$SNe~IIP/RSGs. To produce combined metallicity distributions of SN~IMs/LBVs and SNe~IIP/RSGs, we linearly 
interpolate their observed CDFs (which are marked by a green dashed line and a red solid line in the top-left 
panel of Fig.~\ref{cdf_progIIn}) varying the number of objects in the two samples, while always keeping a total 
of 35 events, which is the number of SNe~IIn that we have in the metallicity CDF (solid black line in the 
top-left panel of Fig.~\ref{cdf_progIIn}). This is done to produce a meaningful comparison to the sample of 
SNe~IIn. The two extreme cases are obtained when the combined distribution contains only SN~IMs/LBVs or only 
SNe~IIP/RSGs. In between, we obtain 33 different distributions with mixed fractions of SNe~IIP/RSGs and SN~IMs/LBVs. We then performed the K-S test between all these distributions and that for SNe~IIn, and plot the 
resulting p-values as a function of the SN~IMs/LBV fraction in the top-right panel of Fig.~\ref{cdf_progIIn}. It is evident 
that the p-value reaches its maximum between $\sim$28\% and $\sim$48\% of SN~IMs/LBVs as 
SN~IIn progenitors, with the rest being formed by SN~IIP/RSGs. This range would change to 20\%--35\% if we exclude the hosts where we adopted the gradient from P04.
The CDF formed by 40\% of SN~IMs and 60\% of 
SNe~IIP/RSGs is shown with a dashed line in the top-left panel of Fig.~\ref{cdf_progIIn}, and it is a good match to the distribution of SNe~IIn. 
We notice that, among SNe~IIn, those resembling SN~1998S show a metallicity distribution similar to those of SNe~IIP and IIL, suggesting that this SN~IIn subclass might preferentially arise from RSGs rather than from LBVs. On the contrary, 1988Z-like SNe show a metallicity distribution consistent with that of the SN~IMs, suggesting a LBV origin.

By using data from H14 and \citet{anderson12}, a similar exercise was repeated with the NCR distributions for SNe~IIn, SN~IMs and SNe~IIP. We found (see the bottom panels of Fig.~\ref{cdf_progIIn}) that the SN~IIn NCR distribution can be reproduced with high statistical significance by a simulated population of SN~IMs (from LBVs, $\sim$40\%) and SNe~IIP (from RSGs, $\sim$60\%), identical to what was found when comparing the metallicity distributions. It is interesting to note that \citet{ofek14} estimate that $\sim$50\% of SNe~IIn show outbursts only a few years before collapse, which is similar to our fraction of SN~IMs/LBVs. This might imply that most LBV progenitors of SNe~IIn display outbursts just before collapse. Indeed it has been found by \citet{moriya14} that SN~IIn progenitors suffer enhanced mass loss as they get closer to collapse.

In reality, the scenario is likely more complicated, especially if we consider that some SN~IMs, i.e. the SN~2008S-like events, are likely produced by super-AGB stars, and not by LBV eruptions.
However, the metallicity distribution of SN~IMs like SN~2008S
is very similar to that of SN~IMs like $\eta$ Carinae (which are likely produced by LBV eruptions). In this perspective, we can consider the estimated 40\% of LBV progenitors as an upper limit. It is more realistic to think that of this 40\% some SN~IIn have for instance super-AGB stars as precursors. SNe~IIn-P (1994W-like events), which indeed show a metallicity distribution similar to SN~IMs (see Fig.~\ref{IInsub}, left-hand panel), might be the result of EC-SNe from super-AGB stars.

If we look at SN rates, SN~II (IIP and IIL) and SNe~IIn form $\sim$55\% and $\sim$9\% of the CC~SNe \citep{smith11_rates}, respectively. In our simplified two-progenitor population scenario, $\sim$60\% of the SNe~IIn
comes from  RSGs with large mass-loss rates (similar to the massive hypergiant VY Canis Majoris).
If we assume a Salpeter initial mass function and single stars as progenitors, 
to reproduce the ratio between SNe~IIn from RSGs and SNe~II ($\sim$10\%) would require that RSG stars with initial mass between 8 and 20 $M_{\odot}$ produce SNe~IIP/L and massive RSG stars with initial mass between 20 and 25 $M_{\odot}$ produce SNe~IIn. These ranges are reasonable, but they do not account for binary systems, which might be important to explain the progenitors of SNe~IIn \citep{smith15asocial} and other CC~SNe.

There are potential biases in the samples of SNe~IIn and SN~IMs, which might affect the comparison between their metallicity distributions. 
Most of the SNe in our samples were discovered by galaxy-targeted surveys, with the exceptions of a few SDSS (Sloan Digital Sky Survey) II SN survey and PTF SNe. Targeted surveys often pitch larger galaxies with more metal rich distributions that the non-targeted surveys \citep[see e.g.,]{sanders12}, so both these distributions are likely biased in that direction.

The SNe~IIn are also located at larger redshifts than the SN~IMs. The latter are more difficult to discover at large distances given that they are typically $\sim$4 magnitudes fainter than SNe~IIn. For the redshifts in this investigation, there is however no metallicity evolution expected (see e.g.~\citealp{savaglio06}). There is also no clear correlation between the SN~IIn luminosity and their host metallicity, with the brighter SNe~IIn (more likely to be discovered at larger distances) showing similar host metallicity to the fainter ones. 

As most SNe~IIn are selected based on the availability of literature data, it could be that the SN~IIn sample is biased towards peculiar objects, which are more often discussed in the literature as compared to the ``normal" ones. It is, however,  unclear if and how this selection would affect the observed metallicity distribution.

An alternative interpretation of the difference between the metallicity distribution of SN~IMs and SNe~IIn might be the following. 
If we instead assume that {\it all} the SNe~IIn are actually produced by LBV stars, then the SN~IMs are preferentially produced by the LBV at 
low metallicity, whereas at relatively high metallicity
LBV would not produce giant outbursts but only exhibit e.g., S-Doradus variability. 
However, we do know that large eruptions like $\eta$ Car and other LBVs are found in both the MW and other nearby metal-rich galaxies (e.g., M~31, M~81, M~101), as well as in nearby metal-poor galaxies such as NGC~2366 and the SMC \citep[see][for a list of LBVs and their metallicities]{weis06}. We therefore do not favour this explanation for the lower metallicity of SN~IMs as compared to that of SNe~IIn. 

\subsection{Implications for the mass loss mechanism}
\label{sec:disc_implimassloss}

The large mass-loss rates derived for SN~IIn progenitors and observed in SN~IMs are not explicable solely in terms of steady, metal line-driven winds \citep{smith14}. This mechanism cannot reproduce $\dot{M}>10^{-4}$~M$_{\odot}$~yr$^{-1}$, which we typically deduce for these transients. Eruptive mechanisms must play an important role.

The relation between metallicity and outbursts has not yet been clarified, but here we present a number of related observational results. We need also to consider that, via steady line-driven winds, the metallicity might anyway have an effect on the appearance of CSI 
transients, as they act for the entire life of their progenitors.

Despite the large spread, in SNe~IIn it is possible to see that higher mass-loss rates and faster CSM velocities correlate with higher metallicity at the SN location (Fig.~\ref{masslossvsmet}). Note that most of the mass-loss rate estimates used assume steady winds as well as shock velocities derived from the width of the broad component of H$\alpha$, even if the progenitor winds may in fact not be steady \citep{dwarkadas11}, and the broad component of the emission lines can be due to Thomson scattering in a dense CSM \citep{chugai02}.

The bottom panel of Fig.~\ref{metVSimpoprop} shows that in SN~IMs, a higher metallicity tend to prompt relatively brighter outbursts in the optical.
$\Delta$M might be produced by an actual variation in the bolometric luminosity of the SN~IM or be the effect of a temperature variation at almost constant bolometric luminosity as is observed for ``S Doradus" LBVs (e.g., \citealp{humphreys94}, \citealp{vink11}). These LBVs present temperatures of several 10$^4$~K in the quiescent state, and then become F-type stars (with temperatures of 8--9$\times$10$^3$~K) during the outburst. In this case, at larger metallicity we would have found the SN~IMs whose photospheric temperatures suffered the stronger cooling during the outburst, i. e., the SN~IMs whose progenitors had the higher temperatures before the outburst. More bolometric information on our SN~IMs are necessary to draw further conclusions.

The fact that we found relatively brighter outbursts at larger metallicity might fit with the mass-loss bi-stability jump scenario discussed by \citet{vink11} for LBVs. When a LBV becomes cooler than 2.5$\times$10$^4$~K, the Fe recombines from \ion{Fe}{iv} to \ion{Fe}{iii}, enhancing the opacity and thus the mass loss. LBVs are likely to cross the bi-stability temperature threshold several times during their lives, inducing variable mass loss. Obviously, large metallicities would favor this mass-loss enhancement and hence the luminosity in the eruptive phase.

Figure~\ref{metVSimpoprop} (central panel) shows that progenitors of SN~IM outbursts at lower metallicity tend to have higher optical luminosity. If we assume similar temperatures for the SN~IM progenitors, this would mean that those with higher bolometric luminosity tend to be more metal poor. 
However, if we instead assume that they have similar bolometric luminosities, the brighter the progenitor in the optical, the lower would its temperature be. Therefore a possible implication of this result is that the cooler progenitors of SN~IMs tend to have lower metallicities. Again, more data are needed to establish the bolometric and color properties of SN~IMs.

The fact that at higher metallicity SN IMs show larger $\Delta$M
but their progenitor show lower absolute magnitudes could also be explained 
if we take into account the presence of dust in SN~IMs. A larger metallicity would favour larger mass loss and thus stronger CSM interaction (and larger $\Delta$M), but at the same time more dust would surround the star and thus the progenitor would suffer of larger extinction. However, there are SN IMs with large $\Delta$M both with (SN~2008S, SN~2010dn, OT2008) and without (SNe~ 1997bs and 2003gm) dust emission \citep{thompson09}.

\section{Conclusions}
\label{sec:concl}
We summarize our main findings as follows:
\begin{itemize}
\item[$\bullet$]{SNe~IIn are located at higher metallicities than SN~IMs, and this difference is statistically significant.}


\item[$\bullet$]{The locations of SNe~IIn-L (1998S-like) exhibit the highest metallicities among SNe~IIn. Their metallicity distribution is similar to those of SNe~IIL and IIP (produced by RSG progenitors). On the other hand, long-lasting SNe~IIn (1988Z-like) are typically metal-poorer and exhibit a metallicity distribution similar to that of SN~IMs (that may be produced by LBV progenitors).}

\item[$\bullet$]{The metallicity distribution of SNe~IIn can be interpreted as a combination of the metallicity distributions of SN~IMs and SNe~IIP, which might be similar to the metallicity distributions of LBVs and RSG stars, respectively. The same is true for the NCR distribution of SNe~IIn. These results on the metallicity and NCR distributions are consistent with a scenario where SNe~IIn have both LBV ($\sim$40\%) and RSG ($\sim$60\%) progenitors. If we also consider the possibility that some SN~IMs come from super-AGB stars, then the estimated $\sim$40\% fraction of LBV progenitors might be considered as an upper limit.}

\item[$\bullet$]{We do not find a significant difference in the metallicity distributions of CSI~SNe with and without dust emission in the MIR, but rather a difference in the distributions of the peak absolute magnitude, with the dust-rich SNe being brighter.}

\item[$\bullet$]{Above $\dot{M}~\sim~$10$^{-3}$~M$_{\odot}$~yr$^{-1}$, SNe~IIn located at higher metallicities
tend to show larger mass-loss rates. There are also indications suggesting larger CSM/wind velocities for SNe~IIn at higher metallicity.}

\item[$\bullet$]{There are hints that at higher(lower) metallicity the difference in optical magnitudes between the SN~IM progenitor and its outburst's peak tend to be larger(smaller) and SN~IM progenitors tend to have lower(higher) optical luminosity.}

\end{itemize}

\begin{acknowledgements}We gratefully acknowledge the support from the Knut and Alice Wallenberg Foundation.
M.~D. Stritzinger gratefully acknowledges generous support provided by the Danish Agency for Science and Technology and Innovation  
realized through a Sapere Aude Level 2 grant.
The Oskar Klein Centre is funded by the Swedish Research Council.
The Nordic Optical Telescope is operated by the Nordic Optical Telescope Scientific Association at the 
Observatorio del Roque de los Muchachos, La Palma, Spain, of the Instituto de Astrofisica de Canarias.
This research has made use of the NASA/IPAC Extragalactic Database (NED) which is operated by the Jet 
Propulsion Laboratory, California Institute of Technology, under contract with the National Aeronautics 
and Space Administration. 
\end{acknowledgements}

\bibliographystyle{aa} 

\onecolumn

\clearpage


\clearpage

\begin{deluxetable}{llllllllllllllll}
\rotate
\tabletypesize{\scriptsize}
\tablewidth{0pt}
\tablecaption{Sample and basic data of 63 (38 SNe~IIn, 
 6 SNe~Ibn, 1 SN~Ia-CSM, 18 SN~IMs) CSI~SN host galaxies.\label{gal}}
\tablehead{
\colhead{SN} &
\colhead{Type} &
\colhead{Data} &
\colhead{Host galaxy} &
\colhead{z}&
\colhead{RA$_{\rm SN}$} &
\colhead{DEC$_{\rm SN}$}&
\colhead{RA$_{\rm host}$}&
\colhead{DEC$_{\rm host}$}&
\colhead{2a}&
\colhead{2b}&
\colhead{t-type}&
\colhead{PA}&
\colhead{reference}&
\colhead{reference}&
\colhead{r$_{\rm SN}$/R$_{25}$}\\
\colhead{} &
\colhead{} &
\colhead{source} &
\colhead{} &
\colhead{}&
\colhead{(hh:mm:ss)}&
\colhead{(dd:mm:ss)}&
\colhead{(hh:mm:ss)}&
\colhead{(dd:mm:ss)}&
\colhead{(\arcmin)}&
\colhead{(\arcmin)}&
\colhead{}&
\colhead{(deg)}&
\colhead{a, b}&
\colhead{PA}&
\colhead{}}
\startdata
  \hline

 1987F   & IIn-88Z   & o & NGC 4615              & 0.015731  & 12:41:38.99   &  +26:04:22.4 & 12:41:37.32  & +26:04:22.2  & 1.5850  &  0.7449   & 6.0    &  125    & NED     & NED    &   0.73     \\
 1988Z   & IIn-88Z   & o & MGC+03-28-022         & 0.022822  & 10:51:50.12   &  +16:00:00.2 & 10:51:49.40  & +16:00:01.5  & 0.7230  &  0.5784   & 10.0   &  163    & NED(R)  & NED(R) &   0.59     \\
 1994W   & IIn-P(94W)& o & NGC 4041              & 0.004116  & 12:02:10.92   &  +62:08:32.7 & 12:02:12.20  & +62:08:14.0  & 2.6917  &  2.5033   & 4.3    &  70     & NED     & SIM(IR)&   0.28     \\
 1994Y   & IIn-L(98S)& l & NGC 5371              & 0.008533  & 13:55:36.90   &  +40:27:53.4 & 13:55:39.94  & +40:27:42.3  & 4.3650  &  3.4483   & 4.2    &  8      & NED     & NED    &   0.36     \\
 1994ak  & IIn       & l & NGC 2782              & 0.008483  & 09:14:01.47   &  +40:06:21.5 & 09:14:05.11  & +40:06:49.3  & 3.4667  &  2.5653   & 1.1    &  15     & NED     & NED(IR)&   0.58     \\
 1995G   & IIn-88Z   & o & NGC 1643              & 0.016269  & 04:43:44.22   &  -05:18:53.8 & 04:43:43.95  & -05:19:10.0  & 1.0967  &  1.0967   & 3.5    &  10     & NED     & NED    &   0.51     \\
 1995N   & IIn-88Z   & o & MGC-02-38-016         & 0.006118  & 14:49:28.31   &  -10:10:14.0 & 14:49:32.95  & -10:09:51.0  & 1.3767  &  0.4681   & 10.0   &  135    & NED(IR) & NED(IR)&   4.90     \\
 1997ab  & IIn-88Z   & l & Anon J095100+2004     & 0.012000  & 09:51:00.40   &  +20:04:24.0 & 09:51:00.45  & +20:04:20.9  & 0.3334  &  0.3334   & \ldots &  \ldots & H97     & \ldots &   0.32     \\
\enddata

\tablenotetext{*}{The full names of the host galaxies of SNe 2006db, 2003ma, 2007va, 2005kj, and LSQ12btw are SDSS J115538.31+442301.8, [R2010] J053101.87-700414.9 and SDSS J142623.25+353529.1, 2MASX J08400884-0536039, and SDSS J101028.68+053212.5, respectively.} 
\tablenotetext{h}{From Hyperleda.}
\tablecomments{The t$-$type is obtained from the ASC. NED$~=~B$ band \citep{devauc91}. NED(B)$=~B$ band (Palomar survey). NED(R)$=~r$ band (SDSS). NED(IR)$=$K band (2MASS). NED(Be)$~=~B$band (ESO survey). HYP$=$Hyperleda. o$=$observed, l$=$literature, a$=$archive.} 
 \end{deluxetable}

\clearpage

\begin{deluxetable}{lll|ll|l}
\tabletypesize{\scriptsize}
\tablewidth{0pt}
\tablecaption{Log of the photometric and spectroscopic observations at the Nordic Optical Telescope with ALFOSC.\label{log}}
\tablehead{
\colhead{SN} &
\colhead{Type} &
\colhead{Galaxy} &
\colhead{Photometry} &
\colhead{}&
\colhead{Spectroscopy}\\
\colhead{}&
\colhead{}&
\colhead{}&
\colhead{Filter}&
\colhead{Integration (sec)}&
\colhead{Integration (sec)}}
\startdata
  \hline                                                                                                    
1987F  &IIn-88Z        &     NGC 4615       & R                &3$\times$40    & 1$\times$1800  \\
       &      &                    & H$\alpha(\#50)$  &3$\times$180                     &                \\      	
1988Z  &IIn-88Z        &  MGC+03-28-022     & R                &3$\times$30    & 1$\times$1800  \\
       &      &                    & H$\alpha(\#51)$  &3$\times$150                     &                \\      	
1994W  &IIn-P(94W)   &      NGC 4041      & R                &3$\times$40    & 1$\times$1800  \\
       &      &                    & H$\alpha(\#49)$  &3$\times$200                     &                \\       
1995G  &IIn-88Z        &  NGC 1643          &                  &               & 2$\times$1800  \\
1995N  &IIn-88Z        &   MGC-02-38-016    & R                &3$\times$40    & 1$\times$1800  \\
\enddata
\tablecomments{Spectroscopic observations were performed with Slit 1.0$\arcsec+$Grism~\#4. We could not image the host galaxies of
the SN~IMs because during the observational campaign of April 2014 several nights were lost due to bad weather, and we decided to spend the remaining nights doing spectroscopy.}
 \end{deluxetable}

\clearpage

\onltab{3}{
\begin{deluxetable}{llcc}
\tabletypesize{\scriptsize}
\tablewidth{0pt}
\tablecaption{Deprojected and normalized distance from the host center and line ratios for the \ion{H}{ii} regions with observed spectrum. \label{tab:lineratios}}
\tablehead{
\colhead{SN} &
\colhead{r/R$_{25}$} &
\colhead{log$_{10}$( F([\ion{N}{ii}]~$\lambda$6584)/F(H$\alpha$) )}&
\colhead{log$_{10}$( F([\ion{O}{iii}]~$\lambda$5007)/F(H$\beta$) )}}
\startdata
2006aa   & 0.018    &    -0.436    &  -0.550*     \\
         & 0.304    &    -0.347    &  -0.464*     \\
         & 0.366    &    -0.434    &  -0.741*     \\                           
\hline
2002ao   & 0.067    &    -0.913    &   0.121      \\
         & 0.073    &    -0.818    &   0.438      \\               
         & 0.204    &    -1.012    &   0.297      \\
         & 0.315    &    -1.016    &   0.162      \\                 
\hline
2005la   & 0.022    &    -0.494    &  -0.178      \\
\enddata

\tablenotetext{*}{Upper limit}
\tablenotetext{***}{Excluded from the metallicity analysis after spectral classification with the BPT diagram.}

 \end{deluxetable}}

\clearpage

\begin{deluxetable}{l|l|l|l|l|l|l|l}
\tabletypesize{\scriptsize}
\tablewidth{0pt}
\tablecaption{Metallicity estimates and Normalised Cumulative Rank (NCR) pixel index for the sample of CSI transients.\label{metal}}
\tablehead{
\colhead{SN} &
\colhead{Type} &
\colhead{Data}&
\colhead{Reference}&
\colhead{12$+$log(O/H)}&
\colhead{\textbf{12$+$log(O/H)}}&
\colhead{12$+$log(O/H)}&
\colhead{NCR}\\
\colhead{}&
\colhead{}&
\colhead{source}&
\colhead{}&
\colhead{(N2 at the}&
\colhead{\textbf{(N2 at r$\mathbf{_{SN}}$})}&
\colhead{gradient}&
\colhead{}\\
\colhead{}&
\colhead{}&
\colhead{}&
\colhead{}&
\colhead{host center)}&
\colhead{}&
\colhead{(R$_{25}^{-1}$)}&
\colhead{}}
\startdata
 1987F   & IIn-88Z                & o & This work                  &8.67(0.07) & 8.47(0.18)     & -0.27(0.21)    &0.352 \\
 1988Z   & IIn-88Z                &o/a& This work + SDSS           &8.47(0.01) & 8.34(0.05)     & -0.23(0.08)    &  \ldots   \\
 1994W   & IIn-P(94W)             & o & This work                  &8.60(0.05) & 8.61(0.08)     & +0.05(0.21)    & 0.795  \\
 1994Y   & IIn-L(98S)             & l & KK12                       &\ldots     & 8.74(0.20)     & -0.47*             &0.000  \\
 1994ak  & IIn                    & l & H14$^c$&\ldots     & 8.74(0.20)     &\ldots              &0.000   \\
 1995G   & IIn-88Z                & o & This work                  &8.39(0.11) & 8.46(0.14)     & +0.14(0.15)    &  \ldots\\
 1995N   & IIn-88Z                & o & This work                  &\ldots     & 8.31(0.18)$^{\#}$     &\ldots   &SN   \\
 1997ab  & IIn-88Z                & l & KK12                       &8.35(0.18) & 8.20(0.20)     & -0.47*             & \ldots \\
\enddata
\tablenotetext{*}{The metallicity gradient is assumed to be -0.47~R$_{25}^{-1}$, from the average of the metallicity gradients presented by P04.}
\tablenotetext{**}{Metallicity from the N2 limit obtained from the spectrum of the closest \ion{H}{ii} region, as the gradient of the galaxy is difficult to obtain given that the host of SN~2005-OT is strongly titled and irregular.}
\tablenotetext{\dagger}{Metallicity obtained from the spectrum of the closest \ion{H}{ii} region, as the gradient of this large galaxy is difficult to obtain given that the host of V1 is larger than the ALFOSC field of view.}
\tablenotetext{\alpha}{Metallicity from the spectrum of the closest \ion{H}{ii} region as measured in S12, as we could not determine the emission line fluxes from our NOT spectra, where the SN is still detected.}
\tablenotetext{\dagger\dagger}{Metallicity from the spectrum of the closest \ion{H}{ii} region, see \citealp{tartaglia14}.}
\tablenotetext{\#}{Metallicity from the spectrum of the closest \ion{H}{ii} region, as SN~1995N sits in a irregular galaxy whose gradient is difficult to determine.}
\tablenotetext{c}{N2 metallicities obtained from O3N2 values converted to the N2 scale using the relation in \citet{kewley08}.}

\tablecomments{We assumed the metallicity gradients measured by P04 for the hosts of eight SN~IMs, and we checked them against our measurements, when available (see Fig.~\ref{allgradSNimpo}). In the last column, ``SN" indicates that the SN is still bright in the optical. All the NCR values are from H14.
For SNe~2002kg and 2003gm there are also metallicity measurements by \citet{maund06}, which are consistent with our results.} 

 \end{deluxetable}
\clearpage

\begin{deluxetable}{llllllllll}
\rotate
\tabletypesize{\scriptsize}
\tablewidth{0pt}
\tablecaption{Apparent ($u/U,B,V,R/r/unf.,I/i$) peak magnitudes, Galactic and host extinction for the SNe~IIn, Ibn and Ia-CSM of our sample. \label{tab:phot}}
\tablehead{
\colhead{SN} &
\colhead{Type} &
\colhead{$U_{\rm max}$}&
\colhead{$B_{\rm max}$}&
\colhead{$V_{\rm max}$}&
\colhead{$R_{\rm max}$}&
\colhead{$I_{\rm max}$}&
\colhead{$A_{V}(MW)$}&
\colhead{$A_{V}(h)$}&
\colhead{Reference}\\
\colhead{}&
\colhead{}&
\colhead{(mag)}&
\colhead{(mag)}&
\colhead{(mag)}&
\colhead{(mag)}&
\colhead{(mag)}&
\colhead{(mag)}&
\colhead{(mag)}&
\colhead{}}
\startdata
  1987F   & IIn-88Z             &   \ldots           &  $<$17.25	      &  $<$15.40	     & \ldots           &   \ldots              &  0.042    &  1.260         &  \citet{wegner96}, $A_{V}(h)$ from Balmer decrement		           \\
 1988Z   & IIn-88Z             &   \ldots           &  16.54	          &  16.23	         & 15.93	        &   \ldots              &  0.074    &  \ldots         &  \citet{turatto93,aretxaga99}		                                               \\
 1994W   & IIn-P(94W)          &   \ldots           &  14.21	          &  14.00	         & 14.00(0.50)          &   \ldots              &  0.049    &  0.527          &  \citet{tsvetkov95} $BV$, \citet{sollerman98} $R$ and $A_{V}(h)$		\\
 1994Y   & IIn-L(98S)          &  \ldots            & 14.20               &  14.20           & 14.20        & \ldots                & 0.027 &                  0     & \citet{tse94Y} \\
 1994ak  & IIn                 & \ldots             & \ldots           & \ldots        & $<$15.9(unf)      & \ldots              &  0.044    &  0             & \citet{rich94ak}             \\
 1995G   & IIn-88Z             & $<$15.96           & $<$16.33         & $<$16.00         & $<$15.73            & $<$15.56            &  0.146    &  0             & \citet{pastorello02}          \\
 1995N   & IIn-88Z             &   $<$18.80          &  $<$19.25	      &  $<$18.50	     & $<$17.00          &   $<$18.20             &  0.316    &   0.025        &  \citet{pastorello05},\citet{fransson02} $A_{V}(h)$		           \\
 1997ab  & IIn-88Z             &   \ldots           &  $<$14.70    &  \ldots          & \ldots           &   \ldots              &  0.063    &  \ldots         &  \citet{hagen97}		                                               \\
\enddata
 \end{deluxetable}

\clearpage

\begin{deluxetable}{llrrrrr}
\tabletypesize{\scriptsize}
\tablewidth{0pt}
\tablecaption{Absolute ($U/u,B,V,R/r/unf.,I/i$) peak magnitudes for the SNe~IIn, Ibn and Ia-CSM of our sample. \label{tab:abs}}
\tablehead{
\colhead{SN} &
\colhead{Type} &
\colhead{$U/u_{\rm max}$}&
\colhead{$B_{\rm max}$}&
\colhead{$V_{\rm max}$}&
\colhead{$R/r/unf._{\rm max}$}&
\colhead{$I/i_{\rm max}$}\\
\colhead{}&
\colhead{}&
\colhead{(mag)}&
\colhead{(mag)}&
\colhead{(mag)}&
\colhead{(mag)}&
\colhead{(mag)}}
\startdata
 1987F   & IIn-88Z              &         \ldots  &      $<-$17.38&      $<-$19.08  &        \ldots &         \ldots    \\
  1988Z   & IIn-88Z              &        \ldots  &       $-$18.40&         $-$18.68  &       $-$18.97 &         \ldots    \\
  1994W   & IIn-P(94W)           &        \ldots  &       $-$17.20&        $-$17.33  &        $-$17.31 &         \ldots    \\
  1994Y   & IIn-L(98S)           &        \ldots  &       $-$18.54&        $-$18.53  &       $-$18.52 &         \ldots    \\
  1994ak  & IIn                  &        \ldots  &        \ldots &         \ldots  &    $<-$16.82 &         \ldots    \\
  1995G   & IIn-88Z              &     $<-$18.37  &    $<-$17.97 &     $<-$18.25  &    $<$  $-$18.50 &     $<-$18.64    \\
  1995N   & IIn-88Z              &     $<-$13.68  &    $<-$13.16 &     $<-$13.80  &    $<-$15.26 &     $<-$13.99    \\
  1997ab  & IIn-88Z              &        \ldots  &    $<-$18.82 &         \ldots  &        \ldots &         \ldots    \\
\enddata
 \end{deluxetable}

\clearpage

\begin{deluxetable}{lllll}
\tabletypesize{\scriptsize}
\tablewidth{0pt}
\tablecaption{Wind velocity and mass-loss rate for a subsample of CSI~SNe (IIn, Ibn, Ia-CSM).\label{tab:spec}}
\tablehead{
\colhead{SN} &
\colhead{Type} &
\colhead{v$_w$}&
\colhead{$\dot{M}$}&
\colhead{Reference}\\
\colhead{}&
\colhead{}&
\colhead{(km~s$^{-1}$)}&
\colhead{(M$_{\odot}$~yr$^{-1}$)}&
\colhead{}}
\startdata
	1987F 	&    IIn-88Z	&	150         &     0.01                   &   \citet{wegner96}, \citet{chugai91}		                    \\
	1988Z 	&    IIn-88Z	&	$<$200	    &     0.0007$-$0.015	       &    \citet{stathakis91}, \citet{chugai94}		                \\
	1994W 	&    IIn-P(94W)	&	1000        &     0.3	                   &   \citet{sollerman98}, \citet{chugai04}		                \\
	1995G   &    IIn-88Z    &  1000         &    0.1                     & \citet{pastorello02}, \citet{kiewe12}, \citet{chugai03}\\
	1995N 	&    IIn-88Z	&	$<$500      &     0.0002                 &   \citet{fransson02}, \citet{zampieri05}		                \\
	1997ab	&    IIn-88Z	&	90          &     0.01                   &   \citet{salamanca98}		                                    \\
	1998S 	&    IIn-L(98S)	&	30$-$100	&     0.0001	               &    Table 9 in \citet{kiewe12}, \citet{anupama01}\tablenotemark{*}       \\
	2003ma	&    IIn-88Z	&	110         &     0.13                   &   \citet{rest11}                                               \\
\enddata
\tablenotetext{*}{\citet{fassia01} report a mass-loss rate of 2$\times$10$^{-5}$~M$_{\odot}$~yr$^{-1}$ for the outer CSM and 3$\times$10$^{-3}$~M$_{\odot}$~yr$^{-1}$ for the inner CSM. The value reported by \citet{anupama01} is in the middle of this range.}
 \end{deluxetable}

\clearpage


\end{document}